\documentclass[%
 reprint, 
 amsmath,amssymb,
 aps,
pra,
floatfix,
]{revtex4-2}

\usepackage{graphicx}
\usepackage{epsfig} 
\usepackage{dcolumn}
\usepackage{bm}
\usepackage{epstopdf}
\usepackage{multirow}
\usepackage{amsmath}
\usepackage{booktabs}
\usepackage{color}
\usepackage{gensymb}
\usepackage{kantlipsum}
\usepackage{hyperref}
\usepackage{braket}
\usepackage{physics}
\usepackage{url}
\usepackage[utf8]{inputenc}
\usepackage[T1]{fontenc}
\usepackage{mathptmx}
\usepackage{lineno}
\usepackage[dvipsnames]{xcolor}
\usepackage[normalem]{ulem}
\usepackage{makecell}

\renewcommand{\figurename}{\textbf{Fig.}}

\renewcommand{\tablename}{\textbf{Table}}

\begin{document}

\title{Trapping of Single Atoms in Metasurface Optical Tweezer Arrays}
\author{Aaron Holman$^{1}$}
\thanks{These authors contributed equally.}
\author{Yuan Xu$^{2}$}
\thanks{These authors contributed equally.}
\author{Ximo Sun$^{1}$}
\author{Jiahao Wu$^{2}$}
\author{Mingxuan Wang$^{1}$}
\author{Zezheng Zhu$^{2}$}
\author{Bojeong Seo$^{1}$}
\author{Nanfang Yu$^{2}$}
\email{ny2214@columbia.edu}
\author{Sebastian Will$^{1}$}
\email{sebastian.will@columbia.edu}

\affiliation{$^{1}$Department of Physics, Columbia University, New York, New York 10027, USA}
\affiliation{$^{2}$Department of Applied Physics and Applied Mathematics, Columbia University, New York, New York 10027, USA}

\date{\today}

\begin{abstract}
Optical tweezer arrays have emerged as a key experimental platform for quantum computation, quantum simulation, and quantum metrology, enabling unprecedented levels of control over single atoms and molecules. However, existing tweezer platforms have fundamental limitations in array geometry, size, and scalability. Here we demonstrate the trapping of single strontium atoms in optical tweezer arrays generated via holographic metasurfaces. We realize two dimensional arrays with more than 1000 trapped atoms, arranged in arbitrary geometries with trap spacings as small as 1.5 \textmu m. The arrays have a high uniformity in terms of trap depth, trap frequency, and positional accuracy, rivaling or surpassing existing approaches. This is enabled by highly efficient holographic metasurfaces fabricated from high-refractive index materials, silicon-rich silicon nitride and titanium dioxide.  Leveraging sub-micrometer pixel sizes and high pixel densities, our platform allows scaling far beyond current capabilities. As a demonstration, we realize an optical tweezer array with 360,000 traps. These advances will facilitate tweezer-array based quantum applications that require large system sizes.
\end{abstract}

\maketitle

Optical tweezer arrays have led to a revolution in the control of ultracold atoms and molecules for quantum applications~\cite{Kaufman2021quantum}. They have broken new ground for quantum simulation~\cite{browaeys2020many} and quantum computation~\cite{Saffman2010quantum, morgado2021quantum}, including the realization of quantum spin systems~\cite{scholl2021quantum, semeghini2021probing}, high-fidelity Rydberg quantum gates~\cite{madjarov2020high, graham2022multi, ma2023high}, and first steps towards error-corrected quantum computation~\cite{bluvstein2024logical}. Recently, dual species atomic arrays~\cite{Singh2022dual,Sheng2022defect} and arrays of dipolar molecules have also been reported~\cite{Zhang2022optical,bao2023dipolar,holland2023demand}. Optical tweezer arrays enable new approaches for quantum metrology, such as optical tweezer clocks~\cite{madjarov2019atomic, young2020half}, and hold great promise for novel experiments in quantum optics, including explorations in cavity quantum electrodynamics~\cite{Yan2023superradiant} and correlated atom-photon interactions~\cite{AsenjoGarcia2017exponential, Holzinger2024harnessing, Masson2024dicke}.

\begin{figure*}
    \centering
    \includegraphics[width=\textwidth]{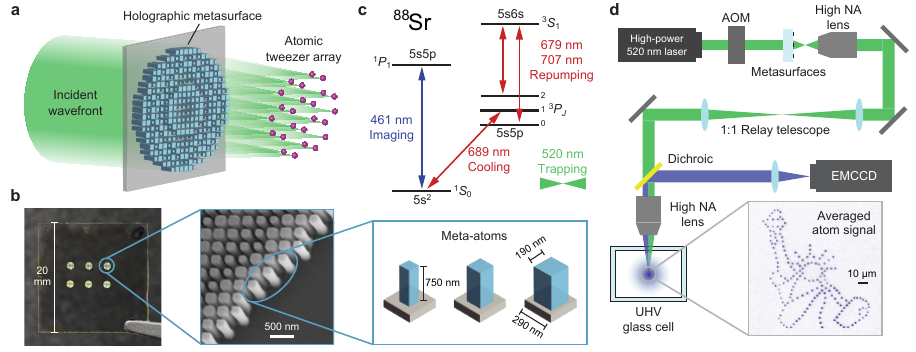}
    \caption{\textbf{Metasurface optical tweezer arrays and integration with ultracold strontium atoms.} \textbf{a,} Illustration of the working principle of a holographic metasurface. The metasurface imprints a phase pattern on an incident collimated Gaussian laser beam and produces a two-dimensional array of tight tweezer traps at the focal plane. In this work, the focal length of the metasurfaces is 0.7 mm and the effective NA is $>0.6$. \textbf{b,} (left) Photograph of a substrate holding 6 metasurfaces, each with a distinct tweezer array pattern. The substrate has a size of 20 mm $\times$ 20 mm. The metasurfaces have a diameter of 1.16 mm. (middle) Scanning electron microscope image of a portion of a metasurface. (right) The meta-atoms have a unit cell size of 290 nm, are 100--190 nm wide, and 750 nm tall. They are chosen from a predetermined library to introduce a wide range of phase delays to the incoming beam without modifying its amplitude. \textbf{c,} Level diagram of $^{88}$Sr showing the optical transitions relevant to this work. Atoms are cooled and imaged on the broad transition at 461 nm in conjunction with the repumpers at 679 nm and 707 nm. Narrow-line cooling on the intercombination line at 689 nm (linewidth 7.5 kHz) creates Sr samples at microkelvin temperatures. \textbf{d,} Schematic of the setup to trap and image atoms. The tweezer laser operates at 520 nm. Its intensity is controlled with an acousto-optic modulator (AOM) before illuminating the metasurface. The metasurface generates and focuses the tweezer array, which is collimated by a high-NA microscope objective, relayed via a 1:1 telescope, and focused down with a high-NA objective into the ultra-high vacuum glass cell to trap atoms. During imaging, fluorescence photons of the atoms are detected with a single-photon sensitive camera.}
    \label{fig:1}
\end{figure*}

The ability to generate high-quality optical tweezer arrays is a central requirement for many experiments. A tweezer array consists of numerous tightly focused laser beams, each constituting a trap for a single particle. Key criteria for the platform include high flexibility in array geometry, trap uniformity, and scalability. In addition, compactness, robustness, and high optical efficiency are desirable, especially with the prospect of deploying tweezer-based quantum devices outside of controlled laboratory environments~\cite{Grotti2018geodesy, Takamoto2020test, Elliott2023quantum}. Currently, optical tweezer arrays are mostly generated via active beam-shaping devices, such as acousto-optical deflectors (AODs)~\cite{endres2016atom, Burgers2022controlling}, liquid crystal-spatial light modulators (SLMs)~\cite{Barredo2016atom, Kim2019large}, or digital micromirror devices (DMDs)~\cite{Wang2020preparation}. These devices require complex control electronics and projection optics with high numerical aperture (NA) to relay the tweezer arrays onto ultracold atoms and molecules. Technical complexity and fundamental limitations constrain array sizes to $\sim10,000$ traps~\cite{manetsch2024tweezer}, which has started to impose a limit on the quantum applications that can be pursued. Alternative techniques, such as amplitude masks~\cite{huft2022simple} and microlens arrays~\cite{pause2024supercharged}, have been explored, but limited beam-shaping capabilities make it challenging to achieve highly uniform arrays.

In recent work~\cite{huang2023metasurface}, we proposed holographic metasurfaces as a new approach to generate versatile and scalable tweezer arrays. Metasurfaces are flat optical devices comprised of sub-micrometer pixels that can imprint an arbitrary phase mask onto an incident laser beam~\cite{fong2010scalar, yu2011light, ni2012broadband}, both generating and focusing an optical tweezer array. Metasurfaces feature high power-handling capabilities~\cite{atikian2022diamond}, diffraction-limited focusing~\cite{arbabi2015subwavelength, capasso16tio2}, and comprehensive polarization control~\cite{balthasarmueller2017metasurface, huang2023int}. A recent experiment demonstrated single atom trapping in a $3 \times 3$ tweezer array that was generated with an AOD and then focused down by a metasurface lens~\cite{hsu2022single}; however, a demonstration of atomic tweezer arrays that utilize the full beam-shaping capabilities of metasurfaces -- integrating array generation and focusing into one device -- has so far been elusive.

In this work, we demonstrate the trapping of single strontium (Sr) atoms in metasurface optical tweezer arrays. Using laser light at a wavelength of 520 nm, we realize two-dimensional atomic arrays with arbitrary geometries, both periodic and non-periodic, with more than 1000 trapped atoms and trap spacings as small as 1.5 \textmu m. We find the trap uniformity to be comparable to state-of-the-art techniques and demonstrate single-atom preparation and detection with high fidelity. We discuss how large-area holographic metasurfaces with subwavelength pixel sizes offer a realistic path towards tweezer arrays with >100,000 atoms. Finally, we experimentally demonstrate a highly uniform optical tweezer array with 360,000 traps, exceeding the current state-of-the-art by two orders of magnitude.

\section*{Metasurfaces for optical tweezers}

Metasurfaces have emerged as a powerful platform for the manipulation of optical waves~\cite{yu2014flat, chen2020flat}. They enable holographic control of incident light fields by manipulating the amplitude and phase of an optical wavefront in the metasurface plane. Our metasurfaces are composed of nanofabricated meta-atoms, each a dielectric nanopillar, a few hundred nanometers in width and height, smaller than the wavelength of the light they manipulate. Meta-atoms with different shapes and sizes are positioned in a two-dimensional (2D) grid with subwavelength spacing to engineer the optical wavefront and generate the desired optical intensity pattern in the imaging plane. Metasurfaces can be fabricated on the millimeter to centimeter scale~\cite{kildishev2013planar, Park2024all} with well over $10^6$ pixels. Due to the subwavelength size and spacing of the meta-atoms, metasurfaces can achieve high optical efficiency and high precision of the generated tweezer pattern.

We use transmitting metasurfaces for visible light at 520 nm (Fig.~\ref{fig:1}\textbf{a}). These metasurfaces are phase-only modulating masks designed using a Gerchberg-Saxton algorithm-based optimization approach (see Methods for details). The optimized metasurface encodes a phase pattern that simultaneously generates and focuses a tweezer array. Multiple metasurfaces are placed on a single substrate (Fig.~\ref{fig:1}\textbf{b}), allowing for easy switching between distinct tweezer arrays by translation of the substrate. The metasurfaces are implemented on two complementary dielectric platforms: silicon-rich silicon nitride (SRN) - allowing for fast CMOS-compatible fabrication - and titanium dioxide ($\text{TiO}_2$) - featuring superior power handling and compatibility with shorter optical wavelengths. The metasurfaces are designed assuming the incidence of a flat phase front and have a circular footprint to reduce diffraction effects of round input beams at their boundaries. The metasurfaces have diameters ranging from 1.2 mm to 3.5 mm. They can handle optical intensities of at least 25 W/mm$^2$ (SRN) or 2,000 W/mm$^2$ ($\text{TiO}_2$) without active cooling, have a diffraction efficiency of $\sim 60 \%$, and an effective NA of $>0.6$. Additional details on the design and optimization of the metasurface hologram, meta-atom library, cleanroom fabrication procedure, and a comparison of holographic metasurfaces with AODs and SLMs are provided in the Methods.

\begin{figure}
\centering
\includegraphics[width=\columnwidth]{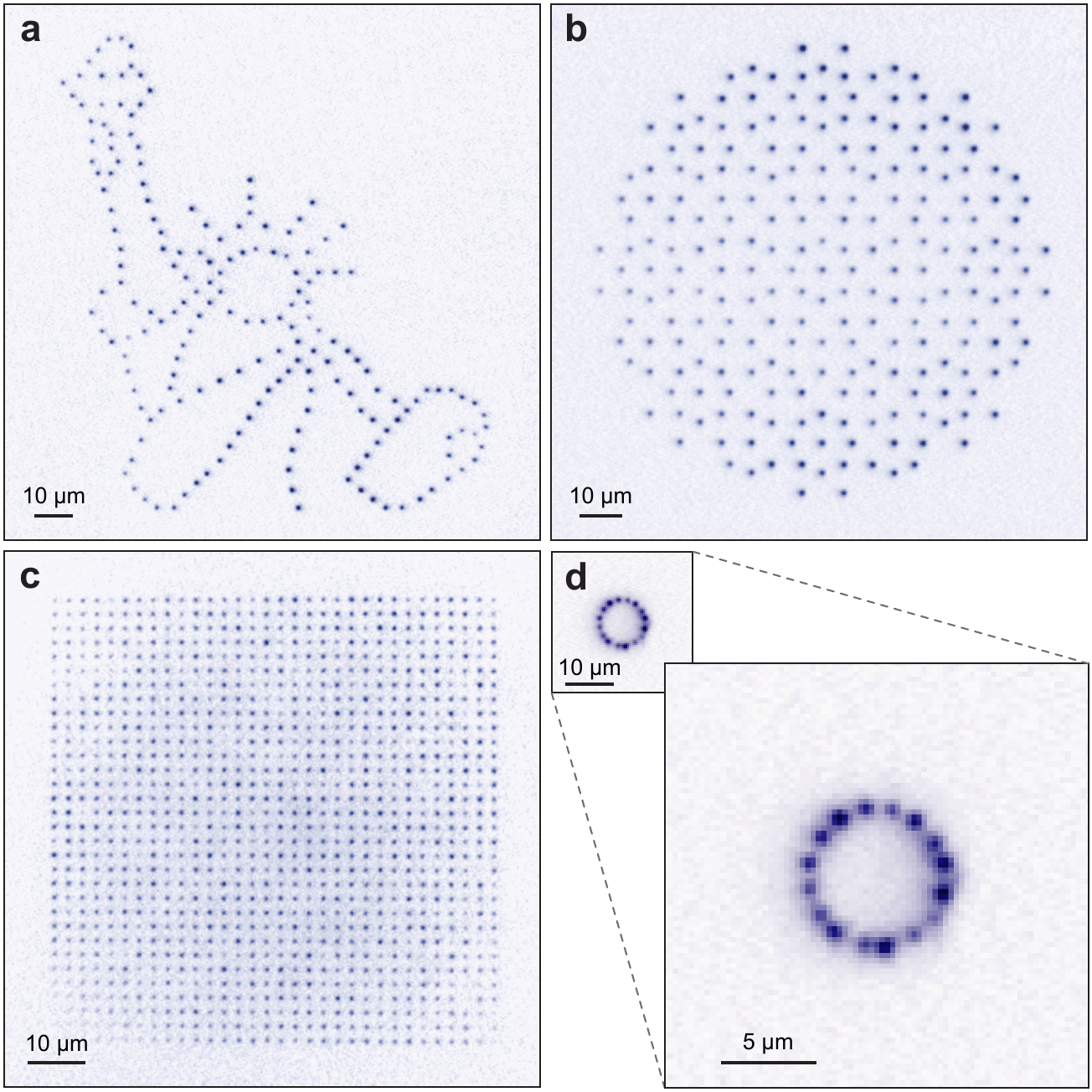}
\caption{\textbf{Fluorescence images of strontium atoms trapped in metasurface-generated optical tweezer arrays.} Each image is an average of 100 individual images without parity projection. \textbf{a,} Statue of Liberty pattern with 183 traps, average spacing of 3 \textmu m. \textbf{b,} Quasicrystal (Penrose tiling) with 225 traps, average spacing of 4 \textmu m. \textbf{c,} Square lattice with 1024 traps, average spacing of 2.5 \textmu m. \textbf{d,} Necklace pattern with 16 traps, average spacing of 1.45 \textmu m.}
\label{fig:2}
\end{figure}

The optical tweezer arrays are projected into the glass cell of an ultrahigh-vacuum chamber (see Methods for details). A schematic of the projection system is shown in Fig.~\ref{fig:1}\textbf{d}. Before illuminating the metasurface, a laser at 520 nm passes through an acousto-optic modulator (AOM) for fast switching and trap depth control. The substrate with the metasurfaces is mounted on a two-axis translation stage, allowing for rapid switching between different array geometries with minimal realignment. The tweezer array, generated by the metasurface at the focal plane, is converted into the optical momentum space by a microscope lens ($\text{NA}= 0.6$), relayed through a 1:1 telescope, and converted back into the tweezer array in the glass cell by an objective lens ($\text{NA}= 0.5$). Tweezer generation with AODs, SLMs, or DMDs typically requires demagnification optics with large-diameter lenses that are prone to aberrations and impose power-handling and field-of-view limitations; in contrast, for our metasurfaces demagnification optics are not necessary due to their intrinsically high NA. In principle, metasurfaces can directly trap atoms at their focal plane (e.g.,~by placing them inside or near the vacuum chamber) without the need for additional relay optics.

We load atoms into the metasurface optical tweezer array from an ultracold cloud of $^{88}$Sr. The atoms are cooled to microkelvin temperatures using standard techniques that leverage strontium's unique level structure (Fig.~\ref{fig:1}\textbf{c}). Subsequently, the trapped atoms are detected via fluorescence imaging (see Methods). Figure~\ref{fig:2} shows fluorescence images of atoms in different metasurface-generated tweezer arrays including a fully arbitrary pattern (Statue of Liberty), a quasicrystal pattern (Penrose tiling), a periodic $32 \times 32$ square lattice pattern, and a necklace pattern with close tweezer spacings on the micrometer-scale. The array sizes are limited to a few hundred trapped atoms solely by the available tweezer laser power of about 1 W in our current setup.

\section*{Single atom trapping and characterization}

\begin{figure}
\centering
\includegraphics[width=\columnwidth]{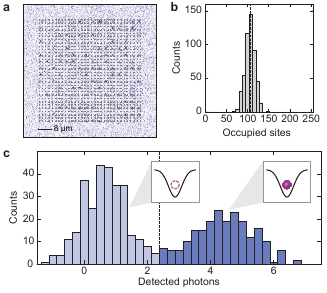}
\caption{\textbf{Single atom preparation and detection in a $\mathbf{16 \times 16}$ metasurface array.} \textbf{a,} Individual fluorescence image. Trap locations are indicated by dashed boxes. \textbf{b,} Histogram of the number of occupied sites after parity projection; the mean occupancy of the array is 106(11), as marked by the dashed line, corresponding to a mean trap occupancy of 41(4)\%. \textbf{c,} Histogram of photon counts for a typical trap in the $16 \times 16$ array, as marked in \textbf{a}. 500 repetitions of the experiment are averaged. The data allows for the distinction between one and zero atoms with high fidelity. The dashed line marks the threshold value, determined as discussed in Methods.}
\label{fig:3}
\end{figure}

We demonstrate single-atom trapping and detection in a $16\times16$ metasurface array. The steps to achieve this are highly sensitive to the quality and uniformity of the optical tweezer potential. The initial loading of the array is statistical; each trap is occupied by at least one atom, but the precise number of atoms is random (see Fig.~\ref{fig:3}\textbf{a}). In the next step, we perform parity projection: traps that initially have an odd (even) number of atoms are turned into sites with one (no) atom. This is achieved via photoassociation into an electronically excited Sr$_2$ molecular state, close to the 689 nm atomic resonance, which induces pairwise atom loss~\cite{Zelevinsky2006narrow}. After parity projection, 41(4)\% of the traps contain an atom, corresponding to $>100$ single atoms in the array, as shown in Fig.~\ref{fig:3}\textbf{b}.

To determine the occupation in the array, we perform fluorescence imaging on the 461 nm transition, while keeping the atoms trapped. Fluorescence photons are collected with a low-noise camera and the photon number in the trap locations is evaluated. To maximize the number of photons scattered per atom, we simultaneously cool via repulsive Sisyphus cooling~\cite{Cooper2018alkaline} on the 689 nm, $m_{J}=\pm1$ transition, counteracting the recoil heating from repeated photon scattering on the 461 nm transition. Figure~\ref{fig:3}\textbf{c} shows a histogram of the detected photon numbers for a typical trap in the array. The histogram shows two peaks: one peak centered on zero photons, corresponding to zero atoms, and a second peak centered on $\sim 4.5$ photons, corresponding to a single atom. The absence of photon counts above the single-atom peak indicates the high efficiency of parity projection. The presence of photon counts between the zero- and single-atom peaks results from loss of Sr atoms during imaging in the 520-nm traps. Similar observations were reported in Ref.~\cite{gyger2024continuous}. We attribute this loss to an ionization process out of the $^3\text{P}_1$ state, which is populated during Sisyphus cooling. This mechanism will be analyzed in further detail in future work. For alternative trapping wavelengths, for example 813 nm~\cite{covey20192000}, such losses are known to be absent. We determine the imaging fidelity to be $>95(3) \%$. When using a smaller $4\times4$ array, not limited by tweezer laser power, we observe a filling fraction of $49(3) \%$ and an imaging fidelity of $99.8(5) \%$ (see Methods).

\begin{figure}
\centering
\includegraphics[width=\columnwidth]{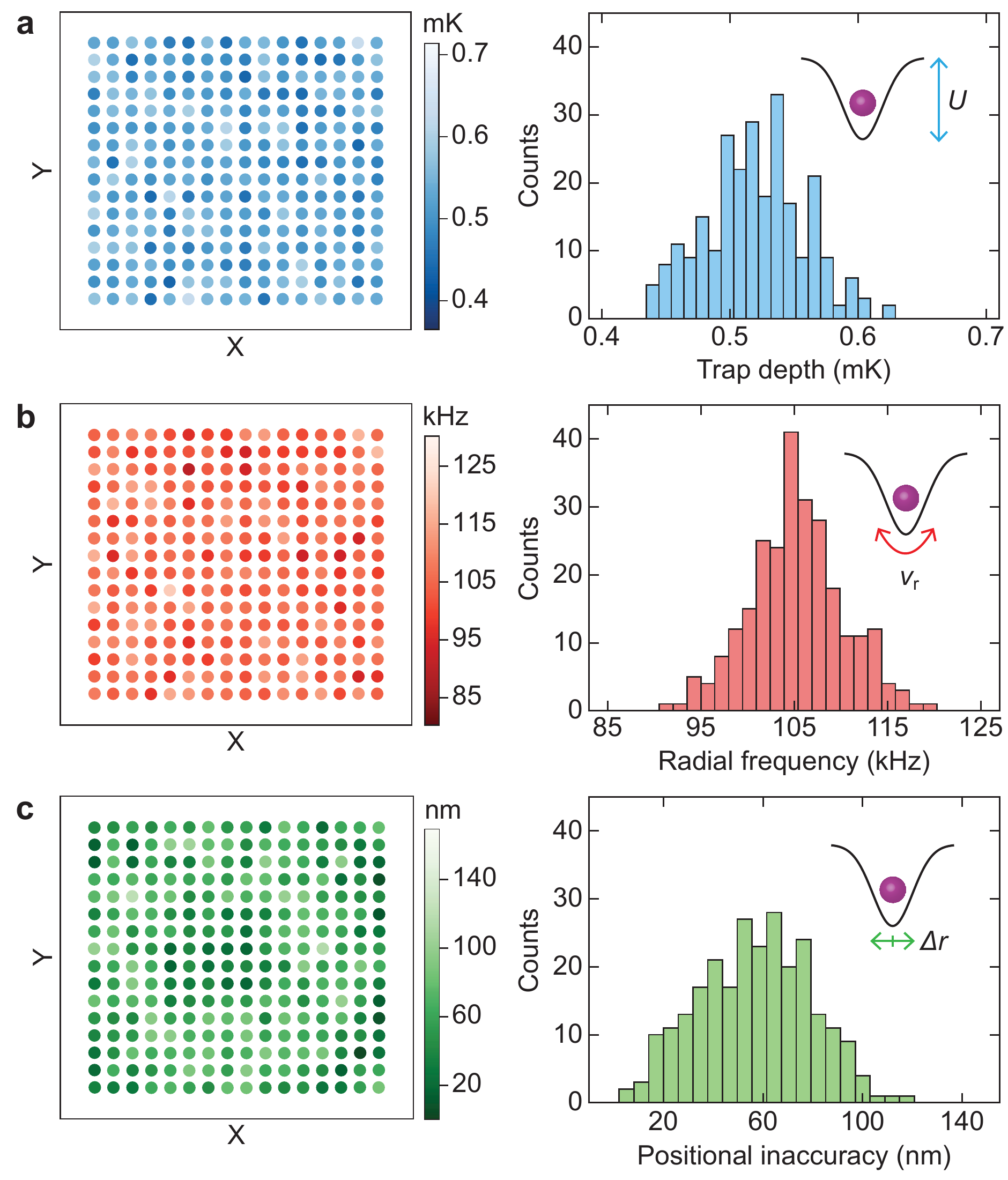}
\caption{\textbf{Characterizing the uniformity of metasurface arrays via atomic response.} A square array with $16\times16$ traps and a 4 \textmu m trap spacing is used. The uniformity in terms of \textbf{a,} trap depth, $U$, \textbf{b,} radial trap frequency, $\nu_{\mathrm{r}}$, and \textbf{c,} trap positions is characterized for each trap in the array (left column). Histograms of the respective quantities observed for each trap (right column).}
\label{fig:4}
\end{figure}

Next, we characterize the uniformity of the $16\times16$ array. A high uniformity in trap depth and frequency ensures that the light shift and on-site vibrational modes are constant across the array; a high accuracy of trap positions is desirable for the precise control of atom-atom interactions. We use the trapped atoms as highly sensitive probes to measure the depth, frequency, and position of each trap. Additional details on the measurements are provided in the Methods. Fig.~\ref{fig:4} shows the results of our characterization. For the trap depth and the radial (axial) trap frequencies we find a standard deviation of $7.5 \%$ and $5 \%$ $(8 \%)$, respectively, across the array. The positional inaccuracy is on the $1.5\%$ level compared to the trap spacing of 4 \textmu m, similar to the extent of the in-trap vibrational wavefunction of Sr.

The data shows that the uniformity of our metasurface arrays rivals or surpasses the performance of existing techniques. For example, arrays generated with liquid crystal SLMs often have trap depth fluctuations of {10\%}~\cite{nogrette2014single, manetsch2024tweezer}. Methods to reduce such fluctuations via feedback are under active research~\cite{Kim2019large, schymik2022situ, chew2024ultra}, but are fundamentally limited for large-scale arrays as discussed in the next section. For the metasurface arrays, we attribute the remaining non-uniformity to imperfections in the relay optics and to fabrication errors of meta-atoms, both of which have considerable room for improvement. Prior to transmission through the relay optics, we measure an intensity non-uniformity at the focal plane of the metasurface as low as 4\% (see Methods).

\begin{figure*}
\centering
\includegraphics[width=0.81\textwidth]{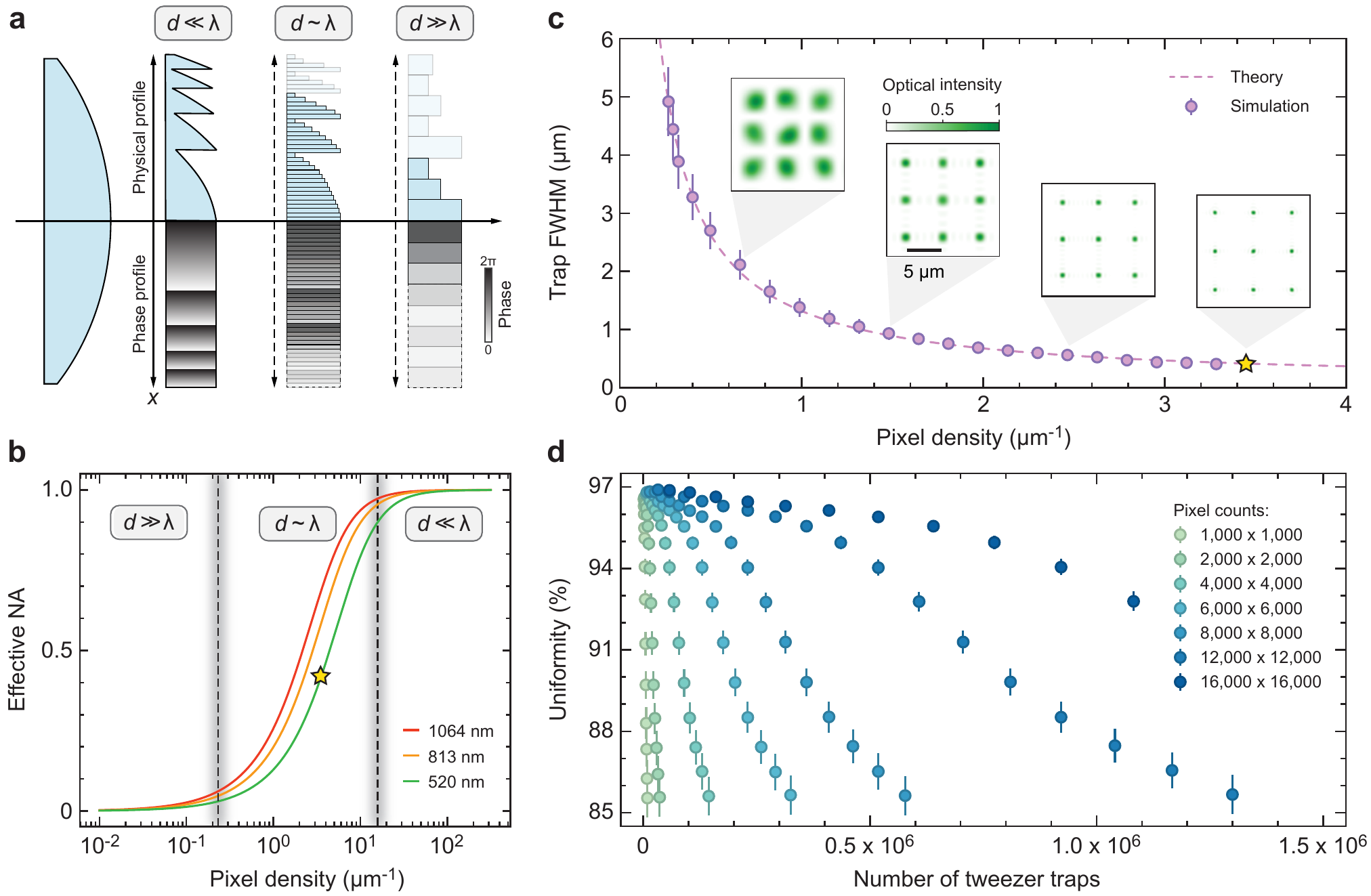}
\caption{\textbf{Performance of pixel-based beam shaping devices.} \textbf{a,} Approximation of the phase profile of a spherical lens with infinitely small pixel size ($d \ll \lambda$), intermediate pixel size ($d \sim \lambda$), and large pixel size ($d \gg \lambda$). For larger pixel sizes, the reproduction of steeper phase gradients $\partial \phi / \partial x$ is limited. This limits the usable diameter of the device and reduces the effective NA. \textbf{b,} Effective NA of a lens generated with a pixel-based device as a function of pixel density $1/d$ for common trapping wavelengths $\lambda$. The dashed vertical lines indicate the approximate separation between the regimes $d \ll \lambda$, $d \sim \lambda$, and $d \gg \lambda$. The yellow star indicates the pixel density for the metasurfaces used in this work. \textbf{c,} Focusing capabilities of pixel-based beam shaping devices. For a laser wavelength of 520 nm, a fixed device resolution of $300 \times 300$ pixels, and varying pixel density $1/d$, an optimized $3 \times 3$ square array with 5 \textmu m spacing is generated (insets) (further details in Methods). Data points show the tightness of the traps, measured as the FWHM. Error bars show the standard deviation across the array. The pixel densities range from state-of-the-art liquid crystal SLMs ($d=4$ \textmu m) to the holographic metasurface used in this work (yellow star, $d=290$ nm). The dashed line shows a fit of the effective NA model (further details in Methods). \textbf{d,} Simulation of the uniformity of trap intensity as a function of the number of tweezer traps for device pixel counts ranging from $1,000 \times 1,000$ (light green) to $16,000 \times 16,000$ (dark blue) pixels (square-shaped device). Uniformity is defined as 100\% minus the standard deviation of the trap intensity across the array (in \%). The simulation assumes a pixel size of $d= 290$ nm (further details in Methods).}
\label{fig:5}
\end{figure*}

\begin{figure*}
\centering
\includegraphics[width=0.85\textwidth]{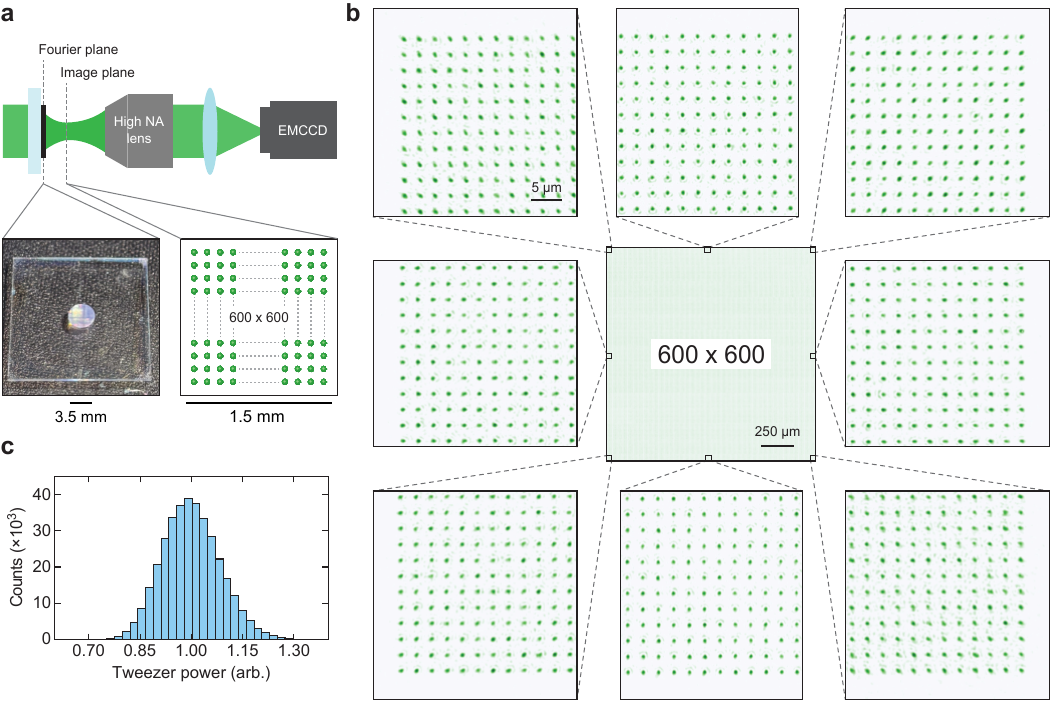}
\caption{\textbf{Realization and characterization of a $\mathbf{600 \times 600}$ tweezer array.} \textbf{a,} (top) Setup for optical characterization including a high NA (0.85) imaging objective. (bottom left) A 3.5 mm diameter TiO$_2$ metasurface in the Fourier plane generates a $600 \times 600$ trap array in the imaging plane, spanning a 1.5 mm $\times$ 1.5 mm area, imaged on a CCD camera. \textbf{b,} Full image of the array, stitched together from 126 individual high-resolution images (see SI). Here, 8 high-resolution images of the edges are shown, illustrating the uniformity and quality of the traps in the array. \textbf{c,} Histogram of the optical power in individual tweezers. The power is determined by summing the intensity in a region around each trap center.}
\label{fig:6}
\end{figure*}

\section*{Scalability of metasurface optical tweezer arrays}

In this section, we discuss the scalability of metasurface optical tweezer arrays and show how their performance compares to other pixel-based beam shaping devices, such as DMDs and liquid crystal SLMs. We find that small pixel sizes and a large pixel number, as provided by metasurfaces, are critically important to enable large and uniform tweezer arrays.

To illustrate this point, we first consider reproducing a simple lens with a pixel-based device, as shown in Fig.~\ref{fig:5}\textbf{a}. The pixel size is denoted by $d$, the pixel density is $1/d$, and the wavelength of the incident laser beam is denoted by $\lambda$. We first convert a spherical lens to a Fresnel lens by taking the phase profile $\phi(x)$ of the former, modulo $2 \pi$. When the phase profile of the Fresnel lens is approximated with a pixel-based device, the finite sampling constrains the steepness of phase gradients that can be reproduced, $\partial \phi/ \partial x \lesssim \pi/2d$. This limits the ability to reproduce steep phase gradients at the edge of the lens. A larger pixel size will reduce the attainable phase gradient, reducing the usable diameter of the device, and effectively reducing the NA that can be achieved. Based on this argument, we derive an approximate analytical expression for the effective numerical aperture attainable for a given pixel-size-to-wavelength ratio, $\mathrm{NA}= 1/\sqrt{(1+(4d/\lambda)^{2})}$ (see Methods for details). In Fig.~\ref{fig:5}\textbf{b}, we plot the effective NA for a broad range of pixel densities and several common tweezer wavelengths in the visible and near-infrared, illustrating that a smaller pixel size relative to the laser wavelength can lead to a dramatically increased effective NA. For a pixel size of several \textmu m (typical for DMDs and liquid cyrstal SLMs), in the regime $d \gg \lambda$, the NA is limited to below 0.05. For subwavelength pixels, in the regime $d \lesssim \lambda$, NAs of 0.5 and higher can be reached.

Going beyond the example of a lens, we simulate optical tweezer arrays that can be generated with a pixel-based device (see Fig.~\ref{fig:5}\textbf{c}). For a smaller pixel size $d$, the effective NA attainable is higher, and thus the individual tweezers can be focused more tightly. This allows the metasurfaces to accommodate more tweezer traps in the same area. In a series of simulations, we vary the pixel size $d$, but keep the device pixel count at $300 \times 300$, while optimizing the phase mask to generate a $3 \times 3$ square array with 5 \textmu m spacing in the focal plane (see Methods). As shown in Fig.~\ref{fig:5}\textbf{c}, the full-width-half-maximum (FWHM) of the traps steadily decreases as the pixel density $1/d$ increases. As a result, holographic metasurfaces with subwavelength pixels can generate optical tweezer arrays that are sufficiently tight at the focal plane for direct trapping of atoms, while devices with larger pixels require additional demagnification optics.

The small pixel size of metasurfaces also allows a large number of pixels to be accommodated within a compact device footprint. This is advantageous as the number of high-quality traps that can be generated is positively correlated with the number of pixels. To quantify this relation, we consider devices with pixel counts ranging from $1,000 \times 1,000$ $(10^6)$ to $16,000 \times 16,000$ $(256 \times 10^6)$ and investigate the uniformity of trap intensity across the array as a function of the number of tweezer traps (see Methods). The results in Fig.~\ref{fig:5}\textbf{d} show that for a fixed pixel count, the uniformity drops monotonically when the number of traps increases. The data suggests a rule-of-thumb that $\sim 300$ pixels are needed to produce one high-quality tweezer trap in an array \footnote{It is important to note this is a fundamental limitation that cannot be improved upon, for example, via active feedback in SLMs or DMDs.}. For example, the number of highly uniform (uniformity $>95 \%$) tweezer traps that can be generated by a top-end SLM with $4000 \times 4000$ pixels will be fundamentally limited to $\sim 50,000$; in contrast, the pixel count of metasurfaces can be well beyond $8000 \times 8000$, allowing the creation of arrays with $>200,000$ traps, provided that sufficient laser power is available.

\section*{360,000 Tweezer Array}

As a demonstration of the high scalability of metasurface arrays, we experimentally realize an array with 360,000 tweezer traps (see Fig.~\ref{fig:6}). The traps are arranged on a $600 \times 600$ square lattice with a spacing of 2.5 \textmu m between nearest-neighbor sites. The metasurface has a diameter of $3.5$ mm, contains approximately 114 million pixels, and is made of titanium dioxide, offering exceptional power-handling capabilities (see Methods).

We record a high-resolution image of the full array and characterize the uniformity of the tweezer intensities. Using the setup shown in Fig.~\ref{fig:6}\textbf{a}, which has a high-resolution field of view of about 190 x 140 \textmu m, we raster scan the array, which has a real space size of 1.5 mm x 1.5 mm, and stich together a composite image from more than 100 individual images. The full image frame, with a $50 \times$ 4K resolution, is provided in the SI. Exemplary high-resolution images of the edges and corners of the array are shown in Fig.~\ref{fig:6}\textbf{b}, demonstrating the high quality and uniformity of the tweezers across the array. From a quantitative analysis of the full array, we find a uniformity of the tweezer power of $92 \%$ (Fig.~\ref{fig:6}\textbf{c}). With a moderate increase of metasurface size, arrays with $>$1,000,000 traps are within realistic reach.

\section*{Conclusions}

In this work, we have demonstrated single-atom trapping in metasurface optical tweezer arrays. We show that the uniformity of the arrays is comparable to that realized using existing methods; additional improvements are expected by optimizing fabrication and system integration. Because of subwavelength pixel sizes, holographic metasurfaces can reach a high effective NA. This allows for the creation of tightly focused tweezer arrays at the metasurface's focal plane, enabling direct trapping of atoms without demagnification or relay optics. With their high pixel counts~\cite{Park2024all} and outstanding power-handling capabilities~\cite{atikian2022diamond}, metasurfaces offer a realistic path towards tweezer arrays with more than $100,000$ atoms. This addresses a critical need for future applications in quantum simulation, quantum computing, quantum sensing, and optical clocks based on atomic arrays.

We envision future extensions of the metasurface optical tweezer array platform. By combining the static arrays demonstrated here with a dynamic sorting beam, it will become possible to rearrange atoms and create unity filled arrays. In addition, metasurfaces can be functionalized in various ways: resonant metasurfaces~\cite{malek2022non} can be designed to only impart a phase pattern on light within a narrow spectral band, while leaving wavelengths outside of this band unchanged. This will enable the creation of multifunctional metasurfaces through which atomic arrays can be generated, sorted, and imaged. In addition, by employing wavelength or polarization multiplexing ~\cite{huang2023metasurface}, metasurfaces can be designed to produce the same tweezer array pattern at different wavelengths, which may prove useful for both single- and dual-species atomic systems ~\cite{Singh2022dual,Sheng2022defect}. Finally, the development of active metasurfaces that allow for feedback and rearrangement of atomic arrays constitute an intriguing frontier for future research~\cite{Shaltout2019}.

\section*{Acknowledgments}
We are grateful to Xiaoyan Huang and Weijun Yuan for contributions in the early development of this project and Stephanie C. Malek for help with the development of the SRN material platform. We thank Chun-Wei Liu and Siwei Zhang for experimental assistance, and Dmytro Filin and Marianna Safronova for providing data on the optical polarizability of Sr atoms. We thank Ana Asenjo-Garcia, Stuart Masson, and Ricardo Gutierrez-Jauregui for fruitful discussions, and Tarik Yefsah for critical reading of the manuscript. This work was supported by the National Science Foundation (Award nos.~1936359, 2040702, and 2004685) and the Air Force Office of Scientific Research (Award nos.~FA9550-16-1-0322 and FA9550-23-1-0404). Device fabrication was carried out at the Columbia Nano Initiative cleanroom, at the Advanced Science Research Center Nanofabrication Facility at the Graduate Center of the City University of New York, and at the Center for Functional Nanomaterials, Brookhaven National Laboratory, supported by the US Department of Energy, Office of Basic Energy Sciences (Contract no. DESC0012704). B.S.~acknowledges support from the National Research Foundation of Korea (Award no.~2021M3H3A1036573). S.W.~acknowledges support from the Alfred P. Sloan Foundation. N.Y. ~acknowledges the Gordon and Betty Moore Foundation Experimental Physics Investigators (EPI) initiative (Award no. 11561).

\section*{Author contributions}
All authors contributed substantially to the work presented in this paper. A.H., X.S., M.W., and B.S.~carried out the atomic experiments. Y.X., J.W., and Z.Z.~designed and fabricated the metasurfaces. N.Y.~and S.W.~supervised the study. All authors contributed to the data analysis and writing of the paper.

\clearpage
\setcounter{figure}{0}
\renewcommand{\figurename}{\textbf{Extended Data Fig.}}
\renewcommand{\tablename}{\textbf{Extended Data Table}}

\section*{Methods}
\appendix

\section*{Material platforms for visible‑spectrum metasurfaces}\label{sec:material}
For the use of metasurfaces in atom trapping, it is beneficial to have several photonic material platforms available to accommodate different laser wavelength requirements, power-handling capabilities, and available fabrication workflows. We implement and benchmark two CMOS‑compatible materials platforms, both supporting the visible spectrum:

\emph{Titanium‑dioxide ($\text{TiO}_2$):} Titanium‑dioxide metasurfaces leverage $\text{TiO}_2$’s high refractive index ($n\gtrsim2.4$ at a wavelength of 520 nm) and negligible absorption to achieve high diffraction efficiencies and low losses across a broad band of wavelengths ~\cite{capasso16tio2, wu19tio2, Chen2023disp}. Established fabrication approaches include (i) conformal atomic‑layer deposition (ALD) of $\text{TiO}_2$ on patterned electron‑beam resist~\cite{capasso16tio2}, and (ii) inductively‑coupled‑plasma (ICP) etching structures in a sputtered $\text{TiO}_2$ thin film using a Cr hard mask~\cite{huang2023metasurface}. While these methods deliver outstanding optical performance, their throughput is limited by the slow ALD growth and by challenges associated with the hard‑mask liftoff, motivating a different platform with faster fabrication for large-scale production.

\emph{Silicon‑rich $\text{Si}_3\text{N}_4$ (SRN):} We introduce silicon‑rich silicon nitride (SRN) as a new metasurface platform~\cite{malek2023}. SRN thin films are deposited by high‑rate plasma‑enhanced chemical‑vapour deposition (PECVD) directly onto fused‑silica wafers. By varying the ratios of the precursor gases ($\text{SiH}_4$, $\text{N}_2$, and $\text{NH}_3$) used in the PECVD process~\cite{sin20}, we can precisely control the real part of the complex refractive index of the SRN films, which can be adjusted from 1.9 to 3.1. This tunability enables versatile designs while retaining fully CMOS‑compatible blanket deposition and etch steps - attributes that have already been exploited for the fabrication of large‑aperture metalenses with $\text{NA}\approx0.98$ in the visible regime~\cite{fan2018sin}. In this work, we use 750‑nm‑thick SRN films with $n=2.3$ and negligible extinction. As detailed below, the resulting SRN metasurfaces achieve high forward‑scattering efficiency and withstand high continuous intensities of at least 25 W/mm$^2$. For applications requiring even higher power handling, $\text{TiO}_2$ offers extreme robustness, tolerating intensities above 2,000 W/mm$^2$, which is especially beneficial for scaling of tweezer arrays beyond 100,000 traps.

\begin{figure}[h]
\centering
\includegraphics[width=.7\columnwidth]{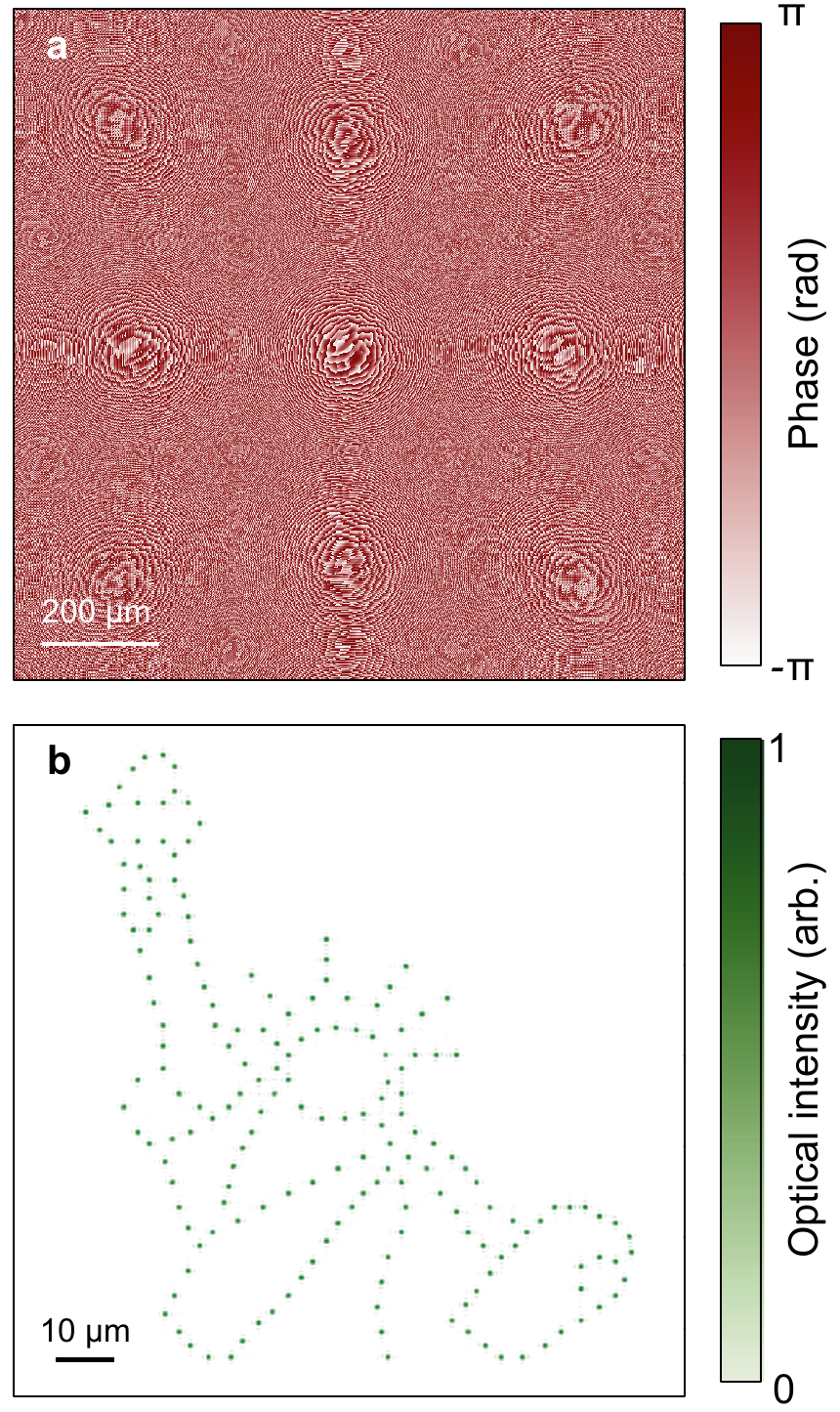}
\caption{\textbf{A calculated phase-only hologram and its corresponding focal-plane trap pattern.} \textbf{a,} A phase-only hologram generated by the modified Gerchberg-Saxton algorithm with $4,000 \times 4,000$ pixels and a pixel size of 290 nm. \textbf{b,} Simulated intensity distribution at the focal plane, showing a Statue of Liberty pattern consisting of 183 traps.}
\label{fig:holofig}
\end{figure}

\section*{Design and optimization of phase-only holograms}\label{sec:hologram}

To design phase-only holograms that create tweezer arrays with a specific focal length and NA, we use an approach based on the Gerchberg-Saxton algorithm~\cite{gs72}. The original algorithm utilizes the Rayleigh-Sommerfeld diffraction integral to iterate between the device plane (in our case, the metasurface plane) and the focal plane by forward and backward light propagation. This iterative process converges to a phase profile that produces the desired focal-plane intensity distribution. In each iteration, the input amplitude over the metasurface plane is set to unity, enforcing a phase-only condition for the algorithm.

\begin{figure}[h]
\centering
\includegraphics[width=0.8\columnwidth]{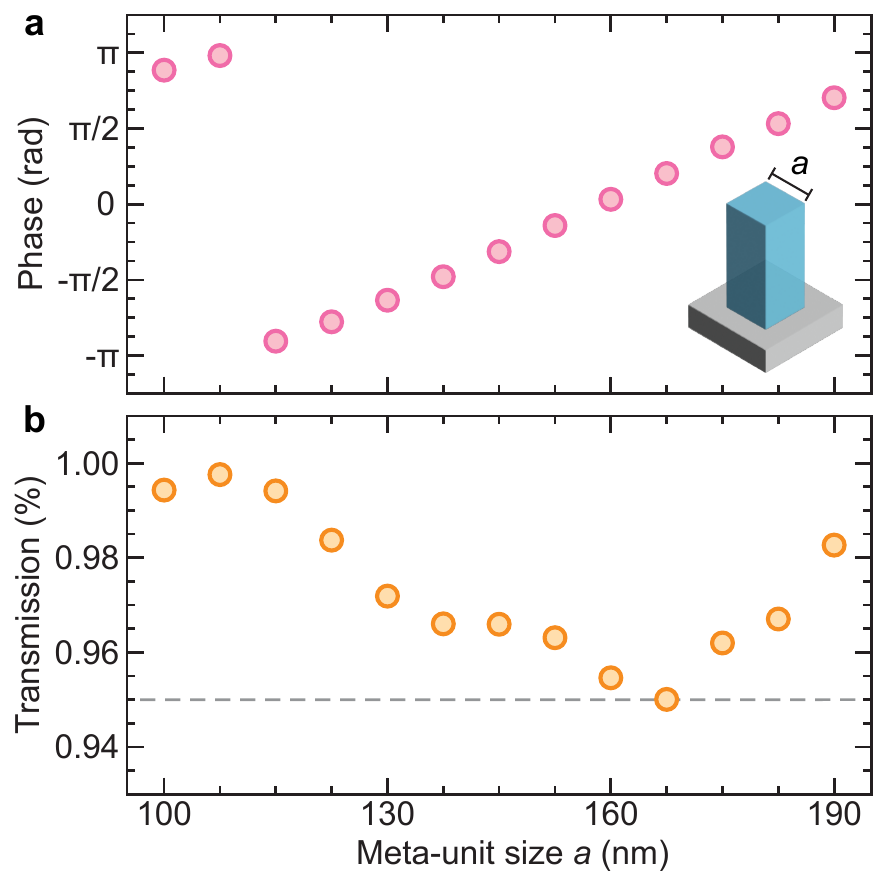}
\caption{\textbf{RCWA-calculated phase response and transmission of the SRN meta-atom library consisting of 13 nanopillars with different cross-sectional sizes.} \textbf{a,} Phase response of the meta-atom library as a function of nanopillar width $a$. \textbf{b,} Transmission or forward scattering efficiency of the meta-atom library. All meta-atoms have a transmission over $95 \%$ (marked as the grey dashed line).}
\label{fig:metaunit}
\end{figure}

The standard Gerchberg-Saxton algorithm is designed for general holography and inevitably introduces speckle and noise into the resulting optical patterns. Although this can be mitigated by using "soft operations" when updating the hologram~\cite{soft00}, achieving high-quality, point-like traps remains a significant challenge. To address this, we employ a weighted Gerchberg-Saxton (GSW) algorithm~\cite{holo07}, modified to account for the subwavelength pixel size and the direct focusing capability of our metasurface holograms. At the $(k+1)^{\text{th}}$ iteration, the target amplitude distribution is determined by the calculated focal plane amplitude and the target amplitude from the previous iteration via
\begin{equation*}
A^{k+1}_{\mathrm{target}}(x,y)=\frac{A^{k}_{\mathrm{target}}(x,y)}{A^{k}_{\mathrm{calculated}}(x,y)},
\end{equation*}
where $A^{k+1}_{\mathrm{target}}(x,y)$ and $A^{k}_{\mathrm{target}}(x,y)$ represent the target amplitude distributions at the $(k+1)^{\text{th}}$ and $k^{\text{th}}$ iterations, respectively, and $A^{k}_{\mathrm{calculated}}$ is the calculated focal plane amplitude. To prevent the target amplitude from producing extreme numerical values, numerical normalization and "soft operations" are performed in each update. Through this approach, we can generate high-quality phase-only holograms capable of producing near-uniform point trap arrays in arbitrary geometries, with a simulated standard deviation below $3 \%$ even for arrays consisting of $>100,000$ traps. An example of a phase-only hologram and the corresponding simulated tweezer array is shown in Extended Data Fig.~\ref{fig:holofig}.

\section*{Design of Meta-atom library}\label{sec:meta-unit}

To construct the meta-atom library for our metasurfaces, we utilize rigorous coupled-wave analysis (RCWA) to calculate the phase and amplitude response of individual SRN or $\text{TiO}_2$ nanopillars. The meta-atoms are non-birefringent and have square-shaped cross-sections. The center-to-center distance between adjacent nanopillars (i.e., the pixel size of the meta-atoms) is set to 290 nm. The height of the nanopillars is set to 750 nm for SRN and 600 nm for $\text{TiO}_2$ and their width ranges from 100 nm to 190 nm to provide a comprehensive phase coverage over the $2\pi$ range at a wavelength of 520 nm, while maintaining near-unity transmission or forward scattering efficiency (Extended Data Fig.~\ref{fig:metaunit}). The shown library achieves a high overall transmission that exceeds 95\% across the entire set of meta-units, resulting in an experimental diffraction efficiency of over 60\%. To prevent defects caused by missing or collapsed pillars during fabrication, the smallest nanopillar has an edge width of 100 nm. The subwavelength width of the nanopillars and size of the meta-atoms allow for high-resolution sampling of the designed metasurface phase profile with an achievable NA of 0.9.

\begin{figure*}[]
\centering
\includegraphics[width=0.75\textwidth]{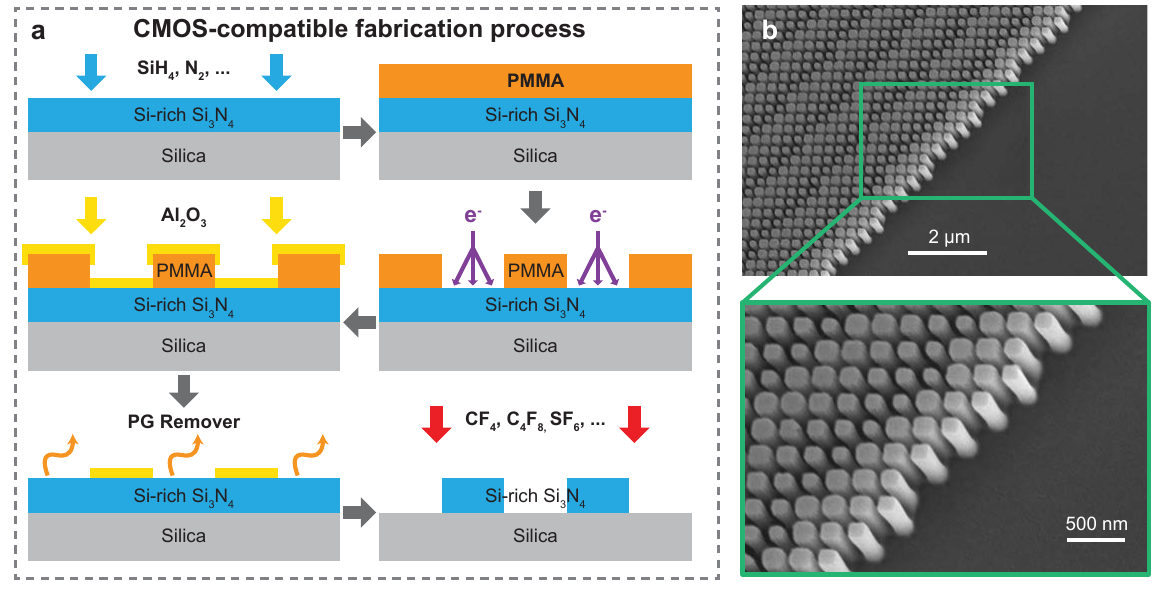}
\caption{\textbf{Cleanroom fabrication process and SEM images of SRN metasurfaces.} \textbf{a,} Illustration of the CMOS-compatible fabrication process. \textbf{b,} SEM images of the fabricated SRN metasurfaces show minimal defects and clean side walls.}
\label{fig:cleanroom}
\end{figure*}

\section*{Fabrication of $\text{TiO}_2$ and SRN metasurfaces}\label{sec:fab}
\emph{Fabrication of $\text{TiO}_2$ metasurfaces:}
Metasurfaces made from ALD-grown TiO\textsubscript{2} have been demonstrated as a promising platform for diffraction‑limited focusing~\cite{capasso16tio2}, broadband phase control~\cite{Fan20tio2, Chen2023disp}, and complex beam shaping in the visible~\cite{Lim23tio2,Jammi24tio2,Zaidi24tio2}. Our $\text{TiO}_2$ metasurfaces are fabricated on 0.5-mm‑thick, double‑side‑polished fused‑silica wafers. A 700-nm ZEP‑520A layer is spin‑coated and baked ($180\ ^{\circ}\text{C}$, 3 min). The thickness of the resist is verified with a stylus profiler (KLA P‑17). After applying an E-spacer layer, the nano‑pillar template is written by 100 kV electron-beam lithography (EBL; Elionix ELS-G100) with a current of 2 nA and a step size of 4 nm.  The resist is developed in chilled o‑Xylene, rinsed in IPA, and nitrogen‑dried, yielding apertures whose depth sets the final $\text{TiO}_2$ pillar height. Amorphous $\text{TiO}_2$ is then conformally deposited at $200\ ^{\circ}\text{C}$ in an ALD reactor (Cambridge NanoTech Savannah) until the apertures are fully filled. Excess $\text{TiO}_2$ material on top is removed by inductively coupled plasma (ICP) etching (CHF\textsubscript{3}/Ar/O\textsubscript{2}, Oxford PlasmaPro 100 Cobra) down to the resist surface.  A final downstream plasma ashing at $220\ ^{\circ}\text{C}$ (Matrix Plasma Asher) removes the resist template, leaving free‑standing $\text{TiO}_2$ nanopillars on the fused‑silica substrate.

\emph{Fabrication of SRN metasurfaces:}
The SRN metasurfaces presented in this work are manufactured through a CMOS-compatible nanofabrication process, as illustrated in Extended Data Fig.~\ref{fig:cleanroom}\textbf{a}. A 750-nm-thick SRN layer, with a designed refractive index of 2.3, is deposited onto diced 500-\textmu m-thick fused silica substrates. A two-layer resist (PMMA 495k A4 and 950k A2) is spun-coated onto the SRN layer, and EBL (Elionix ELS-G100) is conducted with a dose of 770 \textmu$\text{C/cm}^2$ and a current of 2 nA. To prevent the electron charging effect caused by
the non-conductive substrate during the EBL process, a 20-nm E-spacer is spun-coated onto the chip. After exposure, the resist is developed in a mixed solution of IPA:DI water = 3:1 for 2 minutes. The developed resist is then coated with a 25-nm thick $\text{Al}_2\text{O}_3$ layer using electron-beam evaporation to serve as a hard mask for etching. The $\text{Al}_2\text{O}_3$ layer is subsequently lifted off in Remover PG overnight, leaving a patterned mask on the SRN layer. This pattern is etched into the SRN layer to form SRN nanopillars using an ICP etching system (Oxford PlasmaPro 100 Cobra).

\begin{figure}
\centering
\includegraphics[width=.7\columnwidth]{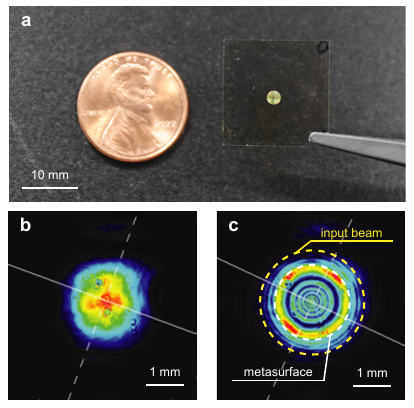}
\caption{\textbf{Images of a metasurface, a laser beam profile, and the beam profile after diffraction by the metasurface.} \textbf{a,} A photo of a 2.32 mm $\times$ 2.32 mm metasurface by the side of an American one-cent coin. \textbf{b,} The profile of a 1.5-mm diameter beam that is incident onto the metasurface. \textbf{c,} A far-field image of non-diffracted light after aligning the metasurface to the beam. This diagnostic, in tandem with maximizing the diffracted power, ensures good optical alignment.}
\label{fig:powerhandling}
\end{figure}

The fabricated $\text{TiO}_2$ and SRN metasurfaces have circular footprints with diameters ranging from 1.16 mm (approximately $4,000\times4,000$ nanopillars or pixels) to 3.48 mm (approximately $12,000\times12,000$ pixels; Extended Data Fig.~\ref{fig:powerhandling}\textbf{a}). Scanning electron microscope (SEM) images of the fabricated metasurfaces (Extended Data Fig.~\ref{fig:cleanroom}\textbf{b}) show that the nanopillars are defect-free and have vertical side walls.

\section*{Optical performance of metasurfaces}\label{sec:power}
The metasurfaces have a circular footprint to facilitate their alignment with the incident beam and to reduce diffraction at the metasurface edges. The portion of the laser beam captured and shaped by the metasurface depends on the input beam size and the efficiency of the metasurface. We choose a beam size slightly larger than the metasurface to ensure a relatively uniform intensity profile and flat optical wavefront across the metasurface, which our metasurface design is based on. For example, the input beam diameter is approximately 1.5 mm for the metasurfaces with a diameter of 1.16 mm. Therefore, some input power transmits through the substrate without being modulated. Excluding the non-overlapping portion of the beam, we observe a diffraction efficiency of $60 \%$ for our metasurfaces. In the future, the phase mask can be designed for a Gaussian input beam, allowing an even larger portion of incident power to be used. 

The high forward scattering efficiency of metasurfaces enables a large number of tweezers for a given incident laser power. Unlike SLMs or AODs, where the 0$^\mathrm{th}$‑order diffraction is not modulated and manifests as a bright specular spot, the non-modulated laser power from a metasurface is diffused over the full $4\pi$ steradians through subwavelength scattering, leaving the atomic plane essentially background‑free. Underlining the high potential of our approach, recent demonstrations of dispersion‑engineered and inverse‑designed metasurfaces have already demonstrated an efficiency above $90 \%$ over a wavelength range of hundreds of nanometers~\cite{Chen2023disp, Dainese2024shape}, and a recent quasi‑3D metasurface even demonstrated $>99 \%$ anomalous‑refraction efficiency~\cite{He2025per}.

The metasurface platform also has a high power-handling capability. We have tested the damage threshold by maximizing the input power and decreasing the waist of the incident beam. We observe that the damage threshold of the SRN metasurfaces is at least $25$ W/mm$^2$, about $5\times$ larger than that of liquid-crystal SLMs. By reducing the "silicon richness" in SRN and thus decreasing the extinction coefficient of the material, it should be possible to further improve the power-handling capability of the SRN metasurfaces. Furthermore, alternative materials promise even better power handling; for TiO$_2$ metasurfaces, we observe no signs of degradation up to 2000 W/mm$^2$. A comparison of the properties of metasurfaces with AODs and SLMs is provided in Extended Data Table~\ref{tab:characteristics}.

Given the measured damage threshold of the SRN metasurfaces, we provide an estimate of how many traps can realistically be generated with the current device parameters. Assuming a 2.3 mm $\times$ 2.3 mm device with $60 \%$ diffraction efficiency, a laser power of 130 W, and about 1 mW laser power per tweezer trap, approximately $80,000$ tweezers can be generated; therefore, metasurface tweezer arrays with $>100,000$ sites are realistic without major improvements over current technology. With a modest improvement on the power-handling capability and an increase in the device footprint to 10 mm $\times$ 10 mm, arrays with $>1,000,000$ traps are within realistic reach even with SRN. The power handling capabilities of TiO$_2$ already far exceed the requirements to generate $>1,000,000$ traps. 

In the main text, we report the non-uniformity of the trap depth as $7.5\%$, measured with atoms. It is worth noting that the non-uniformity of the optical intensity of the traps, measured in the image plane of the metasurface, is significantly lower. By directly measuring the peak optical intensity of each tweezer spot in the image plane behind the metasurface (i.e., without the relay optics), we find a non-uniformity of $4\%$. We attribute the higher non-uniformity in the trap depth of the atomic array to imperfections that are introduced by the relay optics. Removal or improvement of the relay optics should allow for a $4\%$ non-uniformity of the atomic traps. Remarkably, we were able to rapidly improve the intensity non-uniformity from an initial $20\%$ to single digit percentages via improvements in the metasurface design and fabrication, and further potential for improvements is highly likely. The uniformity is insensitive to slight misalignment of the angle or position of the incident beam with respect to the metasurface and is independent of the specific array geometry.

\renewcommand{\arraystretch}{1.3}

\begin{table*}[]
\begin{tabular}{l|l|l|l}
\textbf{Characteristics} & \textbf{AOD} & \textbf{SLM} & \textbf{Metasurface} \\
\hline
\textbf{Type} & Active & Active & Passive \\
\textbf{Relay optics} & Required & Required & Not required \\
\textbf{Wavelength range} & Specified by model & Specified by model & \makecell[tl]{Widely tunable} \\
\textbf{Power handling} & $\sim 10$ W/mm$^2$ & $\sim 2$ W/mm$^2$ & \makecell[tl]{$>25$ W/mm$^2$ - SRN \\ $>2000$ W/mm$^2$ - TiO$_2$} \\
\textbf{Diffraction efficiency} & $\sim50\%$ & $\sim40\%$ & $\sim60\%$ \\
\textbf{Device footprint} & $1-10$ cm$^2$ & $1-10$ cm$^2$ & $1-10$ mm$^2$ \\
\textbf{Pixel size} & NA & $2-10$ \textmu m & $100-500$ nm \\
\textbf{Trap geometry} & \makecell[tl]{2D simple \\ geometry} & \makecell[tl]{Arbitrary pattern \\ in any dimension} & \makecell[tl]{Arbitrary pattern \\ in any dimension}
\end{tabular}
\caption{\textbf{Comparison of beam-shaping devices.}}
\label{tab:characteristics}
\end{table*}

\section*{Preparation and imaging of ultracold strontium}\label{sec:experiment}
The preparation of ultracold strontium atoms starts with the generation of a cold atomic beam with a 2D magneto-optical trap (MOT) that is directly loaded from resistively heated dispensers, as described in previous work~\cite{kwon2023jet}. The atoms are transferred to a glass cell vacuum chamber via a push-beam where they are captured and cooled to about 1 mK by a 3D MOT operating on the 461 nm transition. We simultaneously operate two repump lasers at 679 nm and 707 nm to close a loss channel present in the 461-nm cooling scheme. To further cool the atoms, we operate a second MOT on the narrow-line 689-nm transition. We begin the transfer between MOTs by frequency broadening the 689-nm line to 3 MHz to match the temperature distribution of the 461-nm MOT, before smoothly narrowing down to a single-frequency 689-nm MOT with on average $10^5$ atoms at 1 \textmu K.

From the ultracold strontium gas, we typically load optical tweezer arrays at a trap depth of 100 \textmu K. The trapping light is provided by a 5-W, 520-nm fiber laser with second harmonic generation (Azurlight, ALS-GR-520-5-A-CP-SF), seeded by a home-built extended cavity diode laser operating at 1040 nm (QPhotonics, QLD-1030-100S). Directly after the 520-nm laser output, we use an acousto-optic modulator to dynamically control the trap depth of the tweezers. In front of the metasurface, we use a magnifying telescope to increase the beam waist to be larger than the area of the metasurface. After the metasurface, a high-power-capable microscope lens ($\mathrm{NA}=0.6$, Thorlabs, LMH-50X-532) is used to collimate the generated pattern. The tweezer pattern is relayed through a 1:1 telescope before being focused down onto the atoms via an objective lens ($\mathrm{NA}=0.5$, Mitutoyo, G Plan Apo 50X).

We perform fluorescence imaging to detect the atomic occupation of the tweezer array. We resonantly scatter photons on the 461-nm transition for 50 ms, collect the fluorescence through the high-NA objective that focuses the tweezers, separate the light from the tweezer path via a long-pass dichroic, and image on an EMCCD camera (Andor iXon Ultra 888). We image with a 200-mm lens before the camera, such that a single camera pixel corresponds to a real space size of $260 \text{ nm}\times 260\text{ nm}$.

\begin{figure*}
\centering
\includegraphics[width=0.8\textwidth]{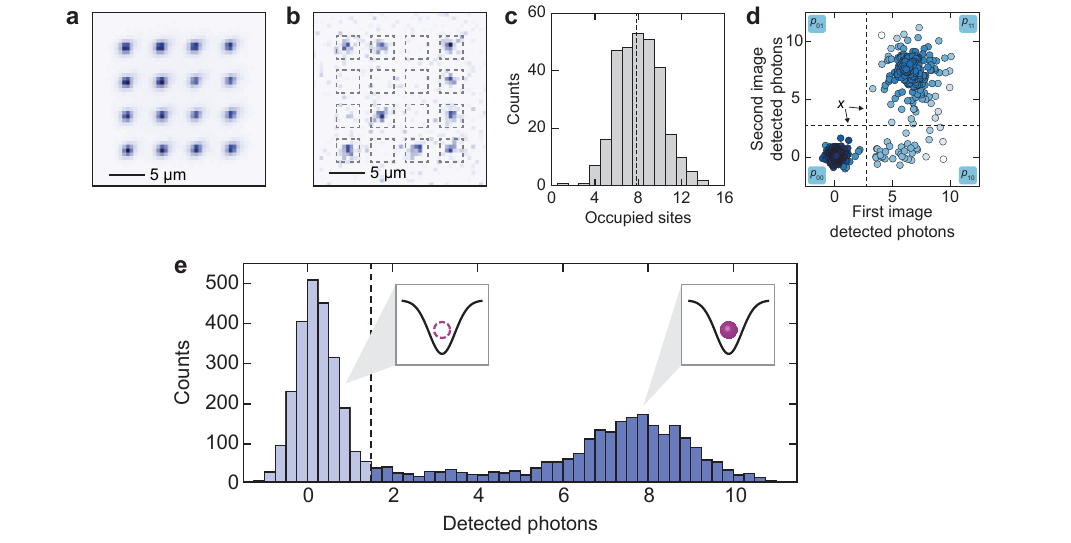}
\caption{\textbf{Single atom preparation and detection in a $\mathbf{4 \times 4}$ metasurface tweezer array.} \textbf{a,} Average of 100 fluorescence images after parity projection. The high uniformity indicates that all traps have an approximately equal chance of being filled with an atom. \textbf{b,} Individual fluorescence image. Trap locations are indicated by dashed boxes. \textbf{c,} Histogram of the number of occupied sites after parity projection; the mean occupancy of a trap, marked by the dashed line, is $49(3) \%$. \textbf{d,} Determination of the photon count threshold $x$ to distinguish between zero and one atom in a trap. Data points show the photon counts for one specific trap in 500 iterations of the experiment. A darker color indicates a higher density of points. The threshold value $x$ divides the data into four quadrants, labeled by $p_{ij}$ with indices $i=0,1$ and $j=0,1$ indicating the absence (presence) of an atom in the first and second image, respectively. \textbf{e,} Histogram of photon counts across the trap locations of the $4 \times 4$ array, as marked in \textbf{b}. 500 repetitions of the experiment are averaged. The data allows for the distinction between sites with one and zero atoms with high fidelity. The dashed line marks the averaged threshold value across the array.}
\label{fig:fidelity}
\end{figure*}

\section*{Single-atom detection fidelity}\label{sec:single}

To determine how well we can distinguish one from zero atoms in a tweezer trap, we use a model-free approach that does not require detailed modeling of the photon counting statistics. Here, we discuss the approach for a $4\times4$ array, with the corresponding data shown in Extended Data Fig. \ref{fig:fidelity}. For the $16\times16$ array in the main text, we use the same methodology. The larger array shows a somewhat reduced single-atom detection fidelity compared to the smaller array which is due to limited available laser power rather than imperfections of the metasurface hologram. Empirically, we find that $\sim 7.5$ mW of power per tweezer trap provide optimal imaging conditions, while only $\sim 4$ mW are available for the $16\times16$ array.

Our methodology is similar to the one used in Ref.~\cite{Norcia2018microscopic}, involving the recording of two fluorescence images of the atom array in short sequence, separated by 25 ms, within the same iteration of the experiment. Besides the detection fidelity, $F$, it yields the initial filling fraction $f$ and survival rate $S$ of atoms between images. From the two images, the atomic fluorescence signal (photon count) at each trap location is recorded. For each trap, there are four possible outcomes and associated probabilities: there is an atom in both images ($p_{11}$), there is an atom in the first but not in the second image ($p_{10}$), there is an atom in the second but not first image ($p_{10}$), and there is no atom present in either image ($p_{00}$). The probability for each event is given by:
\begin{equation*}
    \begin{aligned}
        p_{11} = & f (1-F_0) F_1 (1-S) +(1-f) (1-F_0)^2+f F_1^2 S, \\
        p_{10} = & f F_1 S (1 - F_1) + f F1 (1 - S) F_0 + (1 - f)(1 - F_0) F_0, \\
        p_{01} = & f (1-F_0) (1-F_1) (1-S)+(1-f) (1-F_0) F_0 \\
                 & + f (1-F_1) F_1 S, \\
        p_{00} = & 1 - (p_{11}+p_{10}+p_{01}),
    \end{aligned}
\label{eq:model-free}
\end{equation*}
where $F_0$ denotes the detection fidelity for zero atoms, $F_1$ is the detection fidelity of one atom, $S$ is the survival rate between images, and $f$ is the initial filling fraction. Initially, the threshold photon count $x$ that marks the distinction between zero and one atom is a variable. By iteratively varying $x$ (see Extended Data Fig.~\ref{fig:fidelity}d), we solve the above equations to maximize the total imaging fidelity defined by $F = f F_1 + (1-f)F_0$. By optimizing the threshold $x$ for each individual tweezer, we find a mean detection fidelity across the array of $F=99.8 \%$, a mean survival rate of $84.1 \%$, and a mean filling fraction of $49.2 \%$. The reduced survival rate $S$ is attributed to an enhanced loss process associated with the 520-nm trapping wavelength (see main text).

\section*{Trap characterization}\label{sec:characterization}

\begin{figure}
\centering
\includegraphics[width=\columnwidth]{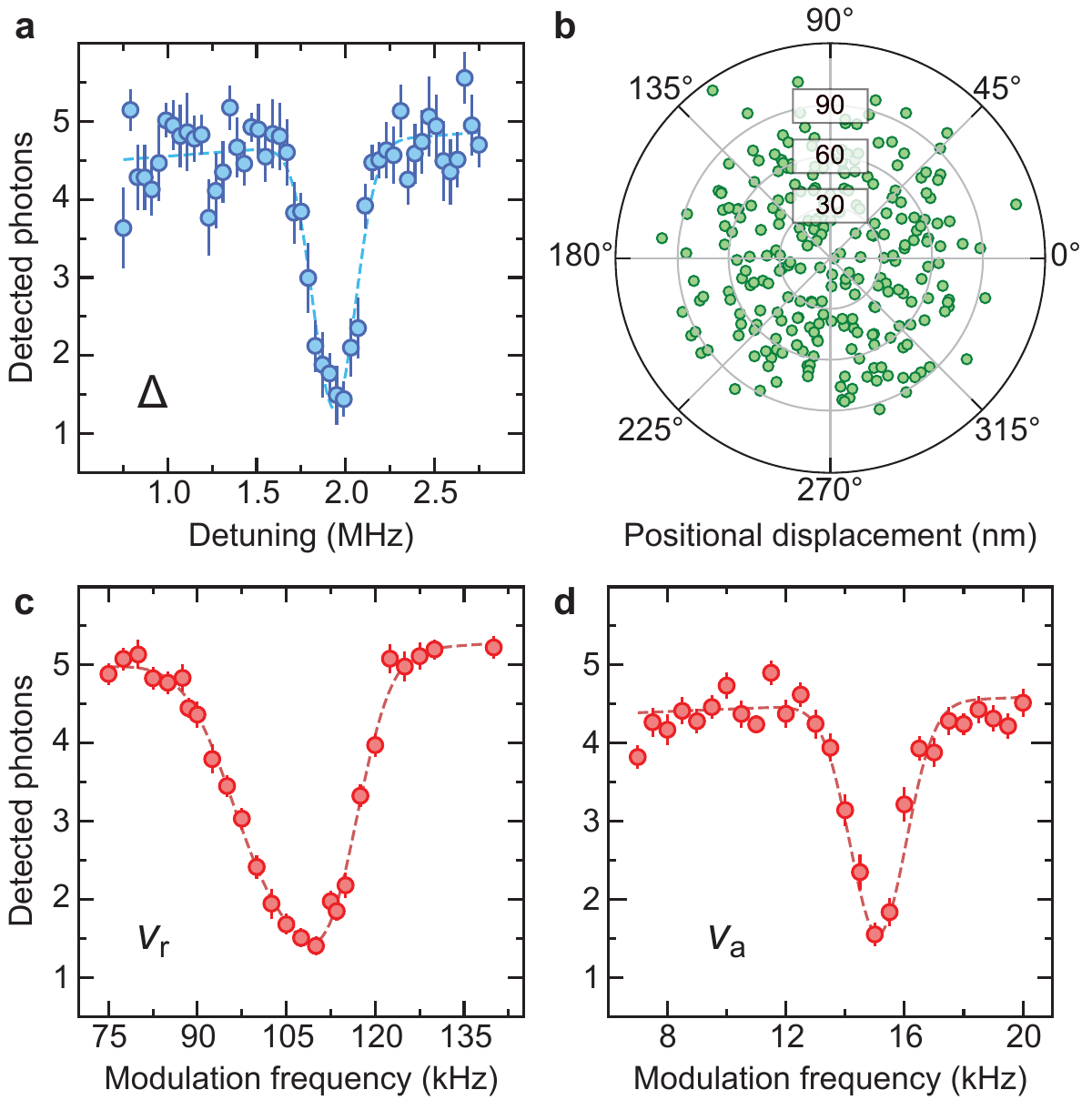}
\caption{\textbf{Measurements of uniformity in the $\mathbf{16\times16}$ array.} \textbf{a,} Measurement of the trap depth via the light shift $\Delta$ of the $^1S_0$ - $^3P_1$ resonance in the presence of the trapping field. Data shows the resonance for a single trap. \textbf{b,} Positional displacement between observed and target trap locations. \textbf{c-d,} show the measurement of the radial and axial trap frequency, respectively, via parametric heating. All error bars in this figure show $1\sigma$ s.e.m. from 20 repetitions of the measurement.}
\label{fig:individual}
\end{figure}

The trap depth is measured via the light shift on the trapped atoms induced by the trapping laser. We probe the light shift on the $m_j=\pm1$ transition between the $^1$S$_0$ and $^3$P$_1$ states using the 689-nm MOT beams. When the probe light is resonant with the shifted resonance, the atoms are heated out of the trap. Extended Data Fig.~\ref{fig:individual}\textbf{a} shows a typical loss resonance in an individual trap. Fitting the resonance feature with a Gaussian model, we determine the detuning $\Delta$ from the free space resonance. From this we obtain the trap depth $U = \Delta \alpha_{^1 \mathrm{S}_0}/(\alpha_{^3 \mathrm{P}_1} - \alpha_{^1 \mathrm{S}_0})$, where $\alpha_{^1 \mathrm{S}_0}$ ($\alpha_{^3 \mathrm{P}_1}$) is the polarization of the $^1$S$_0$ ($^3$P$_1$) state. We report the statistical variance rather than the measurement uncertainty as it is the main contribution to the total uncertainty.

The positional accuracy of the tweezer array is extracted by comparing the measured trap locations with the target trap positions that were used for the design of the metasurface. The trap locations are obtained by fitting each atom's fluorescence signal to a 2D Gaussian distribution. The positional inaccuracy between intended and measured locations is defined as $\Delta r = \sqrt{\Delta x^2 + \Delta y^2}$, where $\Delta x$ ($\Delta y$) is the deviation in the $x$ ($y$) direction. The orientation of the metasurface is not fixed to the lab frame and can rotate freely. We minimize the averaged distance between the measured and target trap locations by rotating and translating the coordinate pattern. Following this optimization, we find that the mean deviation of trap locations is 60 nm. As shown in Extended Data Fig.~\ref{fig:individual}\textbf{b} the deviations between measured and target locations are uniformly distributed in all directions, indicating that there is no systematic bias. The deviation is also comparable to the extent of the ground-state vibrational wavefunctions of Sr of about 30 nm and 90 nm in the radial and axial direction, respectively.

\begin{figure}
\centering
\includegraphics[width=\columnwidth]{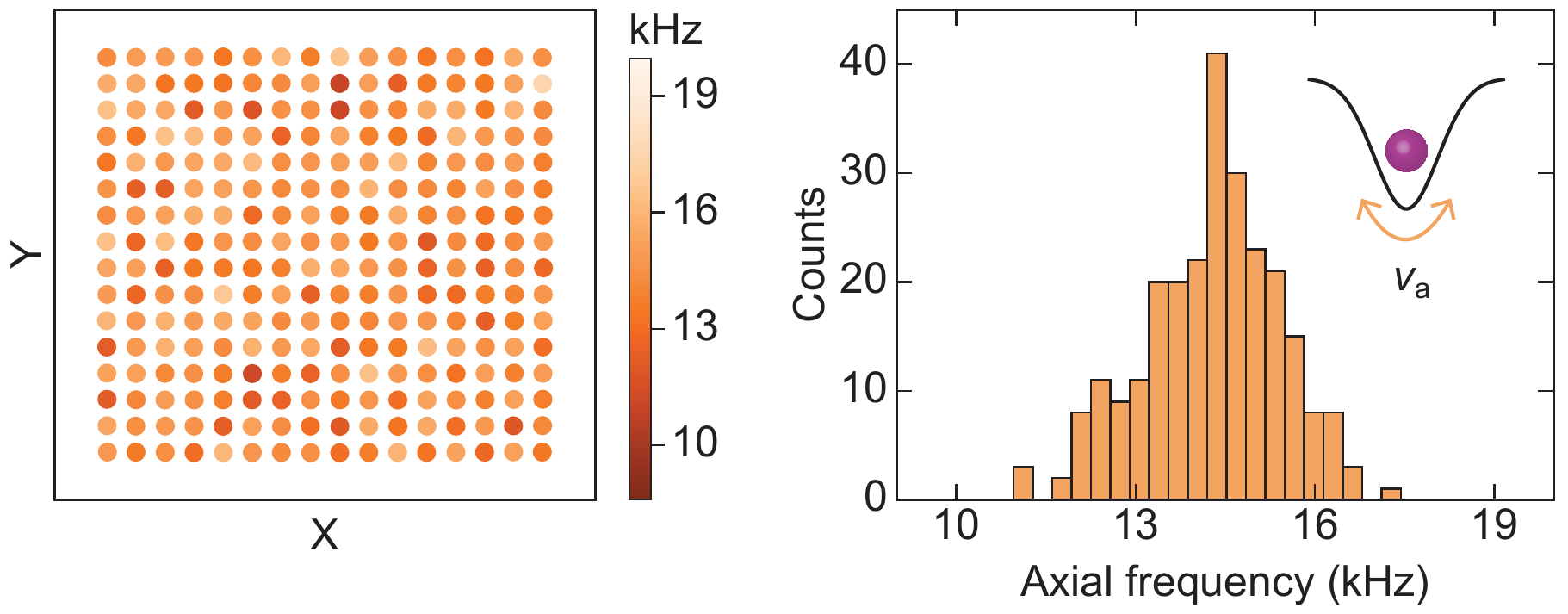}
\caption{\textbf{Axial trap frequency measurements.} The axial trap frequencies measured in a $16 \times 16$ with 4 \textmu m trap spacing (left) and their statistical spread (right).}
\label{fig:axial}
\end{figure}

The trap frequencies are measured via parametric heating. The intensity of the tweezer light is modulated sinusoidally with an amplitude of $5 \%$ for 30 ms while the frequency of the modulation is varied. Individual trap loss features at twice the radial and axial trap frequencies are shown in Extended Data Figs.~\ref{fig:individual}\textbf{c} and \textbf{d}, respectively. The parametric resonance in the radial direction shows a slight asymmetry. There are two distinct trap frequencies in the radial direction, which we attribute to the trapping potential not being perfectly circularly symmetric. We fit the line shape with a double Gaussian distribution and plot the geometric mean as the radial trap frequency. The parametric resonance in the axial direction shows a symmetric dip, which we fit with a single Gaussian distribution. Extended Data Fig.~\ref{fig:axial} shows the axial trap frequencies in each trap and their statistical spread. We report the statistical variance rather than the measurement uncertainty as it is the main contribution to the total uncertainty.

\section*{Effective NA of pixel-based beam shaping devices}\label{sec:NA}

The numerical aperture (NA) measures the angular range within which an optical element, for example, a lens, can accept light from a point source. For a spherical lens in vacuum, the NA is given by $\sin \theta$, where $\theta$ is the half-angle of the acceptance cone of the lens with $\theta = \arctan(D/(2 f))$, where $D$ is the diameter of the lens and $f$ its focal length. 

We use the example of a lens to derive an analytical expression for the effective NA of a pixel-based beam shaping device (see Fig.~\ref{fig:5}\textbf{a}). In the first step, we reduce the phase profile of the lens with focal length $f$ to a Fresnel lens with continuous local phase shifts, modulo $2 \pi$. When we approximate the phase profile of the Fresnel lens $\phi(x)$ with a pixel-based device, the finite sampling limits how well the steep phase gradients at the edge of the lens can be captured. We assume that the phase jumps between neighboring pixels should not be larger than $\pi/2$ to faithfully reproduce the behavior of the lens. For a device with pixel size $d$, this limits the phase gradient to $\partial \phi/ \partial x \lesssim \pi/2d$. The larger the pixel size $d$ relative to wavelength $\lambda$, the smaller the phase gradient that can be faithfully captured. As the phase profile of a lens has the steepest gradients near the edge, the usable diameter of the pixel-based device is effectively reduced when gradients near the edge cannot be reproduced. 

\begin{figure}
\centering
\includegraphics[width=\columnwidth]{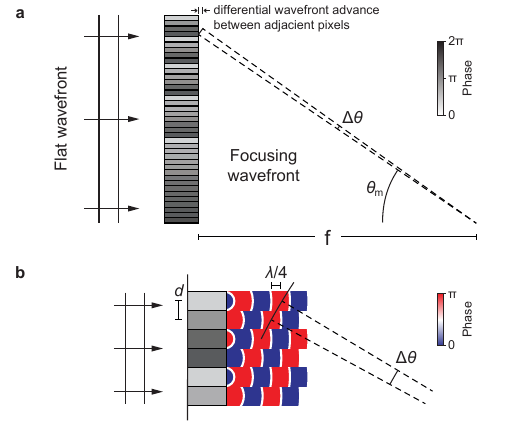}
\caption{\textbf{Calculation of the effective NA of a pixel-based beam shaping device.} \textbf{a,} The phase profile of a lens with focal length $f$ is emulated by a pixelated phase mask. The individual pixels have a size $d$. A flat wavefront impinges on the device and it is converted into a focusing wavefront. $\theta_\mathrm{m}$ is the maximal angle for which the phase advance between neighboring pixels stays smaller than $\pi/2$. The gray scale indicates the phase shift of the metasurface pixels. \textbf{b,} Zooming into the wavefront advance of individual pixels, following Huygens' principle, the angle $\theta_\mathrm{m}$ is determined by the condition where the wavefront advance between neighboring pixels reaches $\lambda/4$ (corresponding to a phase shift of $\pi/2$). $\Delta \theta$ denotes the angular separation between neighboring pixels. The color scale indicates the phase shift of the emerging wavefront behind the metasurface.}
\label{fig:NA}
\end{figure}

To derive the dependence of the effective NA on pixel size $d$ and wavelength $\lambda$, we consider the following (see Extended Data Fig.~\ref{fig:NA}): We treat each individual pixel of the device as a point scatterer which, following Huygens' principle, creates a radial wavefront with a phase shift corresponding to the phase of a given pixel. We define $\theta_\mathrm{m}$ as the maximal angle for which the device can faithfully reproduce the phase gradient near the edge. At this angle the phase difference between neighboring pixels reaches $\pi/2$. In terms of wave propagation, this corresponds to a $\lambda/4$ wavefront advance between the neighboring pixels that are spaced by distance $d$. The effective NA is given by $\sin \theta_\mathrm{m}$. 

Using the construction of Extended Data Figs.~\ref{fig:NA}\textbf{a} and \textbf{b} and defining the angular separation $\Delta \theta$ between neighboring pixels, the following relations hold:
\begin{equation*}
    \begin{aligned}
       f(\cos(\theta_{\mathrm{m}})-\cos(\theta_{\mathrm{m}}+\Delta\theta))& = \lambda/4, \\
       f(\sin(\theta_{\mathrm{m}}+\Delta\theta) - \sin(\theta_{\mathrm{m}}))& = d. 
    \end{aligned}
\end{equation*}
Regrouping the terms yields
\begin{equation*}
    \begin{aligned}
       \cos(\theta_{\mathrm{m}}+\Delta\theta) & = \cos\theta_{\mathrm{m}} - \lambda/(4f), \\
      \sin(\theta_{\mathrm{m}}+\Delta\theta) & = \sin\theta_{\mathrm{m}} + d/f. 
    \end{aligned}
\end{equation*}
Squaring both equations and adding them together gives
\begin{equation*}
    \begin{aligned}
       \left(d/f\right)^2 +  \left(\lambda/(4f)\right)^2 + \left(\lambda/2f\right)\cos\theta_\mathrm{m} - \left(2d/f\right) \sin\theta_\mathrm{m} & =  0.\\
    \end{aligned}
\end{equation*}
As the pixel size and wavelength are much smaller than the focal length, $d, \lambda\ll f$, we neglect the terms quadratic in $d/f$ and $\lambda/f$. From this we obtain the relation $\cos\theta_\mathrm{m} = (4d/\lambda)\sin\theta_\mathrm{m}$. Using the definition for the effective NA, this yields
\begin{equation*}
 \text{NA} = \frac{1}{\sqrt{1+(4d/\lambda)^{2}}}. 
\end{equation*}

Based on this relation, we can distinguish three regimes (see Fig.~\ref{fig:5}\textbf{b}): 

\underline{$d\gg\lambda$}: The pixel size is too large to capture steep phase gradients. The lens is only faithfully reconstructed in the paraxial region at the center, leading to an effective NA close to 0. 

\underline{$d\approx\lambda$}: The effective NA sensitively depends on the ratio of pixel size and wavelength $d/\lambda$. A small reduction of $d/\lambda$ can dramatically increase the effective NA.

\underline{$d\ll\lambda$}: The pixel size is so small that almost any phase gradients can be reproduced, giving rise to an effective NA close to 1. 

For our metasurface, $d/\lambda \sim 0.6$ and the effective NA calculated using this simplified model is 0.41, while the actual NA for our metasurfaces is 0.6. Thanks to the high NA, holographic metasurfaces can directly generate diffraction-limited traps on the micrometer scale in their image plane. For current liquid-crystal SLMs, the effective NA is 0.05 or less due to their relatively large pixel size (4 \textmu m -- 20 \textmu m). They require high-quality demagnification optics with a high NA to realize tweezer arrays on the micrometer scale. Such optics can act as an unintended low-pass filter, cutting off high-frequency spatial Fourier components and thus posing a challenge to generate uniform and well-resolved tweezer arrays. While such issues can be partially mitigated by iterative optimization of the SLM pattern, holographic metasurfaces offer a way to fundamentally circumvent such issues.

\begin{figure}
\centering
\includegraphics[width=\columnwidth]{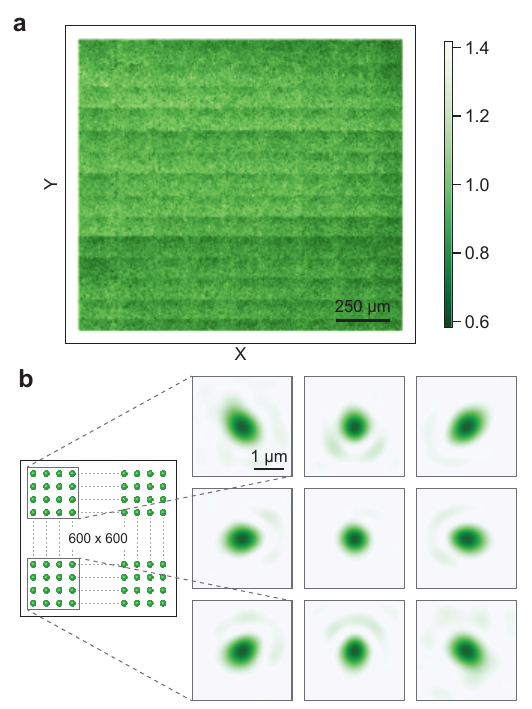}
\caption{\textbf{Imaging the $\mathbf{600\times600}$ tweezer array.} \textbf{a,} Composite image of the full array, stitched together from 126 individual high-resolution images. Each image captures a real-space area of $190\times140$ \textmu m. The individual images show a discontinuity in the measured light intensity, which arises from imperfections in the imaging system and contributes to the reported tweezer non-uniformity. \textbf{b,} Averaged tweezer spots from the center and the edges of the array. Each picture averages $\sim300$ traps. Tweezers at the array edges show a pinching as well as a weak halo oriented towards the center of the array.}
\label{fig:nonuniformity}
\end{figure}

\section*{Array simulations for pixel-based beam shaping devices}\label{sec:pixel_sims}

To numerically simulate the holographic performance of pixel-based beam shaping devices, we employ a fast Fourier transform (FFT)-based numerical method to solve the Rayleigh-Sommerfeld diffraction integral for a given phase-only hologram with specified pixel size and resolution. This method is highly effective in designing and optimizing phase-only holograms, and yields excellent agreement between calculations our experimental results. We analyze holograms with varying pixel sizes, starting with the pixel size of a commercially available liquid-crystal SLM (3.74 \textmu m, GAEA-2 SLM from HOLOEYE Photonics AG) and decreasing to the subwavelength pixel size of the metasurfaces used in this work (290 nm). For each pixel size, the hologram is set to a fixed resolution of $300 \times 300$ pixels. The position of the focal plane is calculated based on the hologram's physical size ($300 d \times 300 d$) and the effective NA. Each hologram is optimized using the modified Gerchberg-Saxton algorithm to generate a $3 \times 3$ square array of traps with a 5-\textmu m spacing in the focal plane without intermediate optics. By averaging the full width at half maximum (FWHM) values along the $x$ and $y$ axes (assuming the optical axis is along $z$ axis) of the optical intensity profile for each trap, we extract the average and range of FWHM values across the entire array, as presented in Fig.~\ref{fig:5}\textbf{c}. Our results demonstrate that subwavelength pixel sizes enable the direct generation of diffraction-limited traps with FWHMs smaller than the working wavelength.

The trends of the numerical data are well captured by the analytic effective NA relation derived above. The FWHM of a single trap is related to the NA via $\text{FWHM} = 0.51\lambda/\text{NA}$. Using the effective NA relation derived above and multiplying by a scaling factor $\alpha$, we obtain the fit function 
\begin{equation*}
 \text{FWHM} = \alpha \times 0.51\lambda \sqrt{{1+(4d/\lambda)^{2}}}. 
\end{equation*}
With the scaling factor $\alpha$ being the only free parameter, this model is in excellent agreement with the numerical simulation results (see Fig.~\ref{fig:5}\textbf{c}).

Furthermore, we numerically investigate the achievable uniformity of trap intensities while increasing the array size. We calculate the normalized standard deviation of trap intensities as a function of the number of generated tweezers in the focal plane for device resolutions ranging from $1000 \times 1000$ to $16,000 \times 16,000$ pixels. The simulation assumes a pixel size of $d = 290$ nm and an $\mathrm{NA}=0.6$, corresponding to the metasurfaces used in this work. The position of the focal plane is determined by the hologram size (resolution $\times$ pixel size) and the NA. The holograms are optimized using the algorithm outlined above, and a final forward propagation is performed using the optimized holograms to calculate the tweezer array formed in the focal plane. We integrate the intensity profile for each tweezer and calculate the standard deviation for the entire array.

As shown in Fig.~\ref{fig:5}\textbf{d}, we find that for each resolution, the uniformity starts to decay rapidly beyond a certain number of tweezers. From the data we find empirically that at least 300 pixels per tweezer trap are needed to maintain a uniformity $>95\%$.

\section*{$\mathbf{600\times600}$ tweezer array characterization}\label{sec:600x600}
The $600 \times 600$ array spans an area of 1.5 mm$^{2}$ in the image plane. Capturing a single image of the entire array while resolving individual tweezers with 2.5 \textmu m spacing is challenging. We perform a raster scan and section the entire array into 126 high-resolution images that are recorded using a high-NA microscope ($\text{NA}= 0.85$). These images are then stitched together to create a composite image of the full array. In order to reduce edge effects of the high-NA microscope, each image overlaps with neighboring ones by several rows and columns. A high-resolution composite image of the entire $600\times600$ array and a video zooming into and out of the array are provided in the Supplementary Information.

As discussed in the main text, we measure an intensity non-uniformity across the array of $8\%$. This is an upper bound for the actual non-uniformity of the array. As can be seen in Fig.~\ref{fig:nonuniformity}\textbf{a}, individual images that contribute to the full image consistently show an intensity gradient that is introduced by the imaging optics rather than the metasurface itself. We do not correct for such imaging-induced imperfections; they are included in the calculation of non-uniformity. To assess the roundness of individual tweezers at different locations in the large array, we show averaged tweezer spots from the array center and the edges, see Fig.~\ref{fig:nonuniformity}\textbf{b}. While tweezers at the edges of the array show a weak halo and some pinching, pointing away from the center of the array, tweezers near the center show a high degree of roundness.


\begin{thebibliography}{72}%
\makeatletter
\providecommand \@ifxundefined [1]{%
 \@ifx{#1\undefined}
}%
\providecommand \@ifnum [1]{%
 \ifnum #1\expandafter \@firstoftwo
 \else \expandafter \@secondoftwo
 \fi
}%
\providecommand \@ifx [1]{%
 \ifx #1\expandafter \@firstoftwo
 \else \expandafter \@secondoftwo
 \fi
}%
\providecommand \natexlab [1]{#1}%
\providecommand \enquote  [1]{``#1''}%
\providecommand \bibnamefont  [1]{#1}%
\providecommand \bibfnamefont [1]{#1}%
\providecommand \citenamefont [1]{#1}%
\providecommand \href@noop [0]{\@secondoftwo}%
\providecommand \href [0]{\begingroup \@sanitize@url \@href}%
\providecommand \@href[1]{\@@startlink{#1}\@@href}%
\providecommand \@@href[1]{\endgroup#1\@@endlink}%
\providecommand \@sanitize@url [0]{\catcode `\\12\catcode `\$12\catcode
  `\&12\catcode `\#12\catcode `\^12\catcode `\_12\catcode `\%12\relax}%
\providecommand \@@startlink[1]{}%
\providecommand \@@endlink[0]{}%
\providecommand \url  [0]{\begingroup\@sanitize@url \@url }%
\providecommand \@url [1]{\endgroup\@href {#1}{\urlprefix }}%
\providecommand \urlprefix  [0]{URL }%
\providecommand \Eprint [0]{\href }%
\providecommand \doibase [0]{https://doi.org/}%
\providecommand \selectlanguage [0]{\@gobble}%
\providecommand \bibinfo  [0]{\@secondoftwo}%
\providecommand \bibfield  [0]{\@secondoftwo}%
\providecommand \translation [1]{[#1]}%
\providecommand \BibitemOpen [0]{}%
\providecommand \bibitemStop [0]{}%
\providecommand \bibitemNoStop [0]{.\EOS\space}%
\providecommand \EOS [0]{\spacefactor3000\relax}%
\providecommand \BibitemShut  [1]{\csname bibitem#1\endcsname}%
\let\auto@bib@innerbib\@empty
\bibitem [{\citenamefont {Kaufman}\ and\ \citenamefont
  {Ni}(2021)}]{Kaufman2021quantum}%
  \BibitemOpen
  \bibfield  {author} {\bibinfo {author} {\bibfnamefont {A.~M.}\ \bibnamefont
  {Kaufman}}\ and\ \bibinfo {author} {\bibfnamefont {K.-K.}\ \bibnamefont
  {Ni}},\ }\bibfield  {title} {\bibinfo {title} {Quantum science with optical
  tweezer arrays of ultracold atoms and molecules},\ }\href
  {http://dx.doi.org/10.1038/s41567-021-01357-2} {\bibfield  {journal}
  {\bibinfo  {journal} {Nat.~Phys.}\ }\textbf {\bibinfo {volume} {17}},\
  \bibinfo {pages} {1324} (\bibinfo {year} {2021})}\BibitemShut {NoStop}%
\bibitem [{\citenamefont {Browaeys}\ and\ \citenamefont
  {Lahaye}(2020)}]{browaeys2020many}%
  \BibitemOpen
  \bibfield  {author} {\bibinfo {author} {\bibfnamefont {A.}~\bibnamefont
  {Browaeys}}\ and\ \bibinfo {author} {\bibfnamefont {T.}~\bibnamefont
  {Lahaye}},\ }\bibfield  {title} {\bibinfo {title} {{Many-body physics with
  individually controlled Rydberg atoms}},\ }\href
  {https://doi.org/10.1038/s41567-019-0733-z} {\bibfield  {journal} {\bibinfo
  {journal} {Nat.~Phys.}\ }\textbf {\bibinfo {volume} {16}},\ \bibinfo {pages}
  {132} (\bibinfo {year} {2020})}\BibitemShut {NoStop}%
\bibitem [{\citenamefont {Saffman}\ \emph {et~al.}(2010)\citenamefont
  {Saffman}, \citenamefont {Walker},\ and\ \citenamefont
  {Mølmer}}]{Saffman2010quantum}%
  \BibitemOpen
  \bibfield  {author} {\bibinfo {author} {\bibfnamefont {M.}~\bibnamefont
  {Saffman}}, \bibinfo {author} {\bibfnamefont {T.~G.}\ \bibnamefont
  {Walker}},\ and\ \bibinfo {author} {\bibfnamefont {K.}~\bibnamefont
  {Mølmer}},\ }\bibfield  {title} {\bibinfo {title} {{Quantum information with
  Rydberg atoms}},\ }\href {http://dx.doi.org/10.1103/RevModPhys.82.2313}
  {\bibfield  {journal} {\bibinfo  {journal} {Rev.~Mod.~Phys.}\ }\textbf
  {\bibinfo {volume} {82}},\ \bibinfo {pages} {2313} (\bibinfo {year}
  {2010})}\BibitemShut {NoStop}%
\bibitem [{\citenamefont {Morgado}\ and\ \citenamefont
  {Whitlock}(2021)}]{morgado2021quantum}%
  \BibitemOpen
  \bibfield  {author} {\bibinfo {author} {\bibfnamefont {M.}~\bibnamefont
  {Morgado}}\ and\ \bibinfo {author} {\bibfnamefont {S.}~\bibnamefont
  {Whitlock}},\ }\bibfield  {title} {\bibinfo {title} {{Quantum simulation and
  computing with Rydberg-interacting qubits}},\ }\href
  {http://dx.doi.org/10.1116/5.0036562} {\bibfield  {journal} {\bibinfo
  {journal} {AVS Quantum Sci.}\ }\textbf {\bibinfo {volume} {3}},\ \bibinfo
  {pages} {023501} (\bibinfo {year} {2021})}\BibitemShut {NoStop}%
\bibitem [{\citenamefont {Scholl}\ \emph {et~al.}(2021)\citenamefont {Scholl},
  \citenamefont {Schuler}, \citenamefont {Williams}, \citenamefont
  {Eberharter}, \citenamefont {Barredo}, \citenamefont {Schymik}, \citenamefont
  {Lienhard}, \citenamefont {Henry}, \citenamefont {Lang}, \citenamefont
  {Lahaye}, \citenamefont {L\"{a}uchli},\ and\ \citenamefont
  {Browaeys}}]{scholl2021quantum}%
  \BibitemOpen
  \bibfield  {author} {\bibinfo {author} {\bibfnamefont {P.}~\bibnamefont
  {Scholl}}, \bibinfo {author} {\bibfnamefont {M.}~\bibnamefont {Schuler}},
  \bibinfo {author} {\bibfnamefont {H.~J.}\ \bibnamefont {Williams}}, \bibinfo
  {author} {\bibfnamefont {A.~A.}\ \bibnamefont {Eberharter}}, \bibinfo
  {author} {\bibfnamefont {D.}~\bibnamefont {Barredo}}, \bibinfo {author}
  {\bibfnamefont {K.-N.}\ \bibnamefont {Schymik}}, \bibinfo {author}
  {\bibfnamefont {V.}~\bibnamefont {Lienhard}}, \bibinfo {author}
  {\bibfnamefont {L.-P.}\ \bibnamefont {Henry}}, \bibinfo {author}
  {\bibfnamefont {T.~C.}\ \bibnamefont {Lang}}, \bibinfo {author}
  {\bibfnamefont {T.}~\bibnamefont {Lahaye}}, \bibinfo {author} {\bibfnamefont
  {A.~M.}\ \bibnamefont {L\"{a}uchli}},\ and\ \bibinfo {author} {\bibfnamefont
  {A.}~\bibnamefont {Browaeys}},\ }\bibfield  {title} {\bibinfo {title}
  {{Quantum simulation of 2D antiferromagnets with hundreds of Rydberg
  atoms}},\ }\href {http://dx.doi.org/10.1038/s41586-021-03585-1} {\bibfield
  {journal} {\bibinfo  {journal} {Nature}\ }\textbf {\bibinfo {volume} {595}},\
  \bibinfo {pages} {233} (\bibinfo {year} {2021})}\BibitemShut {NoStop}%
\bibitem [{\citenamefont {Semeghini}\ \emph {et~al.}(2021)\citenamefont
  {Semeghini}, \citenamefont {Levine}, \citenamefont {Keesling}, \citenamefont
  {Ebadi}, \citenamefont {Wang}, \citenamefont {Bluvstein}, \citenamefont
  {Verresen}, \citenamefont {Pichler}, \citenamefont {Kalinowski},
  \citenamefont {Samajdar}, \citenamefont {Omran}, \citenamefont {Sachdev},
  \citenamefont {Vishwanath}, \citenamefont {Greiner}, \citenamefont
  {Vuletić},\ and\ \citenamefont {Lukin}}]{semeghini2021probing}%
  \BibitemOpen
  \bibfield  {author} {\bibinfo {author} {\bibfnamefont {G.}~\bibnamefont
  {Semeghini}}, \bibinfo {author} {\bibfnamefont {H.}~\bibnamefont {Levine}},
  \bibinfo {author} {\bibfnamefont {A.}~\bibnamefont {Keesling}}, \bibinfo
  {author} {\bibfnamefont {S.}~\bibnamefont {Ebadi}}, \bibinfo {author}
  {\bibfnamefont {T.~T.}\ \bibnamefont {Wang}}, \bibinfo {author}
  {\bibfnamefont {D.}~\bibnamefont {Bluvstein}}, \bibinfo {author}
  {\bibfnamefont {R.}~\bibnamefont {Verresen}}, \bibinfo {author}
  {\bibfnamefont {H.}~\bibnamefont {Pichler}}, \bibinfo {author} {\bibfnamefont
  {M.}~\bibnamefont {Kalinowski}}, \bibinfo {author} {\bibfnamefont
  {R.}~\bibnamefont {Samajdar}}, \bibinfo {author} {\bibfnamefont
  {A.}~\bibnamefont {Omran}}, \bibinfo {author} {\bibfnamefont
  {S.}~\bibnamefont {Sachdev}}, \bibinfo {author} {\bibfnamefont
  {A.}~\bibnamefont {Vishwanath}}, \bibinfo {author} {\bibfnamefont
  {M.}~\bibnamefont {Greiner}}, \bibinfo {author} {\bibfnamefont
  {V.}~\bibnamefont {Vuletić}},\ and\ \bibinfo {author} {\bibfnamefont
  {M.~D.}\ \bibnamefont {Lukin}},\ }\bibfield  {title} {\bibinfo {title}
  {{Probing topological spin liquids on a programmable quantum simulator}},\
  }\href {https://doi.org/10.1126/science.abi8794} {\bibfield  {journal}
  {\bibinfo  {journal} {Science}\ }\textbf {\bibinfo {volume} {374}},\ \bibinfo
  {pages} {1242} (\bibinfo {year} {2021})}\BibitemShut {NoStop}%
\bibitem [{\citenamefont {Madjarov}\ \emph {et~al.}(2020)\citenamefont
  {Madjarov}, \citenamefont {Covey}, \citenamefont {Shaw}, \citenamefont
  {Choi}, \citenamefont {Kale}, \citenamefont {Cooper}, \citenamefont
  {Pichler}, \citenamefont {Schkolnik}, \citenamefont {Williams},\ and\
  \citenamefont {Endres}}]{madjarov2020high}%
  \BibitemOpen
  \bibfield  {author} {\bibinfo {author} {\bibfnamefont {I.~S.}\ \bibnamefont
  {Madjarov}}, \bibinfo {author} {\bibfnamefont {J.~P.}\ \bibnamefont {Covey}},
  \bibinfo {author} {\bibfnamefont {A.~L.}\ \bibnamefont {Shaw}}, \bibinfo
  {author} {\bibfnamefont {J.}~\bibnamefont {Choi}}, \bibinfo {author}
  {\bibfnamefont {A.}~\bibnamefont {Kale}}, \bibinfo {author} {\bibfnamefont
  {A.}~\bibnamefont {Cooper}}, \bibinfo {author} {\bibfnamefont
  {H.}~\bibnamefont {Pichler}}, \bibinfo {author} {\bibfnamefont
  {V.}~\bibnamefont {Schkolnik}}, \bibinfo {author} {\bibfnamefont {J.~R.}\
  \bibnamefont {Williams}},\ and\ \bibinfo {author} {\bibfnamefont
  {M.}~\bibnamefont {Endres}},\ }\bibfield  {title} {\bibinfo {title}
  {{High-fidelity entanglement and detection of alkaline-earth Rydberg
  atoms}},\ }\href {http://dx.doi.org/10.1038/s41567-020-0903-z} {\bibfield
  {journal} {\bibinfo  {journal} {Nat.~Phys.}\ }\textbf {\bibinfo {volume}
  {16}},\ \bibinfo {pages} {857} (\bibinfo {year} {2020})}\BibitemShut
  {NoStop}%
\bibitem [{\citenamefont {Graham}\ \emph {et~al.}(2022)\citenamefont {Graham},
  \citenamefont {Song}, \citenamefont {Scott}, \citenamefont {Poole},
  \citenamefont {Phuttitarn}, \citenamefont {Jooya}, \citenamefont {Eichler},
  \citenamefont {Jiang}, \citenamefont {Marra}, \citenamefont {Grinkemeyer},
  \citenamefont {Kwon}, \citenamefont {Ebert}, \citenamefont {Cherek},
  \citenamefont {Lichtman}, \citenamefont {Gillette}, \citenamefont {Gilbert},
  \citenamefont {Bowman}, \citenamefont {Ballance}, \citenamefont {Campbell},
  \citenamefont {Dahl}, \citenamefont {Crawford}, \citenamefont {Blunt},
  \citenamefont {Rogers}, \citenamefont {Noel},\ and\ \citenamefont
  {Saffman}}]{graham2022multi}%
  \BibitemOpen
  \bibfield  {author} {\bibinfo {author} {\bibfnamefont {T.~M.}\ \bibnamefont
  {Graham}}, \bibinfo {author} {\bibfnamefont {Y.}~\bibnamefont {Song}},
  \bibinfo {author} {\bibfnamefont {J.}~\bibnamefont {Scott}}, \bibinfo
  {author} {\bibfnamefont {C.}~\bibnamefont {Poole}}, \bibinfo {author}
  {\bibfnamefont {L.}~\bibnamefont {Phuttitarn}}, \bibinfo {author}
  {\bibfnamefont {K.}~\bibnamefont {Jooya}}, \bibinfo {author} {\bibfnamefont
  {P.}~\bibnamefont {Eichler}}, \bibinfo {author} {\bibfnamefont
  {X.}~\bibnamefont {Jiang}}, \bibinfo {author} {\bibfnamefont
  {A.}~\bibnamefont {Marra}}, \bibinfo {author} {\bibfnamefont
  {B.}~\bibnamefont {Grinkemeyer}}, \bibinfo {author} {\bibfnamefont
  {M.}~\bibnamefont {Kwon}}, \bibinfo {author} {\bibfnamefont {M.}~\bibnamefont
  {Ebert}}, \bibinfo {author} {\bibfnamefont {J.}~\bibnamefont {Cherek}},
  \bibinfo {author} {\bibfnamefont {M.~T.}\ \bibnamefont {Lichtman}}, \bibinfo
  {author} {\bibfnamefont {M.}~\bibnamefont {Gillette}}, \bibinfo {author}
  {\bibfnamefont {J.}~\bibnamefont {Gilbert}}, \bibinfo {author} {\bibfnamefont
  {D.}~\bibnamefont {Bowman}}, \bibinfo {author} {\bibfnamefont
  {T.}~\bibnamefont {Ballance}}, \bibinfo {author} {\bibfnamefont
  {C.}~\bibnamefont {Campbell}}, \bibinfo {author} {\bibfnamefont {E.~D.}\
  \bibnamefont {Dahl}}, \bibinfo {author} {\bibfnamefont {O.}~\bibnamefont
  {Crawford}}, \bibinfo {author} {\bibfnamefont {N.~S.}\ \bibnamefont {Blunt}},
  \bibinfo {author} {\bibfnamefont {B.}~\bibnamefont {Rogers}}, \bibinfo
  {author} {\bibfnamefont {T.}~\bibnamefont {Noel}},\ and\ \bibinfo {author}
  {\bibfnamefont {M.}~\bibnamefont {Saffman}},\ }\bibfield  {title} {\bibinfo
  {title} {{Multi-qubit entanglement and algorithms on a neutral-atom quantum
  computer}},\ }\href {http://dx.doi.org/10.1038/s41586-022-04603-6} {\bibfield
   {journal} {\bibinfo  {journal} {Nature}\ }\textbf {\bibinfo {volume}
  {604}},\ \bibinfo {pages} {457} (\bibinfo {year} {2022})}\BibitemShut
  {NoStop}%
\bibitem [{\citenamefont {Ma}\ \emph {et~al.}(2023)\citenamefont {Ma},
  \citenamefont {Liu}, \citenamefont {Peng}, \citenamefont {Zhang},
  \citenamefont {Jandura}, \citenamefont {Claes}, \citenamefont {Burgers},
  \citenamefont {Pupillo}, \citenamefont {Puri},\ and\ \citenamefont
  {Thompson}}]{ma2023high}%
  \BibitemOpen
  \bibfield  {author} {\bibinfo {author} {\bibfnamefont {S.}~\bibnamefont
  {Ma}}, \bibinfo {author} {\bibfnamefont {G.}~\bibnamefont {Liu}}, \bibinfo
  {author} {\bibfnamefont {P.}~\bibnamefont {Peng}}, \bibinfo {author}
  {\bibfnamefont {B.}~\bibnamefont {Zhang}}, \bibinfo {author} {\bibfnamefont
  {S.}~\bibnamefont {Jandura}}, \bibinfo {author} {\bibfnamefont
  {J.}~\bibnamefont {Claes}}, \bibinfo {author} {\bibfnamefont {A.~P.}\
  \bibnamefont {Burgers}}, \bibinfo {author} {\bibfnamefont {G.}~\bibnamefont
  {Pupillo}}, \bibinfo {author} {\bibfnamefont {S.}~\bibnamefont {Puri}},\ and\
  \bibinfo {author} {\bibfnamefont {J.~D.}\ \bibnamefont {Thompson}},\
  }\bibfield  {title} {\bibinfo {title} {{High-fidelity gates and mid-circuit
  erasure conversion in an atomic qubit}},\ }\href
  {https://doi.org/10.1038/s41586-023-06438-1} {\bibfield  {journal} {\bibinfo
  {journal} {Nature}\ }\textbf {\bibinfo {volume} {622}},\ \bibinfo {pages}
  {279} (\bibinfo {year} {2023})}\BibitemShut {NoStop}%
\bibitem [{\citenamefont {Bluvstein}\ \emph {et~al.}(2024)\citenamefont
  {Bluvstein}, \citenamefont {Evered}, \citenamefont {Geim}, \citenamefont
  {Li}, \citenamefont {Zhou}, \citenamefont {Manovitz}, \citenamefont {Ebadi},
  \citenamefont {Cain}, \citenamefont {Kalinowski}, \citenamefont {Hangleiter},
  \citenamefont {Bonilla~Ataides}, \citenamefont {Maskara}, \citenamefont
  {Cong}, \citenamefont {Gao}, \citenamefont {Sales~Rodriguez}, \citenamefont
  {Karolyshyn}, \citenamefont {Semeghini}, \citenamefont {Gullans},
  \citenamefont {Greiner}, \citenamefont {Vuletić},\ and\ \citenamefont
  {Lukin}}]{bluvstein2024logical}%
  \BibitemOpen
  \bibfield  {author} {\bibinfo {author} {\bibfnamefont {D.}~\bibnamefont
  {Bluvstein}}, \bibinfo {author} {\bibfnamefont {S.~J.}\ \bibnamefont
  {Evered}}, \bibinfo {author} {\bibfnamefont {A.~A.}\ \bibnamefont {Geim}},
  \bibinfo {author} {\bibfnamefont {S.~H.}\ \bibnamefont {Li}}, \bibinfo
  {author} {\bibfnamefont {H.}~\bibnamefont {Zhou}}, \bibinfo {author}
  {\bibfnamefont {T.}~\bibnamefont {Manovitz}}, \bibinfo {author}
  {\bibfnamefont {S.}~\bibnamefont {Ebadi}}, \bibinfo {author} {\bibfnamefont
  {M.}~\bibnamefont {Cain}}, \bibinfo {author} {\bibfnamefont {M.}~\bibnamefont
  {Kalinowski}}, \bibinfo {author} {\bibfnamefont {D.}~\bibnamefont
  {Hangleiter}}, \bibinfo {author} {\bibfnamefont {J.~P.}\ \bibnamefont
  {Bonilla~Ataides}}, \bibinfo {author} {\bibfnamefont {N.}~\bibnamefont
  {Maskara}}, \bibinfo {author} {\bibfnamefont {I.}~\bibnamefont {Cong}},
  \bibinfo {author} {\bibfnamefont {X.}~\bibnamefont {Gao}}, \bibinfo {author}
  {\bibfnamefont {P.}~\bibnamefont {Sales~Rodriguez}}, \bibinfo {author}
  {\bibfnamefont {T.}~\bibnamefont {Karolyshyn}}, \bibinfo {author}
  {\bibfnamefont {G.}~\bibnamefont {Semeghini}}, \bibinfo {author}
  {\bibfnamefont {M.~J.}\ \bibnamefont {Gullans}}, \bibinfo {author}
  {\bibfnamefont {M.}~\bibnamefont {Greiner}}, \bibinfo {author} {\bibfnamefont
  {V.}~\bibnamefont {Vuletić}},\ and\ \bibinfo {author} {\bibfnamefont
  {M.~D.}\ \bibnamefont {Lukin}},\ }\bibfield  {title} {\bibinfo {title}
  {{Logical quantum processor based on reconfigurable atom arrays}},\ }\href
  {https://doi.org/10.1038/s41586-023-06927-3} {\bibfield  {journal} {\bibinfo
  {journal} {Nature}\ }\textbf {\bibinfo {volume} {626}},\ \bibinfo {pages}
  {58} (\bibinfo {year} {2024})}\BibitemShut {NoStop}%
\bibitem [{\citenamefont {Singh}\ \emph {et~al.}(2022)\citenamefont {Singh},
  \citenamefont {Anand}, \citenamefont {Pocklington}, \citenamefont {Kemp},\
  and\ \citenamefont {Bernien}}]{Singh2022dual}%
  \BibitemOpen
  \bibfield  {author} {\bibinfo {author} {\bibfnamefont {K.}~\bibnamefont
  {Singh}}, \bibinfo {author} {\bibfnamefont {S.}~\bibnamefont {Anand}},
  \bibinfo {author} {\bibfnamefont {A.}~\bibnamefont {Pocklington}}, \bibinfo
  {author} {\bibfnamefont {J.~T.}\ \bibnamefont {Kemp}},\ and\ \bibinfo
  {author} {\bibfnamefont {H.}~\bibnamefont {Bernien}},\ }\bibfield  {title}
  {\bibinfo {title} {{Dual-element, two-dimensional atom array with
  continuous-mode operation}},\ }\href
  {http://dx.doi.org/10.1103/PhysRevX.12.011040} {\bibfield  {journal}
  {\bibinfo  {journal} {Phys.~ Rev.~X}\ }\textbf {\bibinfo {volume} {12}},\
  \bibinfo {pages} {011040} (\bibinfo {year} {2022})}\BibitemShut {NoStop}%
\bibitem [{\citenamefont {Sheng}\ \emph {et~al.}(2022)\citenamefont {Sheng},
  \citenamefont {Hou}, \citenamefont {He}, \citenamefont {Wang}, \citenamefont
  {Guo}, \citenamefont {Zhuang}, \citenamefont {Mamat}, \citenamefont {Xu},
  \citenamefont {Liu}, \citenamefont {Wang},\ and\ \citenamefont
  {Zhan}}]{Sheng2022defect}%
  \BibitemOpen
  \bibfield  {author} {\bibinfo {author} {\bibfnamefont {C.}~\bibnamefont
  {Sheng}}, \bibinfo {author} {\bibfnamefont {J.}~\bibnamefont {Hou}}, \bibinfo
  {author} {\bibfnamefont {X.}~\bibnamefont {He}}, \bibinfo {author}
  {\bibfnamefont {K.}~\bibnamefont {Wang}}, \bibinfo {author} {\bibfnamefont
  {R.}~\bibnamefont {Guo}}, \bibinfo {author} {\bibfnamefont {J.}~\bibnamefont
  {Zhuang}}, \bibinfo {author} {\bibfnamefont {B.}~\bibnamefont {Mamat}},
  \bibinfo {author} {\bibfnamefont {P.}~\bibnamefont {Xu}}, \bibinfo {author}
  {\bibfnamefont {M.}~\bibnamefont {Liu}}, \bibinfo {author} {\bibfnamefont
  {J.}~\bibnamefont {Wang}},\ and\ \bibinfo {author} {\bibfnamefont
  {M.}~\bibnamefont {Zhan}},\ }\bibfield  {title} {\bibinfo {title}
  {{Defect-free arbitrary-geometry assembly of mixed-species atom arrays}},\
  }\href {http://dx.doi.org/10.1103/PhysRevLett.128.083202} {\bibfield
  {journal} {\bibinfo  {journal} {Phys.~Rev.~Lett.}\ }\textbf {\bibinfo
  {volume} {128}},\ \bibinfo {pages} {083202} (\bibinfo {year}
  {2022})}\BibitemShut {NoStop}%
\bibitem [{\citenamefont {Zhang}\ \emph {et~al.}(2022)\citenamefont {Zhang},
  \citenamefont {Picard}, \citenamefont {Cairncross}, \citenamefont {Wang},
  \citenamefont {Yu}, \citenamefont {Fang},\ and\ \citenamefont
  {Ni}}]{Zhang2022optical}%
  \BibitemOpen
  \bibfield  {author} {\bibinfo {author} {\bibfnamefont {J.~T.}\ \bibnamefont
  {Zhang}}, \bibinfo {author} {\bibfnamefont {L.~R.~B.}\ \bibnamefont
  {Picard}}, \bibinfo {author} {\bibfnamefont {W.~B.}\ \bibnamefont
  {Cairncross}}, \bibinfo {author} {\bibfnamefont {K.}~\bibnamefont {Wang}},
  \bibinfo {author} {\bibfnamefont {Y.}~\bibnamefont {Yu}}, \bibinfo {author}
  {\bibfnamefont {F.}~\bibnamefont {Fang}},\ and\ \bibinfo {author}
  {\bibfnamefont {K.-K.}\ \bibnamefont {Ni}},\ }\bibfield  {title} {\bibinfo
  {title} {An optical tweezer array of ground-state polar molecules},\ }\href
  {http://dx.doi.org/10.1088/2058-9565/ac676c} {\bibfield  {journal} {\bibinfo
  {journal} {Quantum Science and Technology}\ }\textbf {\bibinfo {volume}
  {7}},\ \bibinfo {pages} {035006} (\bibinfo {year} {2022})}\BibitemShut
  {NoStop}%
\bibitem [{\citenamefont {Bao}\ \emph {et~al.}(2023)\citenamefont {Bao},
  \citenamefont {Yu}, \citenamefont {Anderegg}, \citenamefont {Chae},
  \citenamefont {Ketterle}, \citenamefont {Ni},\ and\ \citenamefont
  {Doyle}}]{bao2023dipolar}%
  \BibitemOpen
  \bibfield  {author} {\bibinfo {author} {\bibfnamefont {Y.}~\bibnamefont
  {Bao}}, \bibinfo {author} {\bibfnamefont {S.~S.}\ \bibnamefont {Yu}},
  \bibinfo {author} {\bibfnamefont {L.}~\bibnamefont {Anderegg}}, \bibinfo
  {author} {\bibfnamefont {E.}~\bibnamefont {Chae}}, \bibinfo {author}
  {\bibfnamefont {W.}~\bibnamefont {Ketterle}}, \bibinfo {author}
  {\bibfnamefont {K.-K.}\ \bibnamefont {Ni}},\ and\ \bibinfo {author}
  {\bibfnamefont {J.~M.}\ \bibnamefont {Doyle}},\ }\bibfield  {title} {\bibinfo
  {title} {{Dipolar spin-exchange and entanglement between molecules in an
  optical tweezer array}},\ }\href {https://doi.org/10.1126/science.adf8999}
  {\bibfield  {journal} {\bibinfo  {journal} {Science}\ }\textbf {\bibinfo
  {volume} {382}},\ \bibinfo {pages} {1138} (\bibinfo {year}
  {2023})}\BibitemShut {NoStop}%
\bibitem [{\citenamefont {Holland}\ \emph {et~al.}(2023)\citenamefont
  {Holland}, \citenamefont {Lu},\ and\ \citenamefont
  {Cheuk}}]{holland2023demand}%
  \BibitemOpen
  \bibfield  {author} {\bibinfo {author} {\bibfnamefont {C.~M.}\ \bibnamefont
  {Holland}}, \bibinfo {author} {\bibfnamefont {Y.}~\bibnamefont {Lu}},\ and\
  \bibinfo {author} {\bibfnamefont {L.~W.}\ \bibnamefont {Cheuk}},\ }\bibfield
  {title} {\bibinfo {title} {{On-demand entanglement of molecules in a
  reconfigurable optical tweezer array}},\ }\href
  {https://doi.org/10.1126/science.adf4272} {\bibfield  {journal} {\bibinfo
  {journal} {Science}\ }\textbf {\bibinfo {volume} {382}},\ \bibinfo {pages}
  {1143} (\bibinfo {year} {2023})}\BibitemShut {NoStop}%
\bibitem [{\citenamefont {Madjarov}\ \emph {et~al.}(2019)\citenamefont
  {Madjarov}, \citenamefont {Cooper}, \citenamefont {Shaw}, \citenamefont
  {Covey}, \citenamefont {Schkolnik}, \citenamefont {Yoon}, \citenamefont
  {Williams},\ and\ \citenamefont {Endres}}]{madjarov2019atomic}%
  \BibitemOpen
  \bibfield  {author} {\bibinfo {author} {\bibfnamefont {I.~S.}\ \bibnamefont
  {Madjarov}}, \bibinfo {author} {\bibfnamefont {A.}~\bibnamefont {Cooper}},
  \bibinfo {author} {\bibfnamefont {A.~L.}\ \bibnamefont {Shaw}}, \bibinfo
  {author} {\bibfnamefont {J.~P.}\ \bibnamefont {Covey}}, \bibinfo {author}
  {\bibfnamefont {V.}~\bibnamefont {Schkolnik}}, \bibinfo {author}
  {\bibfnamefont {T.~H.}\ \bibnamefont {Yoon}}, \bibinfo {author}
  {\bibfnamefont {J.~R.}\ \bibnamefont {Williams}},\ and\ \bibinfo {author}
  {\bibfnamefont {M.}~\bibnamefont {Endres}},\ }\bibfield  {title} {\bibinfo
  {title} {{An atomic-array optical clock with single-atom readout}},\ }\href
  {https://link.aps.org/doi/10.1103/PhysRevX.9.041052} {\bibfield  {journal}
  {\bibinfo  {journal} {Phys.~Rev.~X}\ }\textbf {\bibinfo {volume} {9}},\
  \bibinfo {pages} {041052} (\bibinfo {year} {2019})}\BibitemShut {NoStop}%
\bibitem [{\citenamefont {Young}\ \emph {et~al.}(2020)\citenamefont {Young},
  \citenamefont {Eckner}, \citenamefont {Milner}, \citenamefont {Kedar},
  \citenamefont {Norcia}, \citenamefont {Oelker}, \citenamefont {Schine},
  \citenamefont {Ye},\ and\ \citenamefont {Kaufman}}]{young2020half}%
  \BibitemOpen
  \bibfield  {author} {\bibinfo {author} {\bibfnamefont {A.~W.}\ \bibnamefont
  {Young}}, \bibinfo {author} {\bibfnamefont {W.~J.}\ \bibnamefont {Eckner}},
  \bibinfo {author} {\bibfnamefont {W.~R.}\ \bibnamefont {Milner}}, \bibinfo
  {author} {\bibfnamefont {D.}~\bibnamefont {Kedar}}, \bibinfo {author}
  {\bibfnamefont {M.~A.}\ \bibnamefont {Norcia}}, \bibinfo {author}
  {\bibfnamefont {E.}~\bibnamefont {Oelker}}, \bibinfo {author} {\bibfnamefont
  {N.}~\bibnamefont {Schine}}, \bibinfo {author} {\bibfnamefont
  {J.}~\bibnamefont {Ye}},\ and\ \bibinfo {author} {\bibfnamefont {A.~M.}\
  \bibnamefont {Kaufman}},\ }\bibfield  {title} {\bibinfo {title}
  {{Half-minute-scale atomic coherence and high relative stability in a tweezer
  clock}},\ }\href {http://dx.doi.org/10.1038/s41586-020-3009-y} {\bibfield
  {journal} {\bibinfo  {journal} {Nature}\ }\textbf {\bibinfo {volume} {588}},\
  \bibinfo {pages} {408} (\bibinfo {year} {2020})}\BibitemShut {NoStop}%
\bibitem [{\citenamefont {Yan}\ \emph {et~al.}(2023)\citenamefont {Yan},
  \citenamefont {Ho}, \citenamefont {Lu}, \citenamefont {Masson}, \citenamefont
  {Asenjo-Garcia},\ and\ \citenamefont {Stamper-Kurn}}]{Yan2023superradiant}%
  \BibitemOpen
  \bibfield  {author} {\bibinfo {author} {\bibfnamefont {Z.}~\bibnamefont
  {Yan}}, \bibinfo {author} {\bibfnamefont {J.}~\bibnamefont {Ho}}, \bibinfo
  {author} {\bibfnamefont {Y.-H.}\ \bibnamefont {Lu}}, \bibinfo {author}
  {\bibfnamefont {S.~J.}\ \bibnamefont {Masson}}, \bibinfo {author}
  {\bibfnamefont {A.}~\bibnamefont {Asenjo-Garcia}},\ and\ \bibinfo {author}
  {\bibfnamefont {D.~M.}\ \bibnamefont {Stamper-Kurn}},\ }\bibfield  {title}
  {\bibinfo {title} {{Superradiant and subradiant cavity scattering by atom
  arrays}},\ }\href {http://dx.doi.org/10.1103/PhysRevLett.131.253603}
  {\bibfield  {journal} {\bibinfo  {journal} {Phys.~Rev.~Lett.}\ }\textbf
  {\bibinfo {volume} {131}},\ \bibinfo {pages} {253603} (\bibinfo {year}
  {2023})}\BibitemShut {NoStop}%
\bibitem [{\citenamefont {Asenjo-Garcia}\ \emph {et~al.}(2017)\citenamefont
  {Asenjo-Garcia}, \citenamefont {Moreno-Cardoner}, \citenamefont {Albrecht},
  \citenamefont {Kimble},\ and\ \citenamefont
  {Chang}}]{AsenjoGarcia2017exponential}%
  \BibitemOpen
  \bibfield  {author} {\bibinfo {author} {\bibfnamefont {A.}~\bibnamefont
  {Asenjo-Garcia}}, \bibinfo {author} {\bibfnamefont {M.}~\bibnamefont
  {Moreno-Cardoner}}, \bibinfo {author} {\bibfnamefont {A.}~\bibnamefont
  {Albrecht}}, \bibinfo {author} {\bibfnamefont {H.~J.}\ \bibnamefont
  {Kimble}},\ and\ \bibinfo {author} {\bibfnamefont {D.~E.}\ \bibnamefont
  {Chang}},\ }\bibfield  {title} {\bibinfo {title} {{Exponential improvement in
  photon storage fidelities using subradiance and “selective radiance” in
  atomic arrays}},\ }\href {http://dx.doi.org/10.1103/PhysRevX.7.031024}
  {\bibfield  {journal} {\bibinfo  {journal} {Phys.~Rev.~X}\ }\textbf {\bibinfo
  {volume} {7}},\ \bibinfo {pages} {031024} (\bibinfo {year}
  {2017})}\BibitemShut {NoStop}%
\bibitem [{\citenamefont {Holzinger}\ \emph {et~al.}(2024)\citenamefont
  {Holzinger}, \citenamefont {Peter}, \citenamefont {Ostermann}, \citenamefont
  {Ritsch},\ and\ \citenamefont {Yelin}}]{Holzinger2024harnessing}%
  \BibitemOpen
  \bibfield  {author} {\bibinfo {author} {\bibfnamefont {R.}~\bibnamefont
  {Holzinger}}, \bibinfo {author} {\bibfnamefont {J.~S.}\ \bibnamefont
  {Peter}}, \bibinfo {author} {\bibfnamefont {S.}~\bibnamefont {Ostermann}},
  \bibinfo {author} {\bibfnamefont {H.}~\bibnamefont {Ritsch}},\ and\ \bibinfo
  {author} {\bibfnamefont {S.}~\bibnamefont {Yelin}},\ }\bibfield  {title}
  {\bibinfo {title} {{Harnessing quantum emitter rings for efficient energy
  transport and trapping}},\ }\href {http://dx.doi.org/10.1364/OPTICAQ.510021}
  {\bibfield  {journal} {\bibinfo  {journal} {Optica Quantum}\ }\textbf
  {\bibinfo {volume} {2}},\ \bibinfo {pages} {57} (\bibinfo {year}
  {2024})}\BibitemShut {NoStop}%
\bibitem [{\citenamefont {Masson}\ \emph {et~al.}(2024)\citenamefont {Masson},
  \citenamefont {Covey}, \citenamefont {Will},\ and\ \citenamefont
  {Asenjo-Garcia}}]{Masson2024dicke}%
  \BibitemOpen
  \bibfield  {author} {\bibinfo {author} {\bibfnamefont {S.~J.}\ \bibnamefont
  {Masson}}, \bibinfo {author} {\bibfnamefont {J.~P.}\ \bibnamefont {Covey}},
  \bibinfo {author} {\bibfnamefont {S.}~\bibnamefont {Will}},\ and\ \bibinfo
  {author} {\bibfnamefont {A.}~\bibnamefont {Asenjo-Garcia}},\ }\bibfield
  {title} {\bibinfo {title} {{Dicke superradiance in ordered arrays of
  multilevel atoms}},\ }\href {http://dx.doi.org/10.1103/PRXQuantum.5.010344}
  {\bibfield  {journal} {\bibinfo  {journal} {PRX Quantum}\ }\textbf {\bibinfo
  {volume} {5}},\ \bibinfo {pages} {010344} (\bibinfo {year}
  {2024})}\BibitemShut {NoStop}%
\bibitem [{\citenamefont {Grotti}\ \emph {et~al.}(2018)\citenamefont {Grotti},
  \citenamefont {Koller}, \citenamefont {Vogt}, \citenamefont {H\"{a}fner},
  \citenamefont {Sterr}, \citenamefont {Lisdat}, \citenamefont {Denker},
  \citenamefont {Voigt}, \citenamefont {Timmen}, \citenamefont {Rolland},
  \citenamefont {Baynes}, \citenamefont {Margolis}, \citenamefont {Zampaolo},
  \citenamefont {Thoumany}, \citenamefont {Pizzocaro}, \citenamefont {Rauf},
  \citenamefont {Bregolin}, \citenamefont {Tampellini}, \citenamefont
  {Barbieri}, \citenamefont {Zucco}, \citenamefont {Costanzo}, \citenamefont
  {Clivati}, \citenamefont {Levi},\ and\ \citenamefont
  {Calonico}}]{Grotti2018geodesy}%
  \BibitemOpen
  \bibfield  {author} {\bibinfo {author} {\bibfnamefont {J.}~\bibnamefont
  {Grotti}}, \bibinfo {author} {\bibfnamefont {S.}~\bibnamefont {Koller}},
  \bibinfo {author} {\bibfnamefont {S.}~\bibnamefont {Vogt}}, \bibinfo {author}
  {\bibfnamefont {S.}~\bibnamefont {H\"{a}fner}}, \bibinfo {author}
  {\bibfnamefont {U.}~\bibnamefont {Sterr}}, \bibinfo {author} {\bibfnamefont
  {C.}~\bibnamefont {Lisdat}}, \bibinfo {author} {\bibfnamefont
  {H.}~\bibnamefont {Denker}}, \bibinfo {author} {\bibfnamefont
  {C.}~\bibnamefont {Voigt}}, \bibinfo {author} {\bibfnamefont
  {L.}~\bibnamefont {Timmen}}, \bibinfo {author} {\bibfnamefont
  {A.}~\bibnamefont {Rolland}}, \bibinfo {author} {\bibfnamefont {F.~N.}\
  \bibnamefont {Baynes}}, \bibinfo {author} {\bibfnamefont {H.~S.}\
  \bibnamefont {Margolis}}, \bibinfo {author} {\bibfnamefont {M.}~\bibnamefont
  {Zampaolo}}, \bibinfo {author} {\bibfnamefont {P.}~\bibnamefont {Thoumany}},
  \bibinfo {author} {\bibfnamefont {M.}~\bibnamefont {Pizzocaro}}, \bibinfo
  {author} {\bibfnamefont {B.}~\bibnamefont {Rauf}}, \bibinfo {author}
  {\bibfnamefont {F.}~\bibnamefont {Bregolin}}, \bibinfo {author}
  {\bibfnamefont {A.}~\bibnamefont {Tampellini}}, \bibinfo {author}
  {\bibfnamefont {P.}~\bibnamefont {Barbieri}}, \bibinfo {author}
  {\bibfnamefont {M.}~\bibnamefont {Zucco}}, \bibinfo {author} {\bibfnamefont
  {G.~A.}\ \bibnamefont {Costanzo}}, \bibinfo {author} {\bibfnamefont
  {C.}~\bibnamefont {Clivati}}, \bibinfo {author} {\bibfnamefont
  {F.}~\bibnamefont {Levi}},\ and\ \bibinfo {author} {\bibfnamefont
  {D.}~\bibnamefont {Calonico}},\ }\bibfield  {title} {\bibinfo {title}
  {{Geodesy and metrology with a transportable optical clock}},\ }\href
  {http://dx.doi.org/10.1038/s41567-017-0042-3} {\bibfield  {journal} {\bibinfo
   {journal} {Nat.~Phys.}\ }\textbf {\bibinfo {volume} {14}},\ \bibinfo {pages}
  {437} (\bibinfo {year} {2018})}\BibitemShut {NoStop}%
\bibitem [{\citenamefont {Takamoto}\ \emph {et~al.}(2020)\citenamefont
  {Takamoto}, \citenamefont {Ushijima}, \citenamefont {Ohmae}, \citenamefont
  {Yahagi}, \citenamefont {Kokado}, \citenamefont {Shinkai},\ and\
  \citenamefont {Katori}}]{Takamoto2020test}%
  \BibitemOpen
  \bibfield  {author} {\bibinfo {author} {\bibfnamefont {M.}~\bibnamefont
  {Takamoto}}, \bibinfo {author} {\bibfnamefont {I.}~\bibnamefont {Ushijima}},
  \bibinfo {author} {\bibfnamefont {N.}~\bibnamefont {Ohmae}}, \bibinfo
  {author} {\bibfnamefont {T.}~\bibnamefont {Yahagi}}, \bibinfo {author}
  {\bibfnamefont {K.}~\bibnamefont {Kokado}}, \bibinfo {author} {\bibfnamefont
  {H.}~\bibnamefont {Shinkai}},\ and\ \bibinfo {author} {\bibfnamefont
  {H.}~\bibnamefont {Katori}},\ }\bibfield  {title} {\bibinfo {title} {{Test of
  general relativity by a pair of transportable optical lattice clocks}},\
  }\href {http://dx.doi.org/10.1038/s41566-020-0619-8} {\bibfield  {journal}
  {\bibinfo  {journal} {Nat.~Photonics}\ }\textbf {\bibinfo {volume} {14}},\
  \bibinfo {pages} {411} (\bibinfo {year} {2020})}\BibitemShut {NoStop}%
\bibitem [{\citenamefont {Elliott}\ \emph {et~al.}(2023)\citenamefont
  {Elliott}, \citenamefont {Aveline}, \citenamefont {Bigelow}, \citenamefont
  {Boegel}, \citenamefont {Botsi}, \citenamefont {Charron}, \citenamefont
  {D’Incao}, \citenamefont {Engels}, \citenamefont {Estrampes}, \citenamefont
  {Gaaloul}, \citenamefont {Kellogg}, \citenamefont {Kohel}, \citenamefont
  {Lay}, \citenamefont {Lundblad}, \citenamefont {Meister}, \citenamefont
  {Mossman}, \citenamefont {M\"{u}ller}, \citenamefont {M\"{u}ller},
  \citenamefont {Oudrhiri}, \citenamefont {Phillips}, \citenamefont {Pichery},
  \citenamefont {Rasel}, \citenamefont {Sackett}, \citenamefont {Sbroscia},
  \citenamefont {Schleich}, \citenamefont {Thompson},\ and\ \citenamefont
  {Williams}}]{Elliott2023quantum}%
  \BibitemOpen
  \bibfield  {author} {\bibinfo {author} {\bibfnamefont {E.~R.}\ \bibnamefont
  {Elliott}}, \bibinfo {author} {\bibfnamefont {D.~C.}\ \bibnamefont
  {Aveline}}, \bibinfo {author} {\bibfnamefont {N.~P.}\ \bibnamefont
  {Bigelow}}, \bibinfo {author} {\bibfnamefont {P.}~\bibnamefont {Boegel}},
  \bibinfo {author} {\bibfnamefont {S.}~\bibnamefont {Botsi}}, \bibinfo
  {author} {\bibfnamefont {E.}~\bibnamefont {Charron}}, \bibinfo {author}
  {\bibfnamefont {J.~P.}\ \bibnamefont {D’Incao}}, \bibinfo {author}
  {\bibfnamefont {P.}~\bibnamefont {Engels}}, \bibinfo {author} {\bibfnamefont
  {T.}~\bibnamefont {Estrampes}}, \bibinfo {author} {\bibfnamefont
  {N.}~\bibnamefont {Gaaloul}}, \bibinfo {author} {\bibfnamefont {J.~R.}\
  \bibnamefont {Kellogg}}, \bibinfo {author} {\bibfnamefont {J.~M.}\
  \bibnamefont {Kohel}}, \bibinfo {author} {\bibfnamefont {N.~E.}\ \bibnamefont
  {Lay}}, \bibinfo {author} {\bibfnamefont {N.}~\bibnamefont {Lundblad}},
  \bibinfo {author} {\bibfnamefont {M.}~\bibnamefont {Meister}}, \bibinfo
  {author} {\bibfnamefont {M.~E.}\ \bibnamefont {Mossman}}, \bibinfo {author}
  {\bibfnamefont {G.}~\bibnamefont {M\"{u}ller}}, \bibinfo {author}
  {\bibfnamefont {H.}~\bibnamefont {M\"{u}ller}}, \bibinfo {author}
  {\bibfnamefont {K.}~\bibnamefont {Oudrhiri}}, \bibinfo {author}
  {\bibfnamefont {L.~E.}\ \bibnamefont {Phillips}}, \bibinfo {author}
  {\bibfnamefont {A.}~\bibnamefont {Pichery}}, \bibinfo {author} {\bibfnamefont
  {E.~M.}\ \bibnamefont {Rasel}}, \bibinfo {author} {\bibfnamefont {C.~A.}\
  \bibnamefont {Sackett}}, \bibinfo {author} {\bibfnamefont {M.}~\bibnamefont
  {Sbroscia}}, \bibinfo {author} {\bibfnamefont {W.~P.}\ \bibnamefont
  {Schleich}}, \bibinfo {author} {\bibfnamefont {R.~J.}\ \bibnamefont
  {Thompson}},\ and\ \bibinfo {author} {\bibfnamefont {J.~R.}\ \bibnamefont
  {Williams}},\ }\bibfield  {title} {\bibinfo {title} {{Quantum gas mixtures
  and dual-species atom interferometry in space}},\ }\href
  {http://dx.doi.org/10.1038/s41586-023-06645-w} {\bibfield  {journal}
  {\bibinfo  {journal} {Nature}\ }\textbf {\bibinfo {volume} {623}},\ \bibinfo
  {pages} {502} (\bibinfo {year} {2023})}\BibitemShut {NoStop}%
\bibitem [{\citenamefont {Endres}\ \emph {et~al.}(2016)\citenamefont {Endres},
  \citenamefont {Bernien}, \citenamefont {Keesling}, \citenamefont {Levine},
  \citenamefont {Anschuetz}, \citenamefont {Krajenbrink}, \citenamefont
  {Senko}, \citenamefont {Vuletic}, \citenamefont {Greiner},\ and\
  \citenamefont {Lukin}}]{endres2016atom}%
  \BibitemOpen
  \bibfield  {author} {\bibinfo {author} {\bibfnamefont {M.}~\bibnamefont
  {Endres}}, \bibinfo {author} {\bibfnamefont {H.}~\bibnamefont {Bernien}},
  \bibinfo {author} {\bibfnamefont {A.}~\bibnamefont {Keesling}}, \bibinfo
  {author} {\bibfnamefont {H.}~\bibnamefont {Levine}}, \bibinfo {author}
  {\bibfnamefont {E.~R.}\ \bibnamefont {Anschuetz}}, \bibinfo {author}
  {\bibfnamefont {A.}~\bibnamefont {Krajenbrink}}, \bibinfo {author}
  {\bibfnamefont {C.}~\bibnamefont {Senko}}, \bibinfo {author} {\bibfnamefont
  {V.}~\bibnamefont {Vuletic}}, \bibinfo {author} {\bibfnamefont
  {M.}~\bibnamefont {Greiner}},\ and\ \bibinfo {author} {\bibfnamefont {M.~D.}\
  \bibnamefont {Lukin}},\ }\bibfield  {title} {\bibinfo {title} {{Atom-by-atom
  assembly of defect-free one-dimensional cold atom arrays}},\ }\href
  {http://dx.doi.org/10.1126/science.aah3752} {\bibfield  {journal} {\bibinfo
  {journal} {Science}\ }\textbf {\bibinfo {volume} {354}},\ \bibinfo {pages}
  {1024} (\bibinfo {year} {2016})}\BibitemShut {NoStop}%
\bibitem [{\citenamefont {Burgers}\ \emph {et~al.}(2022)\citenamefont
  {Burgers}, \citenamefont {Ma}, \citenamefont {Saskin}, \citenamefont
  {Wilson}, \citenamefont {Alarcón}, \citenamefont {Greene},\ and\
  \citenamefont {Thompson}}]{Burgers2022controlling}%
  \BibitemOpen
  \bibfield  {author} {\bibinfo {author} {\bibfnamefont {A.~P.}\ \bibnamefont
  {Burgers}}, \bibinfo {author} {\bibfnamefont {S.}~\bibnamefont {Ma}},
  \bibinfo {author} {\bibfnamefont {S.}~\bibnamefont {Saskin}}, \bibinfo
  {author} {\bibfnamefont {J.}~\bibnamefont {Wilson}}, \bibinfo {author}
  {\bibfnamefont {M.~A.}\ \bibnamefont {Alarcón}}, \bibinfo {author}
  {\bibfnamefont {C.~H.}\ \bibnamefont {Greene}},\ and\ \bibinfo {author}
  {\bibfnamefont {J.~D.}\ \bibnamefont {Thompson}},\ }\bibfield  {title}
  {\bibinfo {title} {{Controlling Rydberg excitations using ion-core
  transitions in alkaline-earth atom-tweezer arrays}},\ }\href
  {https://link.aps.org/doi/10.1103/PRXQuantum.3.020326} {\bibfield  {journal}
  {\bibinfo  {journal} {PRX Quantum}\ }\textbf {\bibinfo {volume} {3}},\
  \bibinfo {pages} {020326} (\bibinfo {year} {2022})}\BibitemShut {NoStop}%
\bibitem [{\citenamefont {Barredo}\ \emph {et~al.}(2016)\citenamefont
  {Barredo}, \citenamefont {de~Léséleuc}, \citenamefont {Lienhard},
  \citenamefont {Lahaye},\ and\ \citenamefont {Browaeys}}]{Barredo2016atom}%
  \BibitemOpen
  \bibfield  {author} {\bibinfo {author} {\bibfnamefont {D.}~\bibnamefont
  {Barredo}}, \bibinfo {author} {\bibfnamefont {S.}~\bibnamefont
  {de~Léséleuc}}, \bibinfo {author} {\bibfnamefont {V.}~\bibnamefont
  {Lienhard}}, \bibinfo {author} {\bibfnamefont {T.}~\bibnamefont {Lahaye}},\
  and\ \bibinfo {author} {\bibfnamefont {A.}~\bibnamefont {Browaeys}},\
  }\bibfield  {title} {\bibinfo {title} {{An atom-by-atom assembler of
  defect-free arbitrary two-dimensional atomic arrays}},\ }\href
  {http://dx.doi.org/10.1126/science.aah3778} {\bibfield  {journal} {\bibinfo
  {journal} {Science}\ }\textbf {\bibinfo {volume} {354}},\ \bibinfo {pages}
  {1021} (\bibinfo {year} {2016})}\BibitemShut {NoStop}%
\bibitem [{\citenamefont {Kim}\ \emph {et~al.}(2019)\citenamefont {Kim},
  \citenamefont {Keesling}, \citenamefont {Omran}, \citenamefont {Levine},
  \citenamefont {Bernien}, \citenamefont {Greiner}, \citenamefont {Lukin},\
  and\ \citenamefont {Englund}}]{Kim2019large}%
  \BibitemOpen
  \bibfield  {author} {\bibinfo {author} {\bibfnamefont {D.}~\bibnamefont
  {Kim}}, \bibinfo {author} {\bibfnamefont {A.}~\bibnamefont {Keesling}},
  \bibinfo {author} {\bibfnamefont {A.}~\bibnamefont {Omran}}, \bibinfo
  {author} {\bibfnamefont {H.}~\bibnamefont {Levine}}, \bibinfo {author}
  {\bibfnamefont {H.}~\bibnamefont {Bernien}}, \bibinfo {author} {\bibfnamefont
  {M.}~\bibnamefont {Greiner}}, \bibinfo {author} {\bibfnamefont {M.~D.}\
  \bibnamefont {Lukin}},\ and\ \bibinfo {author} {\bibfnamefont {D.~R.}\
  \bibnamefont {Englund}},\ }\bibfield  {title} {\bibinfo {title} {{Large-scale
  uniform optical focus array generation with a phase spatial light
  modulator}},\ }\href {http://dx.doi.org/10.1364/OL.44.003178} {\bibfield
  {journal} {\bibinfo  {journal} {Opt.~Lett.}\ }\textbf {\bibinfo {volume}
  {44}},\ \bibinfo {pages} {3178} (\bibinfo {year} {2019})}\BibitemShut
  {NoStop}%
\bibitem [{\citenamefont {Wang}\ \emph {et~al.}(2020)\citenamefont {Wang},
  \citenamefont {Shevate}, \citenamefont {Wintermantel}, \citenamefont
  {Morgado}, \citenamefont {Lochead},\ and\ \citenamefont
  {Whitlock}}]{Wang2020preparation}%
  \BibitemOpen
  \bibfield  {author} {\bibinfo {author} {\bibfnamefont {Y.}~\bibnamefont
  {Wang}}, \bibinfo {author} {\bibfnamefont {S.}~\bibnamefont {Shevate}},
  \bibinfo {author} {\bibfnamefont {T.~M.}\ \bibnamefont {Wintermantel}},
  \bibinfo {author} {\bibfnamefont {M.}~\bibnamefont {Morgado}}, \bibinfo
  {author} {\bibfnamefont {G.}~\bibnamefont {Lochead}},\ and\ \bibinfo {author}
  {\bibfnamefont {S.}~\bibnamefont {Whitlock}},\ }\bibfield  {title} {\bibinfo
  {title} {{Preparation of hundreds of microscopic atomic ensembles in optical
  tweezer arrays}},\ }\href {http://dx.doi.org/10.1038/s41534-020-0285-1}
  {\bibfield  {journal} {\bibinfo  {journal} {npj Quantum Inf.}\ }\textbf
  {\bibinfo {volume} {6}},\ \bibinfo {pages} {54} (\bibinfo {year}
  {2020})}\BibitemShut {NoStop}%
\bibitem [{\citenamefont {Manetsch}\ \emph {et~al.}(2024)\citenamefont
  {Manetsch}, \citenamefont {Nomura}, \citenamefont {Bataille}, \citenamefont
  {Leung}, \citenamefont {Lv},\ and\ \citenamefont
  {Endres}}]{manetsch2024tweezer}%
  \BibitemOpen
  \bibfield  {author} {\bibinfo {author} {\bibfnamefont {H.~J.}\ \bibnamefont
  {Manetsch}}, \bibinfo {author} {\bibfnamefont {G.}~\bibnamefont {Nomura}},
  \bibinfo {author} {\bibfnamefont {E.}~\bibnamefont {Bataille}}, \bibinfo
  {author} {\bibfnamefont {K.~H.}\ \bibnamefont {Leung}}, \bibinfo {author}
  {\bibfnamefont {X.}~\bibnamefont {Lv}},\ and\ \bibinfo {author}
  {\bibfnamefont {M.}~\bibnamefont {Endres}},\ }\bibfield  {title} {\bibinfo
  {title} {{A tweezer array with 6100 highly coherent atomic qubits}},\ }\href
  {https://doi.org/10.48550/arXiv.2403.12021} {\bibfield  {journal} {\bibinfo
  {journal} {arXiv:2403.12021}\ } (\bibinfo {year} {2024})}\BibitemShut
  {NoStop}%
\bibitem [{\citenamefont {Huft}\ \emph {et~al.}(2022)\citenamefont {Huft},
  \citenamefont {Song}, \citenamefont {Graham}, \citenamefont {Jooya},
  \citenamefont {Deshpande}, \citenamefont {Fang}, \citenamefont {Kats},\ and\
  \citenamefont {Saffman}}]{huft2022simple}%
  \BibitemOpen
  \bibfield  {author} {\bibinfo {author} {\bibfnamefont {P.}~\bibnamefont
  {Huft}}, \bibinfo {author} {\bibfnamefont {Y.}~\bibnamefont {Song}}, \bibinfo
  {author} {\bibfnamefont {T.~M.}\ \bibnamefont {Graham}}, \bibinfo {author}
  {\bibfnamefont {K.}~\bibnamefont {Jooya}}, \bibinfo {author} {\bibfnamefont
  {S.}~\bibnamefont {Deshpande}}, \bibinfo {author} {\bibfnamefont
  {C.}~\bibnamefont {Fang}}, \bibinfo {author} {\bibfnamefont {M.}~\bibnamefont
  {Kats}},\ and\ \bibinfo {author} {\bibfnamefont {M.}~\bibnamefont
  {Saffman}},\ }\bibfield  {title} {\bibinfo {title} {{Simple, passive design
  for large optical trap arrays for single atoms}},\ }\href
  {https://doi.org/10.1103/PhysRevA.105.063111} {\bibfield  {journal} {\bibinfo
   {journal} {Phys.~Rev.~A}\ }\textbf {\bibinfo {volume} {105}},\ \bibinfo
  {pages} {063111} (\bibinfo {year} {2022})}\BibitemShut {NoStop}%
\bibitem [{\citenamefont {Pause}\ \emph {et~al.}(2024)\citenamefont {Pause},
  \citenamefont {Sturm}, \citenamefont {Mittenb\"{u}hler}, \citenamefont
  {Amann}, \citenamefont {Preuschoff}, \citenamefont {Sch\"{a}ffner},
  \citenamefont {Schlosser},\ and\ \citenamefont
  {Birkl}}]{pause2024supercharged}%
  \BibitemOpen
  \bibfield  {author} {\bibinfo {author} {\bibfnamefont {L.}~\bibnamefont
  {Pause}}, \bibinfo {author} {\bibfnamefont {L.}~\bibnamefont {Sturm}},
  \bibinfo {author} {\bibfnamefont {M.}~\bibnamefont {Mittenb\"{u}hler}},
  \bibinfo {author} {\bibfnamefont {S.}~\bibnamefont {Amann}}, \bibinfo
  {author} {\bibfnamefont {T.}~\bibnamefont {Preuschoff}}, \bibinfo {author}
  {\bibfnamefont {D.}~\bibnamefont {Sch\"{a}ffner}}, \bibinfo {author}
  {\bibfnamefont {M.}~\bibnamefont {Schlosser}},\ and\ \bibinfo {author}
  {\bibfnamefont {G.}~\bibnamefont {Birkl}},\ }\bibfield  {title} {\bibinfo
  {title} {{Supercharged two-dimensional tweezer array with more than 1000
  atomic qubits}},\ }\href {http://dx.doi.org/10.1364/OPTICA.513551} {\bibfield
   {journal} {\bibinfo  {journal} {Optica}\ }\textbf {\bibinfo {volume} {11}},\
  \bibinfo {pages} {222} (\bibinfo {year} {2024})}\BibitemShut {NoStop}%
\bibitem [{\citenamefont {Huang}\ \emph
  {et~al.}(2023{\natexlab{a}})\citenamefont {Huang}, \citenamefont {Yuan},
  \citenamefont {Holman}, \citenamefont {Kwon}, \citenamefont {Masson},
  \citenamefont {Gutierrez-Jauregui}, \citenamefont {Asenjo-Garcia},
  \citenamefont {Will},\ and\ \citenamefont {Yu}}]{huang2023metasurface}%
  \BibitemOpen
  \bibfield  {author} {\bibinfo {author} {\bibfnamefont {X.}~\bibnamefont
  {Huang}}, \bibinfo {author} {\bibfnamefont {W.}~\bibnamefont {Yuan}},
  \bibinfo {author} {\bibfnamefont {A.}~\bibnamefont {Holman}}, \bibinfo
  {author} {\bibfnamefont {M.}~\bibnamefont {Kwon}}, \bibinfo {author}
  {\bibfnamefont {S.~J.}\ \bibnamefont {Masson}}, \bibinfo {author}
  {\bibfnamefont {R.}~\bibnamefont {Gutierrez-Jauregui}}, \bibinfo {author}
  {\bibfnamefont {A.}~\bibnamefont {Asenjo-Garcia}}, \bibinfo {author}
  {\bibfnamefont {S.}~\bibnamefont {Will}},\ and\ \bibinfo {author}
  {\bibfnamefont {N.}~\bibnamefont {Yu}},\ }\bibfield  {title} {\bibinfo
  {title} {{Metasurface holographic optical traps for ultracold atoms}},\
  }\href {http://dx.doi.org/10.1016/j.pquantelec.2023.100470} {\bibfield
  {journal} {\bibinfo  {journal} {Prog.~Quantum Electron.}\ }\textbf {\bibinfo
  {volume} {89}},\ \bibinfo {pages} {100470} (\bibinfo {year}
  {2023}{\natexlab{a}})}\BibitemShut {NoStop}%
\bibitem [{\citenamefont {Fong}\ \emph {et~al.}(2010)\citenamefont {Fong},
  \citenamefont {Colburn}, \citenamefont {Ottusch}, \citenamefont {Visher},\
  and\ \citenamefont {Sievenpiper}}]{fong2010scalar}%
  \BibitemOpen
  \bibfield  {author} {\bibinfo {author} {\bibfnamefont {B.~H.}\ \bibnamefont
  {Fong}}, \bibinfo {author} {\bibfnamefont {J.~S.}\ \bibnamefont {Colburn}},
  \bibinfo {author} {\bibfnamefont {J.~J.}\ \bibnamefont {Ottusch}}, \bibinfo
  {author} {\bibfnamefont {J.~L.}\ \bibnamefont {Visher}},\ and\ \bibinfo
  {author} {\bibfnamefont {D.~F.}\ \bibnamefont {Sievenpiper}},\ }\bibfield
  {title} {\bibinfo {title} {{Scalar and tensor holographic artificial
  impedance surfaces}},\ }\href {http://dx.doi.org/10.1109/TAP.2010.2055812}
  {\bibfield  {journal} {\bibinfo  {journal} {IEEE Trans.~Antennas Propag.}\
  }\textbf {\bibinfo {volume} {58}},\ \bibinfo {pages} {3212} (\bibinfo {year}
  {2010})}\BibitemShut {NoStop}%
\bibitem [{\citenamefont {Yu}\ \emph {et~al.}(2011)\citenamefont {Yu},
  \citenamefont {Genevet}, \citenamefont {Kats}, \citenamefont {Aieta},
  \citenamefont {Tetienne}, \citenamefont {Capasso},\ and\ \citenamefont
  {Gaburro}}]{yu2011light}%
  \BibitemOpen
  \bibfield  {author} {\bibinfo {author} {\bibfnamefont {N.}~\bibnamefont
  {Yu}}, \bibinfo {author} {\bibfnamefont {P.}~\bibnamefont {Genevet}},
  \bibinfo {author} {\bibfnamefont {M.~A.}\ \bibnamefont {Kats}}, \bibinfo
  {author} {\bibfnamefont {F.}~\bibnamefont {Aieta}}, \bibinfo {author}
  {\bibfnamefont {J.-P.}\ \bibnamefont {Tetienne}}, \bibinfo {author}
  {\bibfnamefont {F.}~\bibnamefont {Capasso}},\ and\ \bibinfo {author}
  {\bibfnamefont {Z.}~\bibnamefont {Gaburro}},\ }\bibfield  {title} {\bibinfo
  {title} {{Light propagation with phase discontinuities: generalized laws of
  reflection and refraction}},\ }\href
  {http://dx.doi.org/10.1126/science.1210713} {\bibfield  {journal} {\bibinfo
  {journal} {Science}\ }\textbf {\bibinfo {volume} {334}},\ \bibinfo {pages}
  {333} (\bibinfo {year} {2011})}\BibitemShut {NoStop}%
\bibitem [{\citenamefont {Ni}\ \emph {et~al.}(2012)\citenamefont {Ni},
  \citenamefont {Emani}, \citenamefont {Kildishev}, \citenamefont
  {Boltasseva},\ and\ \citenamefont {Shalaev}}]{ni2012broadband}%
  \BibitemOpen
  \bibfield  {author} {\bibinfo {author} {\bibfnamefont {X.}~\bibnamefont
  {Ni}}, \bibinfo {author} {\bibfnamefont {N.~K.}\ \bibnamefont {Emani}},
  \bibinfo {author} {\bibfnamefont {A.~V.}\ \bibnamefont {Kildishev}}, \bibinfo
  {author} {\bibfnamefont {A.}~\bibnamefont {Boltasseva}},\ and\ \bibinfo
  {author} {\bibfnamefont {V.~M.}\ \bibnamefont {Shalaev}},\ }\bibfield
  {title} {\bibinfo {title} {{Broadband light bending with plasmonic
  nanoantennas}},\ }\href {http://dx.doi.org/10.1126/science.1214686}
  {\bibfield  {journal} {\bibinfo  {journal} {Science}\ }\textbf {\bibinfo
  {volume} {335}},\ \bibinfo {pages} {427} (\bibinfo {year}
  {2012})}\BibitemShut {NoStop}%
\bibitem [{\citenamefont {Atikian}\ \emph {et~al.}(2022)\citenamefont
  {Atikian}, \citenamefont {Sinclair}, \citenamefont {Latawiec}, \citenamefont
  {Xiong}, \citenamefont {Meesala}, \citenamefont {Gauthier}, \citenamefont
  {Wintz}, \citenamefont {Randi}, \citenamefont {Bernot}, \citenamefont
  {DeFrances}, \citenamefont {Thomas}, \citenamefont {Roman}, \citenamefont
  {Durrant}, \citenamefont {Capasso},\ and\ \citenamefont
  {Lončar}}]{atikian2022diamond}%
  \BibitemOpen
  \bibfield  {author} {\bibinfo {author} {\bibfnamefont {H.~A.}\ \bibnamefont
  {Atikian}}, \bibinfo {author} {\bibfnamefont {N.}~\bibnamefont {Sinclair}},
  \bibinfo {author} {\bibfnamefont {P.}~\bibnamefont {Latawiec}}, \bibinfo
  {author} {\bibfnamefont {X.}~\bibnamefont {Xiong}}, \bibinfo {author}
  {\bibfnamefont {S.}~\bibnamefont {Meesala}}, \bibinfo {author} {\bibfnamefont
  {S.}~\bibnamefont {Gauthier}}, \bibinfo {author} {\bibfnamefont
  {D.}~\bibnamefont {Wintz}}, \bibinfo {author} {\bibfnamefont
  {J.}~\bibnamefont {Randi}}, \bibinfo {author} {\bibfnamefont
  {D.}~\bibnamefont {Bernot}}, \bibinfo {author} {\bibfnamefont
  {S.}~\bibnamefont {DeFrances}}, \bibinfo {author} {\bibfnamefont
  {J.}~\bibnamefont {Thomas}}, \bibinfo {author} {\bibfnamefont
  {M.}~\bibnamefont {Roman}}, \bibinfo {author} {\bibfnamefont
  {S.}~\bibnamefont {Durrant}}, \bibinfo {author} {\bibfnamefont
  {F.}~\bibnamefont {Capasso}},\ and\ \bibinfo {author} {\bibfnamefont
  {M.}~\bibnamefont {Lončar}},\ }\bibfield  {title} {\bibinfo {title}
  {{Diamond mirrors for high-power continuous-wave lasers}},\ }\href
  {http://dx.doi.org/10.1038/s41467-022-30335-2} {\bibfield  {journal}
  {\bibinfo  {journal} {Nat.~Commun.}\ }\textbf {\bibinfo {volume} {13}},\
  \bibinfo {pages} {2610} (\bibinfo {year} {2022})}\BibitemShut {NoStop}%
\bibitem [{\citenamefont {Arbabi}\ \emph {et~al.}(2015)\citenamefont {Arbabi},
  \citenamefont {Horie}, \citenamefont {Ball}, \citenamefont {Bagheri},\ and\
  \citenamefont {Faraon}}]{arbabi2015subwavelength}%
  \BibitemOpen
  \bibfield  {author} {\bibinfo {author} {\bibfnamefont {A.}~\bibnamefont
  {Arbabi}}, \bibinfo {author} {\bibfnamefont {Y.}~\bibnamefont {Horie}},
  \bibinfo {author} {\bibfnamefont {A.~J.}\ \bibnamefont {Ball}}, \bibinfo
  {author} {\bibfnamefont {M.}~\bibnamefont {Bagheri}},\ and\ \bibinfo {author}
  {\bibfnamefont {A.}~\bibnamefont {Faraon}},\ }\bibfield  {title} {\bibinfo
  {title} {{Subwavelength-thick lenses with high numerical apertures and large
  efficiency based on high-contrast transmitarrays}},\ }\href
  {http://dx.doi.org/10.1038/ncomms8069} {\bibfield  {journal} {\bibinfo
  {journal} {Nat.~Commun.}\ }\textbf {\bibinfo {volume} {6}},\ \bibinfo {pages}
  {7069} (\bibinfo {year} {2015})}\BibitemShut {NoStop}%
\bibitem [{\citenamefont {Khorasaninejad}\ \emph {et~al.}(2016)\citenamefont
  {Khorasaninejad}, \citenamefont {Chen}, \citenamefont {Devlin}, \citenamefont
  {Oh}, \citenamefont {Zhu},\ and\ \citenamefont {Capasso}}]{capasso16tio2}%
  \BibitemOpen
  \bibfield  {author} {\bibinfo {author} {\bibfnamefont {M.}~\bibnamefont
  {Khorasaninejad}}, \bibinfo {author} {\bibfnamefont {W.~T.}\ \bibnamefont
  {Chen}}, \bibinfo {author} {\bibfnamefont {R.~C.}\ \bibnamefont {Devlin}},
  \bibinfo {author} {\bibfnamefont {J.}~\bibnamefont {Oh}}, \bibinfo {author}
  {\bibfnamefont {A.~Y.}\ \bibnamefont {Zhu}},\ and\ \bibinfo {author}
  {\bibfnamefont {F.}~\bibnamefont {Capasso}},\ }\bibfield  {title} {\bibinfo
  {title} {{Metalenses at visible wavelengths: diffraction-limited focusing and
  subwavelength resolution imaging}},\ }\href
  {https://www.science.org/doi/abs/10.1126/science.aaf6644} {\bibfield
  {journal} {\bibinfo  {journal} {Science}\ }\textbf {\bibinfo {volume}
  {352}},\ \bibinfo {pages} {1190} (\bibinfo {year} {2016})}\BibitemShut
  {NoStop}%
\bibitem [{\citenamefont {Balthasar~Mueller}\ \emph {et~al.}(2017)\citenamefont
  {Balthasar~Mueller}, \citenamefont {Rubin}, \citenamefont {Devlin},
  \citenamefont {Groever},\ and\ \citenamefont
  {Capasso}}]{balthasarmueller2017metasurface}%
  \BibitemOpen
  \bibfield  {author} {\bibinfo {author} {\bibfnamefont {J.~P.}\ \bibnamefont
  {Balthasar~Mueller}}, \bibinfo {author} {\bibfnamefont {N.~A.}\ \bibnamefont
  {Rubin}}, \bibinfo {author} {\bibfnamefont {R.~C.}\ \bibnamefont {Devlin}},
  \bibinfo {author} {\bibfnamefont {B.}~\bibnamefont {Groever}},\ and\ \bibinfo
  {author} {\bibfnamefont {F.}~\bibnamefont {Capasso}},\ }\bibfield  {title}
  {\bibinfo {title} {{Metasurface polarization optics: independent phase
  control of arbitrary orthogonal states of polarization}},\ }\href
  {http://dx.doi.org/10.1103/PhysRevLett.118.113901} {\bibfield  {journal}
  {\bibinfo  {journal} {Phys.~Rev.~Lett.}\ }\textbf {\bibinfo {volume} {118}},\
  \bibinfo {pages} {113901} (\bibinfo {year} {2017})}\BibitemShut {NoStop}%
\bibitem [{\citenamefont {Huang}\ \emph
  {et~al.}(2023{\natexlab{b}})\citenamefont {Huang}, \citenamefont {Overvig},
  \citenamefont {Xu}, \citenamefont {Malek}, \citenamefont {Tsai},
  \citenamefont {Alù},\ and\ \citenamefont {Yu}}]{huang2023int}%
  \BibitemOpen
  \bibfield  {author} {\bibinfo {author} {\bibfnamefont {H.}~\bibnamefont
  {Huang}}, \bibinfo {author} {\bibfnamefont {A.~C.}\ \bibnamefont {Overvig}},
  \bibinfo {author} {\bibfnamefont {Y.}~\bibnamefont {Xu}}, \bibinfo {author}
  {\bibfnamefont {S.~C.}\ \bibnamefont {Malek}}, \bibinfo {author}
  {\bibfnamefont {C.-C.}\ \bibnamefont {Tsai}}, \bibinfo {author}
  {\bibfnamefont {A.}~\bibnamefont {Alù}},\ and\ \bibinfo {author}
  {\bibfnamefont {N.}~\bibnamefont {Yu}},\ }\bibfield  {title} {\bibinfo
  {title} {{Leaky-wave metasurfaces for integrated photonics}},\ }\href
  {https://doi.org/10.1038/s41565-023-01360-z} {\bibfield  {journal} {\bibinfo
  {journal} {Nat.~Nanotechnol.}\ }\textbf {\bibinfo {volume} {18}},\ \bibinfo
  {pages} {580} (\bibinfo {year} {2023}{\natexlab{b}})}\BibitemShut {NoStop}%
\bibitem [{\citenamefont {Hsu}\ \emph {et~al.}(2022)\citenamefont {Hsu},
  \citenamefont {Zhu}, \citenamefont {Thiele}, \citenamefont {Brown},
  \citenamefont {Papp}, \citenamefont {Agrawal},\ and\ \citenamefont
  {Regal}}]{hsu2022single}%
  \BibitemOpen
  \bibfield  {author} {\bibinfo {author} {\bibfnamefont {T.-W.}\ \bibnamefont
  {Hsu}}, \bibinfo {author} {\bibfnamefont {W.}~\bibnamefont {Zhu}}, \bibinfo
  {author} {\bibfnamefont {T.}~\bibnamefont {Thiele}}, \bibinfo {author}
  {\bibfnamefont {M.~O.}\ \bibnamefont {Brown}}, \bibinfo {author}
  {\bibfnamefont {S.~B.}\ \bibnamefont {Papp}}, \bibinfo {author}
  {\bibfnamefont {A.}~\bibnamefont {Agrawal}},\ and\ \bibinfo {author}
  {\bibfnamefont {C.~A.}\ \bibnamefont {Regal}},\ }\bibfield  {title} {\bibinfo
  {title} {{Single-atom trapping in a metasurface-lens optical tweezer}},\
  }\href {https://link.aps.org/doi/10.1103/PRXQuantum.3.030316} {\bibfield
  {journal} {\bibinfo  {journal} {PRX Quantum}\ }\textbf {\bibinfo {volume}
  {3}},\ \bibinfo {pages} {030316} (\bibinfo {year} {2022})}\BibitemShut
  {NoStop}%
\bibitem [{\citenamefont {Yu}\ and\ \citenamefont
  {Capasso}(2014)}]{yu2014flat}%
  \BibitemOpen
  \bibfield  {author} {\bibinfo {author} {\bibfnamefont {N.}~\bibnamefont
  {Yu}}\ and\ \bibinfo {author} {\bibfnamefont {F.}~\bibnamefont {Capasso}},\
  }\bibfield  {title} {\bibinfo {title} {{Flat optics with designer
  metasurfaces}},\ }\href {http://dx.doi.org/10.1038/nmat3839} {\bibfield
  {journal} {\bibinfo  {journal} {Nat.~Mater.}\ }\textbf {\bibinfo {volume}
  {13}},\ \bibinfo {pages} {139} (\bibinfo {year} {2014})}\BibitemShut
  {NoStop}%
\bibitem [{\citenamefont {Chen}\ \emph {et~al.}(2020)\citenamefont {Chen},
  \citenamefont {Zhu},\ and\ \citenamefont {Capasso}}]{chen2020flat}%
  \BibitemOpen
  \bibfield  {author} {\bibinfo {author} {\bibfnamefont {W.~T.}\ \bibnamefont
  {Chen}}, \bibinfo {author} {\bibfnamefont {A.~Y.}\ \bibnamefont {Zhu}},\ and\
  \bibinfo {author} {\bibfnamefont {F.}~\bibnamefont {Capasso}},\ }\bibfield
  {title} {\bibinfo {title} {{Flat optics with dispersion-engineered
  metasurfaces}},\ }\href {https://doi.org/10.1038/s41578-020-0203-3}
  {\bibfield  {journal} {\bibinfo  {journal} {Nat.~Rev.~Mater.}\ }\textbf
  {\bibinfo {volume} {5}},\ \bibinfo {pages} {604} (\bibinfo {year}
  {2020})}\BibitemShut {NoStop}%
\bibitem [{\citenamefont {Kildishev}\ \emph {et~al.}(2013)\citenamefont
  {Kildishev}, \citenamefont {Boltasseva},\ and\ \citenamefont
  {Shalaev}}]{kildishev2013planar}%
  \BibitemOpen
  \bibfield  {author} {\bibinfo {author} {\bibfnamefont {A.~V.}\ \bibnamefont
  {Kildishev}}, \bibinfo {author} {\bibfnamefont {A.}~\bibnamefont
  {Boltasseva}},\ and\ \bibinfo {author} {\bibfnamefont {V.~M.}\ \bibnamefont
  {Shalaev}},\ }\bibfield  {title} {\bibinfo {title} {{Planar photonics with
  metasurfaces}},\ }\href {http://dx.doi.org/10.1126/science.1232009}
  {\bibfield  {journal} {\bibinfo  {journal} {Science}\ }\textbf {\bibinfo
  {volume} {339}},\ \bibinfo {pages} {1232009} (\bibinfo {year}
  {2013})}\BibitemShut {NoStop}%
\bibitem [{\citenamefont {Park}\ \emph {et~al.}(2024)\citenamefont {Park},
  \citenamefont {Lim}, \citenamefont {Amirzhan}, \citenamefont {Kang},
  \citenamefont {Karrfalt}, \citenamefont {Kim}, \citenamefont {Leger},
  \citenamefont {Urbas}, \citenamefont {Ossiander}, \citenamefont {Li},\ and\
  \citenamefont {Capasso}}]{Park2024all}%
  \BibitemOpen
  \bibfield  {author} {\bibinfo {author} {\bibfnamefont {J.-S.}\ \bibnamefont
  {Park}}, \bibinfo {author} {\bibfnamefont {S.~W.~D.}\ \bibnamefont {Lim}},
  \bibinfo {author} {\bibfnamefont {A.}~\bibnamefont {Amirzhan}}, \bibinfo
  {author} {\bibfnamefont {H.}~\bibnamefont {Kang}}, \bibinfo {author}
  {\bibfnamefont {K.}~\bibnamefont {Karrfalt}}, \bibinfo {author}
  {\bibfnamefont {D.}~\bibnamefont {Kim}}, \bibinfo {author} {\bibfnamefont
  {J.}~\bibnamefont {Leger}}, \bibinfo {author} {\bibfnamefont
  {A.}~\bibnamefont {Urbas}}, \bibinfo {author} {\bibfnamefont
  {M.}~\bibnamefont {Ossiander}}, \bibinfo {author} {\bibfnamefont
  {Z.}~\bibnamefont {Li}},\ and\ \bibinfo {author} {\bibfnamefont
  {F.}~\bibnamefont {Capasso}},\ }\bibfield  {title} {\bibinfo {title}
  {{All-glass 100 mm diameter visible metalens for imaging the cosmos}},\
  }\href {http://dx.doi.org/10.1021/acsnano.3c09462} {\bibfield  {journal}
  {\bibinfo  {journal} {ACS Nano}\ }\textbf {\bibinfo {volume} {18}},\ \bibinfo
  {pages} {3187} (\bibinfo {year} {2024})}\BibitemShut {NoStop}%
\bibitem [{\citenamefont {Zelevinsky}\ \emph {et~al.}(2006)\citenamefont
  {Zelevinsky}, \citenamefont {Boyd}, \citenamefont {Ludlow}, \citenamefont
  {Ido}, \citenamefont {Ye}, \citenamefont {Ciuryło}, \citenamefont {Naidon},\
  and\ \citenamefont {Julienne}}]{Zelevinsky2006narrow}%
  \BibitemOpen
  \bibfield  {author} {\bibinfo {author} {\bibfnamefont {T.}~\bibnamefont
  {Zelevinsky}}, \bibinfo {author} {\bibfnamefont {M.~M.}\ \bibnamefont
  {Boyd}}, \bibinfo {author} {\bibfnamefont {A.~D.}\ \bibnamefont {Ludlow}},
  \bibinfo {author} {\bibfnamefont {T.}~\bibnamefont {Ido}}, \bibinfo {author}
  {\bibfnamefont {J.}~\bibnamefont {Ye}}, \bibinfo {author} {\bibfnamefont
  {R.}~\bibnamefont {Ciuryło}}, \bibinfo {author} {\bibfnamefont
  {P.}~\bibnamefont {Naidon}},\ and\ \bibinfo {author} {\bibfnamefont {P.~S.}\
  \bibnamefont {Julienne}},\ }\bibfield  {title} {\bibinfo {title} {{Narrow
  line photoassociation in an optical lattice}},\ }\href
  {http://dx.doi.org/10.1103/PhysRevLett.96.203201} {\bibfield  {journal}
  {\bibinfo  {journal} {Phys.~Rev.~Lett.}\ }\textbf {\bibinfo {volume} {96}},\
  \bibinfo {pages} {203201} (\bibinfo {year} {2006})}\BibitemShut {NoStop}%
\bibitem [{\citenamefont {Cooper}\ \emph {et~al.}(2018)\citenamefont {Cooper},
  \citenamefont {Covey}, \citenamefont {Madjarov}, \citenamefont {Porsev},
  \citenamefont {Safronova},\ and\ \citenamefont
  {Endres}}]{Cooper2018alkaline}%
  \BibitemOpen
  \bibfield  {author} {\bibinfo {author} {\bibfnamefont {A.}~\bibnamefont
  {Cooper}}, \bibinfo {author} {\bibfnamefont {J.~P.}\ \bibnamefont {Covey}},
  \bibinfo {author} {\bibfnamefont {I.~S.}\ \bibnamefont {Madjarov}}, \bibinfo
  {author} {\bibfnamefont {S.~G.}\ \bibnamefont {Porsev}}, \bibinfo {author}
  {\bibfnamefont {M.~S.}\ \bibnamefont {Safronova}},\ and\ \bibinfo {author}
  {\bibfnamefont {M.}~\bibnamefont {Endres}},\ }\bibfield  {title} {\bibinfo
  {title} {{Alkaline-earth atoms in optical tweezers}},\ }\href
  {http://dx.doi.org/10.1103/PhysRevX.8.041055} {\bibfield  {journal} {\bibinfo
   {journal} {Phys.~Rev.~X}\ }\textbf {\bibinfo {volume} {8}},\ \bibinfo
  {pages} {041055} (\bibinfo {year} {2018})}\BibitemShut {NoStop}%
\bibitem [{\citenamefont {Gyger}\ \emph {et~al.}(2024)\citenamefont {Gyger},
  \citenamefont {Ammenwerth}, \citenamefont {Tao}, \citenamefont {Timme},
  \citenamefont {Snigirev}, \citenamefont {Bloch},\ and\ \citenamefont
  {Zeiher}}]{gyger2024continuous}%
  \BibitemOpen
  \bibfield  {author} {\bibinfo {author} {\bibfnamefont {F.}~\bibnamefont
  {Gyger}}, \bibinfo {author} {\bibfnamefont {M.}~\bibnamefont {Ammenwerth}},
  \bibinfo {author} {\bibfnamefont {R.}~\bibnamefont {Tao}}, \bibinfo {author}
  {\bibfnamefont {H.}~\bibnamefont {Timme}}, \bibinfo {author} {\bibfnamefont
  {S.}~\bibnamefont {Snigirev}}, \bibinfo {author} {\bibfnamefont
  {I.}~\bibnamefont {Bloch}},\ and\ \bibinfo {author} {\bibfnamefont
  {J.}~\bibnamefont {Zeiher}},\ }\bibfield  {title} {\bibinfo {title}
  {{Continuous operation of large-scale atom arrays in optical lattices}},\
  }\href {https://link.aps.org/doi/10.1103/PhysRevResearch.6.033104} {\bibfield
   {journal} {\bibinfo  {journal} {Phys.~Rev.~Res.}\ }\textbf {\bibinfo
  {volume} {6}},\ \bibinfo {pages} {033104} (\bibinfo {year}
  {2024})}\BibitemShut {NoStop}%
\bibitem [{\citenamefont {Covey}\ \emph {et~al.}(2019)\citenamefont {Covey},
  \citenamefont {Madjarov}, \citenamefont {Cooper},\ and\ \citenamefont
  {Endres}}]{covey20192000}%
  \BibitemOpen
  \bibfield  {author} {\bibinfo {author} {\bibfnamefont {J.~P.}\ \bibnamefont
  {Covey}}, \bibinfo {author} {\bibfnamefont {I.~S.}\ \bibnamefont {Madjarov}},
  \bibinfo {author} {\bibfnamefont {A.}~\bibnamefont {Cooper}},\ and\ \bibinfo
  {author} {\bibfnamefont {M.}~\bibnamefont {Endres}},\ }\bibfield  {title}
  {\bibinfo {title} {{2000-times repeated imaging of strontium atoms in
  clock-magic tweezer arrays}},\ }\href
  {https://link.aps.org/doi/10.1103/PhysRevLett.122.173201} {\bibfield
  {journal} {\bibinfo  {journal} {Phys.~Rev.~Lett.}\ }\textbf {\bibinfo
  {volume} {122}},\ \bibinfo {pages} {173201} (\bibinfo {year}
  {2019})}\BibitemShut {NoStop}%
\bibitem [{\citenamefont {Nogrette}\ \emph {et~al.}(2014)\citenamefont
  {Nogrette}, \citenamefont {Labuhn}, \citenamefont {Ravets}, \citenamefont
  {Barredo}, \citenamefont {B{\'e}guin}, \citenamefont {Vernier}, \citenamefont
  {Lahaye},\ and\ \citenamefont {Browaeys}}]{nogrette2014single}%
  \BibitemOpen
  \bibfield  {author} {\bibinfo {author} {\bibfnamefont {F.}~\bibnamefont
  {Nogrette}}, \bibinfo {author} {\bibfnamefont {H.}~\bibnamefont {Labuhn}},
  \bibinfo {author} {\bibfnamefont {S.}~\bibnamefont {Ravets}}, \bibinfo
  {author} {\bibfnamefont {D.}~\bibnamefont {Barredo}}, \bibinfo {author}
  {\bibfnamefont {L.}~\bibnamefont {B{\'e}guin}}, \bibinfo {author}
  {\bibfnamefont {A.}~\bibnamefont {Vernier}}, \bibinfo {author} {\bibfnamefont
  {T.}~\bibnamefont {Lahaye}},\ and\ \bibinfo {author} {\bibfnamefont
  {A.}~\bibnamefont {Browaeys}},\ }\bibfield  {title} {\bibinfo {title}
  {{Single-atom trapping in holographic 2D arrays of microtraps with arbitrary
  geometries}},\ }\href {https://doi.org/10.1103/PhysRevX.4.021034} {\bibfield
  {journal} {\bibinfo  {journal} {Phys.~Rev.~X}\ }\textbf {\bibinfo {volume}
  {4}},\ \bibinfo {pages} {021034} (\bibinfo {year} {2014})}\BibitemShut
  {NoStop}%
\bibitem [{\citenamefont {Schymik}\ \emph {et~al.}(2022)\citenamefont
  {Schymik}, \citenamefont {Ximenez}, \citenamefont {Bloch}, \citenamefont
  {Dreon}, \citenamefont {Signoles}, \citenamefont {Nogrette}, \citenamefont
  {Barredo}, \citenamefont {Browaeys},\ and\ \citenamefont
  {Lahaye}}]{schymik2022situ}%
  \BibitemOpen
  \bibfield  {author} {\bibinfo {author} {\bibfnamefont {K.-N.}\ \bibnamefont
  {Schymik}}, \bibinfo {author} {\bibfnamefont {B.}~\bibnamefont {Ximenez}},
  \bibinfo {author} {\bibfnamefont {E.}~\bibnamefont {Bloch}}, \bibinfo
  {author} {\bibfnamefont {D.}~\bibnamefont {Dreon}}, \bibinfo {author}
  {\bibfnamefont {A.}~\bibnamefont {Signoles}}, \bibinfo {author}
  {\bibfnamefont {F.}~\bibnamefont {Nogrette}}, \bibinfo {author}
  {\bibfnamefont {D.}~\bibnamefont {Barredo}}, \bibinfo {author} {\bibfnamefont
  {A.}~\bibnamefont {Browaeys}},\ and\ \bibinfo {author} {\bibfnamefont
  {T.}~\bibnamefont {Lahaye}},\ }\bibfield  {title} {\bibinfo {title} {{In situ
  equalization of single-atom loading in large-scale optical tweezer arrays}},\
  }\href {https://doi.org/10.1103/PhysRevA.106.022611} {\bibfield  {journal}
  {\bibinfo  {journal} {Phys.~Rev.~A}\ }\textbf {\bibinfo {volume} {106}},\
  \bibinfo {pages} {022611} (\bibinfo {year} {2022})}\BibitemShut {NoStop}%
\bibitem [{\citenamefont {Chew}\ \emph {et~al.}(2024)\citenamefont {Chew},
  \citenamefont {Poitrinal}, \citenamefont {Tomita}, \citenamefont {Kitade},
  \citenamefont {Mauricio}, \citenamefont {Ohmori},\ and\ \citenamefont
  {de~L{\'e}s{\'e}leuc}}]{chew2024ultra}%
  \BibitemOpen
  \bibfield  {author} {\bibinfo {author} {\bibfnamefont {Y.~T.}\ \bibnamefont
  {Chew}}, \bibinfo {author} {\bibfnamefont {M.}~\bibnamefont {Poitrinal}},
  \bibinfo {author} {\bibfnamefont {T.}~\bibnamefont {Tomita}}, \bibinfo
  {author} {\bibfnamefont {S.}~\bibnamefont {Kitade}}, \bibinfo {author}
  {\bibfnamefont {J.}~\bibnamefont {Mauricio}}, \bibinfo {author}
  {\bibfnamefont {K.}~\bibnamefont {Ohmori}},\ and\ \bibinfo {author}
  {\bibfnamefont {S.}~\bibnamefont {de~L{\'e}s{\'e}leuc}},\ }\bibfield  {title}
  {\bibinfo {title} {{Ultra-precise holographic optical tweezers array}},\
  }\href {https://doi.org/10.48550/arXiv.2407.20699} {\bibfield  {journal}
  {\bibinfo  {journal} {arXiv:2407.20699}\ } (\bibinfo {year}
  {2024})}\BibitemShut {NoStop}%
\bibitem [{Note1()}]{Note1}%
  \BibitemOpen
  \bibinfo {note} {It is important to note this is a fundamental limitation
  that cannot be improved upon, for example, via active feedback in SLMs or
  DMDs.}\BibitemShut {Stop}%
\bibitem [{\citenamefont {Malek}\ \emph {et~al.}(2022)\citenamefont {Malek},
  \citenamefont {Overvig}, \citenamefont {Alù},\ and\ \citenamefont
  {Yu}}]{malek2022non}%
  \BibitemOpen
  \bibfield  {author} {\bibinfo {author} {\bibfnamefont {S.~C.}\ \bibnamefont
  {Malek}}, \bibinfo {author} {\bibfnamefont {A.~C.}\ \bibnamefont {Overvig}},
  \bibinfo {author} {\bibfnamefont {A.}~\bibnamefont {Alù}},\ and\ \bibinfo
  {author} {\bibfnamefont {N.}~\bibnamefont {Yu}},\ }\bibfield  {title}
  {\bibinfo {title} {{Multifunctional resonant wavefront-shaping meta-optics
  based on multilayer and multi-perturbation nonlocal metasurfaces}},\ }\href
  {https://doi.org/10.1038/s41377-022-00905-6} {\bibfield  {journal} {\bibinfo
  {journal} {Light Sci.~Appl.}\ }\textbf {\bibinfo {volume} {11}},\ \bibinfo
  {pages} {246} (\bibinfo {year} {2022})}\BibitemShut {NoStop}%
\bibitem [{\citenamefont {Shaltout}\ \emph {et~al.}(2019)\citenamefont
  {Shaltout}, \citenamefont {Shalaev},\ and\ \citenamefont
  {Brongersma}}]{Shaltout2019}%
  \BibitemOpen
  \bibfield  {author} {\bibinfo {author} {\bibfnamefont {A.~M.}\ \bibnamefont
  {Shaltout}}, \bibinfo {author} {\bibfnamefont {V.~M.}\ \bibnamefont
  {Shalaev}},\ and\ \bibinfo {author} {\bibfnamefont {M.~L.}\ \bibnamefont
  {Brongersma}},\ }\bibfield  {title} {\bibinfo {title} {Spatiotemporal light
  control with active metasurfaces},\ }\href
  {http://dx.doi.org/10.1126/science.aat3100} {\bibfield  {journal} {\bibinfo
  {journal} {Science}\ }\textbf {\bibinfo {volume} {364}} (\bibinfo {year}
  {2019})}\BibitemShut {NoStop}%
\bibitem [{\citenamefont {Wu}\ \emph {et~al.}(2019)\citenamefont {Wu},
  \citenamefont {Yang}, \citenamefont {Fan}, \citenamefont {Song},\ and\
  \citenamefont {Xiao}}]{wu19tio2}%
  \BibitemOpen
  \bibfield  {author} {\bibinfo {author} {\bibfnamefont {Y.}~\bibnamefont
  {Wu}}, \bibinfo {author} {\bibfnamefont {W.}~\bibnamefont {Yang}}, \bibinfo
  {author} {\bibfnamefont {Y.}~\bibnamefont {Fan}}, \bibinfo {author}
  {\bibfnamefont {Q.}~\bibnamefont {Song}},\ and\ \bibinfo {author}
  {\bibfnamefont {S.}~\bibnamefont {Xiao}},\ }\bibfield  {title} {\bibinfo
  {title} {{TiO$_2$ metasurfaces: from visible planar photonics to
  photochemistry}},\ }\href
  {https://www.science.org/doi/abs/10.1126/sciadv.aax0939} {\bibfield
  {journal} {\bibinfo  {journal} {Sci.~Adv.}\ }\textbf {\bibinfo {volume}
  {5}},\ \bibinfo {pages} {eaax0939} (\bibinfo {year} {2019})}\BibitemShut
  {NoStop}%
\bibitem [{\citenamefont {Chen}\ \emph {et~al.}(2023)\citenamefont {Chen},
  \citenamefont {Park}, \citenamefont {Marchioni}, \citenamefont {Millay},
  \citenamefont {Yousef},\ and\ \citenamefont {Capasso}}]{Chen2023disp}%
  \BibitemOpen
  \bibfield  {author} {\bibinfo {author} {\bibfnamefont {W.~T.}\ \bibnamefont
  {Chen}}, \bibinfo {author} {\bibfnamefont {J.-S.}\ \bibnamefont {Park}},
  \bibinfo {author} {\bibfnamefont {J.}~\bibnamefont {Marchioni}}, \bibinfo
  {author} {\bibfnamefont {S.}~\bibnamefont {Millay}}, \bibinfo {author}
  {\bibfnamefont {K.~M.~A.}\ \bibnamefont {Yousef}},\ and\ \bibinfo {author}
  {\bibfnamefont {F.}~\bibnamefont {Capasso}},\ }\bibfield  {title} {\bibinfo
  {title} {Dispersion-engineered metasurfaces reaching broadband 90
  diffraction efficiency},\ }\href {https://doi.org/10.1038/s41467-023-38185-2}
  {\bibfield  {journal} {\bibinfo  {journal} {Nat.~Commun.}\ }\textbf {\bibinfo
  {volume} {14}},\ \bibinfo {pages} {2544} (\bibinfo {year}
  {2023})}\BibitemShut {NoStop}%
\bibitem [{\citenamefont {Malek}\ \emph {et~al.}(2023)\citenamefont {Malek},
  \citenamefont {Xu},\ and\ \citenamefont {Yu}}]{malek2023}%
  \BibitemOpen
  \bibfield  {author} {\bibinfo {author} {\bibfnamefont {S.~C.}\ \bibnamefont
  {Malek}}, \bibinfo {author} {\bibfnamefont {Y.}~\bibnamefont {Xu}},\ and\
  \bibinfo {author} {\bibfnamefont {N.}~\bibnamefont {Yu}},\ }\bibfield
  {title} {\bibinfo {title} {{Visible-spectrum wavelength-selective metalenses
  based on quasi-bound states in the continuum}},\ }in\ \href
  {http://dx.doi.org/10.1364/CLEO_FS.2023.FTh5B.8} {\emph {\bibinfo {booktitle}
  {CLEO 2023}}}\ (\bibinfo  {publisher} {IEEE},\ \bibinfo {year} {2023})\ pp.\
  \bibinfo {pages} {1--2}\BibitemShut {NoStop}%
\bibitem [{\citenamefont {Nejadriahi}\ \emph {et~al.}(2020)\citenamefont
  {Nejadriahi}, \citenamefont {Friedman}, \citenamefont {Sharma}, \citenamefont
  {Pappert}, \citenamefont {Fainman},\ and\ \citenamefont {Yu}}]{sin20}%
  \BibitemOpen
  \bibfield  {author} {\bibinfo {author} {\bibfnamefont {H.}~\bibnamefont
  {Nejadriahi}}, \bibinfo {author} {\bibfnamefont {A.}~\bibnamefont
  {Friedman}}, \bibinfo {author} {\bibfnamefont {R.}~\bibnamefont {Sharma}},
  \bibinfo {author} {\bibfnamefont {S.}~\bibnamefont {Pappert}}, \bibinfo
  {author} {\bibfnamefont {Y.}~\bibnamefont {Fainman}},\ and\ \bibinfo {author}
  {\bibfnamefont {P.}~\bibnamefont {Yu}},\ }\bibfield  {title} {\bibinfo
  {title} {{Thermo-optic properties of silicon-rich silicon nitride for on-chip
  applications}},\ }\href
  {https://opg.optica.org/oe/abstract.cfm?URI=oe-28-17-24951} {\bibfield
  {journal} {\bibinfo  {journal} {Opt.~Express}\ }\textbf {\bibinfo {volume}
  {28}},\ \bibinfo {pages} {24951} (\bibinfo {year} {2020})}\BibitemShut
  {NoStop}%
\bibitem [{\citenamefont {Fan}\ \emph {et~al.}(2018)\citenamefont {Fan},
  \citenamefont {Shao}, \citenamefont {Xie}, \citenamefont {Pang},
  \citenamefont {Ruan}, \citenamefont {Zhao}, \citenamefont {Chen},
  \citenamefont {Yu},\ and\ \citenamefont {Dong}}]{fan2018sin}%
  \BibitemOpen
  \bibfield  {author} {\bibinfo {author} {\bibfnamefont {Z.-B.}\ \bibnamefont
  {Fan}}, \bibinfo {author} {\bibfnamefont {Z.-K.}\ \bibnamefont {Shao}},
  \bibinfo {author} {\bibfnamefont {M.-Y.}\ \bibnamefont {Xie}}, \bibinfo
  {author} {\bibfnamefont {X.-N.}\ \bibnamefont {Pang}}, \bibinfo {author}
  {\bibfnamefont {W.-S.}\ \bibnamefont {Ruan}}, \bibinfo {author}
  {\bibfnamefont {F.-L.}\ \bibnamefont {Zhao}}, \bibinfo {author}
  {\bibfnamefont {Y.-J.}\ \bibnamefont {Chen}}, \bibinfo {author}
  {\bibfnamefont {S.-Y.}\ \bibnamefont {Yu}},\ and\ \bibinfo {author}
  {\bibfnamefont {J.-W.}\ \bibnamefont {Dong}},\ }\bibfield  {title} {\bibinfo
  {title} {Silicon nitride metalenses for close-to-one numerical aperture and
  wide-angle visible imaging},\ }\href
  {https://doi.org/10.1103/PhysRevApplied.10.014005} {\bibfield  {journal}
  {\bibinfo  {journal} {Phys.~Rev.~Appl.}\ }\textbf {\bibinfo {volume} {10}},\
  \bibinfo {pages} {014005} (\bibinfo {year} {2018})}\BibitemShut {NoStop}%
\bibitem [{\citenamefont {Gerchberg}\ and\ \citenamefont
  {Saxton}(1972)}]{gs72}%
  \BibitemOpen
  \bibfield  {author} {\bibinfo {author} {\bibfnamefont {R.~W.}\ \bibnamefont
  {Gerchberg}}\ and\ \bibinfo {author} {\bibfnamefont {W.~O.}\ \bibnamefont
  {Saxton}},\ }\bibfield  {title} {\bibinfo {title} {{A practical algorithm for
  the determination of phase from image and diffraction plane pictures}},\
  }\href {https://ci.nii.ac.jp/naid/10025518647/en/} {\bibfield  {journal}
  {\bibinfo  {journal} {Optik}\ }\textbf {\bibinfo {volume} {35}},\ \bibinfo
  {pages} {237} (\bibinfo {year} {1972})}\BibitemShut {NoStop}%
\bibitem [{\citenamefont {Johansson}\ and\ \citenamefont
  {Bengtsson}(2000)}]{soft00}%
  \BibitemOpen
  \bibfield  {author} {\bibinfo {author} {\bibfnamefont {M.}~\bibnamefont
  {Johansson}}\ and\ \bibinfo {author} {\bibfnamefont {J.}~\bibnamefont
  {Bengtsson}},\ }\bibfield  {title} {\bibinfo {title} {{Robust design method
  for highly efficient beam-shaping diffractive optical elements using an
  iterative-Fourier-transform algorithm with soft operations}},\ }\href
  {https://doi.org/10.1080/09500340008235111} {\bibfield  {journal} {\bibinfo
  {journal} {J.~Mod.~Opt.}\ }\textbf {\bibinfo {volume} {47}},\ \bibinfo
  {pages} {1385} (\bibinfo {year} {2000})}\BibitemShut {NoStop}%
\bibitem [{\citenamefont {Di~Leonardo}\ \emph {et~al.}(2007)\citenamefont
  {Di~Leonardo}, \citenamefont {Ianni},\ and\ \citenamefont {Ruocco}}]{holo07}%
  \BibitemOpen
  \bibfield  {author} {\bibinfo {author} {\bibfnamefont {R.}~\bibnamefont
  {Di~Leonardo}}, \bibinfo {author} {\bibfnamefont {F.}~\bibnamefont {Ianni}},\
  and\ \bibinfo {author} {\bibfnamefont {G.}~\bibnamefont {Ruocco}},\
  }\bibfield  {title} {\bibinfo {title} {{Computer generation of optimal
  holograms for optical trap arrays}},\ }\href
  {http://www.opticsexpress.org/abstract.cfm?URI=oe-15-4-1913} {\bibfield
  {journal} {\bibinfo  {journal} {Opt.~Express}\ }\textbf {\bibinfo {volume}
  {15}},\ \bibinfo {pages} {1913} (\bibinfo {year} {2007})}\BibitemShut
  {NoStop}%
\bibitem [{\citenamefont {Fan}\ \emph {et~al.}(2020)\citenamefont {Fan},
  \citenamefont {Liu}, \citenamefont {Zhang}, \citenamefont {Zhu},
  \citenamefont {Wang}, \citenamefont {Lin}, \citenamefont {Yan}, \citenamefont
  {Chen}, \citenamefont {Lezec},\ and\ \citenamefont {Lu}}]{Fan20tio2}%
  \BibitemOpen
  \bibfield  {author} {\bibinfo {author} {\bibfnamefont {Q.}~\bibnamefont
  {Fan}}, \bibinfo {author} {\bibfnamefont {M.}~\bibnamefont {Liu}}, \bibinfo
  {author} {\bibfnamefont {C.}~\bibnamefont {Zhang}}, \bibinfo {author}
  {\bibfnamefont {W.}~\bibnamefont {Zhu}}, \bibinfo {author} {\bibfnamefont
  {Y.}~\bibnamefont {Wang}}, \bibinfo {author} {\bibfnamefont {P.}~\bibnamefont
  {Lin}}, \bibinfo {author} {\bibfnamefont {F.}~\bibnamefont {Yan}}, \bibinfo
  {author} {\bibfnamefont {L.}~\bibnamefont {Chen}}, \bibinfo {author}
  {\bibfnamefont {H.~J.}\ \bibnamefont {Lezec}},\ and\ \bibinfo {author}
  {\bibfnamefont {Y.}~\bibnamefont {Lu}},\ }\bibfield  {title} {\bibinfo
  {title} {Independent amplitude control of arbitrary orthogonal states of
  polarization via dielectric metasurfaces},\ }\href@noop {} {\bibfield
  {journal} {\bibinfo  {journal} {Phys.~Rev.~Lett.}\ }\textbf {\bibinfo
  {volume} {125}},\ \bibinfo {pages} {267402} (\bibinfo {year}
  {2020})}\BibitemShut {NoStop}%
\bibitem [{\citenamefont {Lim}\ \emph {et~al.}(2023)\citenamefont {Lim},
  \citenamefont {Park}, \citenamefont {Kazakov}, \citenamefont {Spägele},
  \citenamefont {Dorrah}, \citenamefont {Meretska},\ and\ \citenamefont
  {Capasso}}]{Lim23tio2}%
  \BibitemOpen
  \bibfield  {author} {\bibinfo {author} {\bibfnamefont {S.~W.~D.}\
  \bibnamefont {Lim}}, \bibinfo {author} {\bibfnamefont {J.-S.}\ \bibnamefont
  {Park}}, \bibinfo {author} {\bibfnamefont {D.}~\bibnamefont {Kazakov}},
  \bibinfo {author} {\bibfnamefont {C.~M.}\ \bibnamefont {Spägele}}, \bibinfo
  {author} {\bibfnamefont {A.~H.}\ \bibnamefont {Dorrah}}, \bibinfo {author}
  {\bibfnamefont {M.~L.}\ \bibnamefont {Meretska}},\ and\ \bibinfo {author}
  {\bibfnamefont {F.}~\bibnamefont {Capasso}},\ }\bibfield  {title} {\bibinfo
  {title} {Point singularity array with metasurfaces},\ }\href
  {https://doi.org/10.1038/s41467-023-39072-6} {\bibfield  {journal} {\bibinfo
  {journal} {Nat.~Commun.}\ }\textbf {\bibinfo {volume} {14}},\ \bibinfo
  {pages} {3237} (\bibinfo {year} {2023})}\BibitemShut {NoStop}%
\bibitem [{\citenamefont {Jammi}\ \emph {et~al.}(2024)\citenamefont {Jammi},
  \citenamefont {Ferdinand}, \citenamefont {Luo}, \citenamefont {Newman},
  \citenamefont {Spektor}, \citenamefont {Song}, \citenamefont {Koksal},
  \citenamefont {Rakholia}, \citenamefont {Lunden}, \citenamefont {Sheredy},
  \citenamefont {Patel}, \citenamefont {Boyd}, \citenamefont {Zhu},
  \citenamefont {Agrawal}, \citenamefont {Briles},\ and\ \citenamefont
  {Papp}}]{Jammi24tio2}%
  \BibitemOpen
  \bibfield  {author} {\bibinfo {author} {\bibfnamefont {S.}~\bibnamefont
  {Jammi}}, \bibinfo {author} {\bibfnamefont {A.~R.}\ \bibnamefont
  {Ferdinand}}, \bibinfo {author} {\bibfnamefont {Z.}~\bibnamefont {Luo}},
  \bibinfo {author} {\bibfnamefont {Z.~L.}\ \bibnamefont {Newman}}, \bibinfo
  {author} {\bibfnamefont {G.}~\bibnamefont {Spektor}}, \bibinfo {author}
  {\bibfnamefont {J.}~\bibnamefont {Song}}, \bibinfo {author} {\bibfnamefont
  {O.}~\bibnamefont {Koksal}}, \bibinfo {author} {\bibfnamefont {A.~V.}\
  \bibnamefont {Rakholia}}, \bibinfo {author} {\bibfnamefont {W.}~\bibnamefont
  {Lunden}}, \bibinfo {author} {\bibfnamefont {D.}~\bibnamefont {Sheredy}},
  \bibinfo {author} {\bibfnamefont {P.~B.}\ \bibnamefont {Patel}}, \bibinfo
  {author} {\bibfnamefont {M.~M.}\ \bibnamefont {Boyd}}, \bibinfo {author}
  {\bibfnamefont {W.}~\bibnamefont {Zhu}}, \bibinfo {author} {\bibfnamefont
  {A.}~\bibnamefont {Agrawal}}, \bibinfo {author} {\bibfnamefont {T.~C.}\
  \bibnamefont {Briles}},\ and\ \bibinfo {author} {\bibfnamefont {S.~B.}\
  \bibnamefont {Papp}},\ }\bibfield  {title} {\bibinfo {title}
  {Three-dimensional, multi-wavelength beam formation with integrated
  metasurface optics for sr laser cooling},\ }\href
  {https://doi.org/10.1364/OL.526056} {\bibfield  {journal} {\bibinfo
  {journal} {Optics Letters}\ }\textbf {\bibinfo {volume} {49}},\ \bibinfo
  {pages} {6013} (\bibinfo {year} {2024})}\BibitemShut {NoStop}%
\bibitem [{\citenamefont {Zaidi}\ \emph {et~al.}(2024)\citenamefont {Zaidi},
  \citenamefont {Rubin}, \citenamefont {Meretska}, \citenamefont {Li},
  \citenamefont {Dorrah}, \citenamefont {Park},\ and\ \citenamefont
  {Capasso}}]{Zaidi24tio2}%
  \BibitemOpen
  \bibfield  {author} {\bibinfo {author} {\bibfnamefont {A.}~\bibnamefont
  {Zaidi}}, \bibinfo {author} {\bibfnamefont {N.~A.}\ \bibnamefont {Rubin}},
  \bibinfo {author} {\bibfnamefont {M.~L.}\ \bibnamefont {Meretska}}, \bibinfo
  {author} {\bibfnamefont {L.~W.}\ \bibnamefont {Li}}, \bibinfo {author}
  {\bibfnamefont {A.~H.}\ \bibnamefont {Dorrah}}, \bibinfo {author}
  {\bibfnamefont {J.-S.}\ \bibnamefont {Park}},\ and\ \bibinfo {author}
  {\bibfnamefont {F.}~\bibnamefont {Capasso}},\ }\bibfield  {title} {\bibinfo
  {title} {Metasurface-enabled single-shot and complete mueller matrix
  imaging},\ }\bibfield  {journal} {\bibinfo  {journal} {Nat.~Photonics}\
  }\href {https://doi.org/10.1038/s41566-024-01426-x}
  {10.1038/s41566-024-01426-x} (\bibinfo {year} {2024})\BibitemShut {NoStop}%
\bibitem [{\citenamefont {Dainese}\ \emph {et~al.}(2024)\citenamefont
  {Dainese}, \citenamefont {Marra}, \citenamefont {Cassara}, \citenamefont
  {Portes}, \citenamefont {Oh}, \citenamefont {Yang}, \citenamefont {Palmieri},
  \citenamefont {Rodrigues}, \citenamefont {Dorrah},\ and\ \citenamefont
  {Capasso}}]{Dainese2024shape}%
  \BibitemOpen
  \bibfield  {author} {\bibinfo {author} {\bibfnamefont {P.}~\bibnamefont
  {Dainese}}, \bibinfo {author} {\bibfnamefont {L.}~\bibnamefont {Marra}},
  \bibinfo {author} {\bibfnamefont {D.}~\bibnamefont {Cassara}}, \bibinfo
  {author} {\bibfnamefont {A.}~\bibnamefont {Portes}}, \bibinfo {author}
  {\bibfnamefont {J.}~\bibnamefont {Oh}}, \bibinfo {author} {\bibfnamefont
  {J.}~\bibnamefont {Yang}}, \bibinfo {author} {\bibfnamefont {A.}~\bibnamefont
  {Palmieri}}, \bibinfo {author} {\bibfnamefont {J.~R.}\ \bibnamefont
  {Rodrigues}}, \bibinfo {author} {\bibfnamefont {A.~H.}\ \bibnamefont
  {Dorrah}},\ and\ \bibinfo {author} {\bibfnamefont {F.}~\bibnamefont
  {Capasso}},\ }\bibfield  {title} {\bibinfo {title} {Shape optimization for
  high efficiency metasurfaces: theory and implementation},\ }\href
  {https://doi.org/10.1038/s41377-024-01629-5} {\bibfield  {journal} {\bibinfo
  {journal} {Light: Science \& Applications}\ }\textbf {\bibinfo {volume}
  {13}},\ \bibinfo {pages} {300} (\bibinfo {year} {2024})}\BibitemShut
  {NoStop}%
\bibitem [{\citenamefont {He}\ \emph {et~al.}(2025)\citenamefont {He},
  \citenamefont {Li}, \citenamefont {Li}, \citenamefont {Liang}, \citenamefont
  {Feng}, \citenamefont {Zhu}, \citenamefont {Xie}, \citenamefont {Dong},
  \citenamefont {Shi}, \citenamefont {Dun}, \citenamefont {Wei}, \citenamefont
  {Wang},\ and\ \citenamefont {Cheng}}]{He2025per}%
  \BibitemOpen
  \bibfield  {author} {\bibinfo {author} {\bibfnamefont {T.}~\bibnamefont
  {He}}, \bibinfo {author} {\bibfnamefont {D.}~\bibnamefont {Li}}, \bibinfo
  {author} {\bibfnamefont {C.}~\bibnamefont {Li}}, \bibinfo {author}
  {\bibfnamefont {H.}~\bibnamefont {Liang}}, \bibinfo {author} {\bibfnamefont
  {C.}~\bibnamefont {Feng}}, \bibinfo {author} {\bibfnamefont {J.}~\bibnamefont
  {Zhu}}, \bibinfo {author} {\bibfnamefont {L.}~\bibnamefont {Xie}}, \bibinfo
  {author} {\bibfnamefont {S.}~\bibnamefont {Dong}}, \bibinfo {author}
  {\bibfnamefont {Y.}~\bibnamefont {Shi}}, \bibinfo {author} {\bibfnamefont
  {X.}~\bibnamefont {Dun}}, \bibinfo {author} {\bibfnamefont {Z.}~\bibnamefont
  {Wei}}, \bibinfo {author} {\bibfnamefont {Z.}~\bibnamefont {Wang}},\ and\
  \bibinfo {author} {\bibfnamefont {X.}~\bibnamefont {Cheng}},\ }\bibfield
  {title} {\bibinfo {title} {Perfect anomalous refraction metasurfaces
  empowered half-space optical beam scanning},\ }\href
  {https://doi.org/10.1038/s41467-025-58502-1} {\bibfield  {journal} {\bibinfo
  {journal} {Nat.~Commun.}\ }\textbf {\bibinfo {volume} {16}},\ \bibinfo
  {pages} {3115} (\bibinfo {year} {2025})}\BibitemShut {NoStop}%
\bibitem [{\citenamefont {Kwon}\ \emph {et~al.}(2023)\citenamefont {Kwon},
  \citenamefont {Holman}, \citenamefont {Gan}, \citenamefont {Liu},
  \citenamefont {Molinelli}, \citenamefont {Stevenson},\ and\ \citenamefont
  {Will}}]{kwon2023jet}%
  \BibitemOpen
  \bibfield  {author} {\bibinfo {author} {\bibfnamefont {M.}~\bibnamefont
  {Kwon}}, \bibinfo {author} {\bibfnamefont {A.}~\bibnamefont {Holman}},
  \bibinfo {author} {\bibfnamefont {Q.}~\bibnamefont {Gan}}, \bibinfo {author}
  {\bibfnamefont {C.-W.}\ \bibnamefont {Liu}}, \bibinfo {author} {\bibfnamefont
  {M.}~\bibnamefont {Molinelli}}, \bibinfo {author} {\bibfnamefont
  {I.}~\bibnamefont {Stevenson}},\ and\ \bibinfo {author} {\bibfnamefont
  {S.}~\bibnamefont {Will}},\ }\bibfield  {title} {\bibinfo {title}
  {{Jet-loaded cold atomic beam source for strontium}},\ }\href
  {http://dx.doi.org/10.1063/5.0131429} {\bibfield  {journal} {\bibinfo
  {journal} {Rev.~Sci.~Instrum.}\ }\textbf {\bibinfo {volume} {94}},\ \bibinfo
  {pages} {013202} (\bibinfo {year} {2023})}\BibitemShut {NoStop}%
\bibitem [{\citenamefont {Norcia}\ \emph {et~al.}(2018)\citenamefont {Norcia},
  \citenamefont {Young},\ and\ \citenamefont
  {Kaufman}}]{Norcia2018microscopic}%
  \BibitemOpen
  \bibfield  {author} {\bibinfo {author} {\bibfnamefont {M.~A.}\ \bibnamefont
  {Norcia}}, \bibinfo {author} {\bibfnamefont {A.~W.}\ \bibnamefont {Young}},\
  and\ \bibinfo {author} {\bibfnamefont {A.~M.}\ \bibnamefont {Kaufman}},\
  }\bibfield  {title} {\bibinfo {title} {{Microscopic control and detection of
  ultracold strontium in optical-tweezer arrays}},\ }\href
  {http://dx.doi.org/10.1103/PhysRevX.8.041054} {\bibfield  {journal} {\bibinfo
   {journal} {Phys.~Rev.~X}\ }\textbf {\bibinfo {volume} {8}},\ \bibinfo
  {pages} {041054} (\bibinfo {year} {2018})}\BibitemShut {NoStop}%
\end{thebibliography}
\end{document}